\documentclass[%
reprint, 
 amsmath,amssymb,
 aps,
prd,
]{revtex4-1}

\pdfoutput=1

\usepackage{atlasphysics}
\usepackage{graphicx}
\usepackage{subfigure}
\usepackage{dcolumn}
\usepackage{bm}
\usepackage[hyperindex,breaklinks]{hyperref} 

\usepackage{preprintcover}  
\PreprintCoverPaperTitle{A search for $t\bar{t}$ resonances in the lepton plus jets final state with ATLAS using 4.7~fb$^{-1}$ of $pp$ collisions at $\sqrt{s} = 7$~TeV}  
\PreprintIdNumber{CERN-PH-EP-2013-032}  
\PreprintCoverAbstract{A search for new particles that decay into top quark pairs (\ttbar) is performed with the ATLAS experiment at the LHC using an integrated luminosity of 4.7\,\ifb\ of proton--proton ($pp$) collision data collected at a center-of-mass energy \mbox{$\sqrt{s}=7$~\TeV}. In the $\ttbar \rightarrow WbWb$ decay, the lepton plus jets final state is used, where one $W$ boson decays leptonically and the other hadronically. The \ttbar\ system is reconstructed using both small-radius and large-radius jets, the latter being supplemented by a jet substructure analysis. A search for local excesses in the number of data events compared to the Standard Model expectation in the \ttbar\ invariant mass spectrum is performed. No evidence for a \ttbar\ resonance is found and 95\% credibility-level limits on the production rate are determined for massive states predicted in two benchmark models. The upper limits on the cross section times branching ratio of a narrow \Zprime\ resonance range from 5.1~pb for a boson mass of 0.5~\TeV\ to 0.03~pb for a mass of 3~\TeV. A narrow leptophobic topcolor \Zprime\ resonance with a mass below 1.74~\TeV\ is excluded. Limits are also derived for a broad color-octet resonance with $\Gamma/m =$~15.3\%. A Kaluza--Klein excitation of the gluon in a Randall--Sundrum model is excluded for masses below 2.07~\TeV.}  
\PreprintJournalName{Physical Review D}  

\newcommand{\kommentar}[1]{}
 % ...
\kommentar{%
   the text in the brackets is interpreted as comment...
  }%

\newcommand{\akt}{anti-$k_{t}$}
\newcommand{\kt}{\ensuremath{k_{t}}}
\newcommand{\dsplit}{\ensuremath{\sqrt{d_{12}}}}
\newcommand{\mt}{\ensuremath{m_{\mathrm{T}}}}
\newcommand{\mttbar}{\ensuremath{m_{\ttbar}}}
\newcommand{\mttbarreco}{\ensuremath{m_{\ttbar}^{\mathrm{reco}}}}
\newcommand{\mthad}{\ensuremath{m^{t,\,\mathrm{had}}}}
\newcommand{\mtlep}{\ensuremath{m^{t,\,\mathrm{lep}}}}

\newcommand{\Pythia}{{\sc Pythia}}
\newcommand{\Herwig}{{\sc Herwig}}
\newcommand{\AcerMC}{{\sc AcerMC}}
\newcommand{\Powheg}{{\sc Powheg}}

\newcommand{\gkk}{\ensuremath{g_{\mathrm{KK}}}}

\newcommand{\totlumi}{4.7\,\ifb}
\newcommand{\totlumiwitherr}{$4.7\pm 0.2$\,\ifb}

\newcommand{\ZpexcludedMass}{1.74~\TeV}
\newcommand{\kkgexcludedMass}{2.07~\TeV}

\newcommand{\Zpxseclow}{5.1~pb}
\newcommand{\Zpxsechigh}{0.03~pb}
\newcommand{\kkgxseclow}{5.0~pb}
\newcommand{\kkgxsechigh}{0.08~pb}

\newcommand{\NdataResolvedTot}{$61931$}

\newcommand{\NdataOverlap}{701}

\newcommand{\NdataBoostedTot}{1078}

\begin{document}

\preprint{APS/123-QED}

\title{A search for \ttbar\ resonances in the lepton plus jets final state with ATLAS using 4.7~fb$^{-1}$ of $pp$ collisions at $\sqrt{s} = 7$~\TeV}
\author{The ATLAS Collaboration}






\date{\today}

\begin{abstract}
A search for new particles that decay into top quark pairs (\ttbar) is performed with the ATLAS experiment at the LHC using an integrated luminosity of \totlumi\ of proton--proton ($pp$) collision data collected at a center-of-mass energy \mbox{$\sqrt{s}=7$~\TeV}. 
In the $\ttbar \rightarrow WbWb$ decay, the lepton plus jets final state is used, where one $W$ boson decays leptonically and the other hadronically. 
The \ttbar\ system is reconstructed using both small-radius and large-radius jets, the latter being supplemented by a jet substructure analysis. 
A search for local excesses in the number of data events compared to the Standard Model expectation in the \ttbar\ invariant mass spectrum is performed. 
No evidence for a \ttbar\ resonance is found and 95\% credibility-level limits on the production rate are determined for massive states predicted in two benchmark models. 
The upper limits on the cross section times branching ratio of a narrow \Zprime\ resonance range from \Zpxseclow\ for a boson mass of 0.5~\TeV\ to \Zpxsechigh\ for a mass of 3~\TeV. 
A narrow leptophobic topcolor \Zprime\ resonance with a mass below \ZpexcludedMass\ is excluded. 
Limits are also derived for a broad color-octet resonance with $\Gamma/m =$~15.3\%. 
A Kaluza--Klein excitation of the gluon in a Randall--Sundrum model is excluded for masses below \kkgexcludedMass.

\end{abstract}

\pacs{Valid PACS appear here}
\maketitle


\section{Introduction}

Despite its many successes, the Standard Model (SM) of particle physics is believed to be an effective field theory valid only for energies up to the \TeV\ scale. 
Due to its large mass, the top quark is of particular interest for the electroweak symmetry breaking mechanism and could potentially be connected with new phenomena. 
Several proposed extensions to the SM predict the existence of heavy particles that decay primarily to top quark pairs. 

This paper presents the results of a search for production of new particles decaying to top quark pairs using a dataset corresponding to an integrated luminosity of \totlumi\ of 7~\TeV~center-of-mass energy proton--proton ($pp$) collisions collected by the ATLAS experiment in 2011.
The search is carried out using the lepton plus jets decay channel where the $W$ boson from one top quark decays leptonically (to an electron or a muon, and a neutrino) and the other top quark decays hadronically.  
This search uses a combination of \emph{resolved} and \emph{boosted} reconstruction schemes, defined by the cases when the top-quark decay products are well-separated or merged in the detector, respectively. 
In the resolved reconstruction scheme, the hadronically decaying top quark is identified by two or three distinct small-radius jets, while in the boosted reconstruction scheme one large-radius jet with substructure consistent with jets from a $W$ boson and a $b$-quark is used. 
The boosted reconstruction scheme is more suitable for high-mass \ttbar\ resonances as the top-quark decay products become more collimated. 

Examples of hypothetical models that contain high-mass \ttbar\ resonances are the topcolor assisted technicolor model (TC2)~\cite{Hill:1994hp,topcolor2,Harris:2011ez}, which predicts a leptophobic \Zprime\ particle~\footnote{In common with other experimental searches, the specific model used is the leptophobic scenario, model IV in Ref.~\cite{topcolor2} with $f_1 =1$ and $f_2=0$. The corrections to the Lagrangian discussed in Ref.~\cite{Harris:2011ez} are included.},  
and a Randall--Sundrum (RS) warped extra-dimension model, which predicts a bulk Kaluza--Klein (KK) gluon~\cite{Lillie:2007yh,Lillie:2007ve,Agashe:2006hk,Djouadi:2007eg,Agashe:2007zd}, a color-octet. 
Two specific benchmark models are chosen and are used throughout the rest of the paper.
In the first model, a leptophobic topcolor \Zprime\ particle of width $\Gamma_{\Zprime}/m_{\Zprime}=1.2\%$ is considered as a resonance that is narrow with respect to the detector resolution of typically 10\%. 
In the second model, a KK gluon (\gkk) of width $\Gamma_{\gkk}/m_{\gkk}=15.3\%$ is considered as a resonance that is broad with respect to the detector resolution.    

Previous searches for \ttbar\ resonances were carried out by ATLAS in the lepton plus jets final state with 2~\ifb\ of integrated luminosity at \mbox{$\sqrt{s}=7$~\TeV} using resolved and boosted reconstruction techniques separately~\cite{Aad:2012wm,Arce:1431929}.  
With a resolved reconstruction technique, a \Zprime\ resonance is excluded for \mbox{$0.50<m_{\Zprime} < 0.88$~\TeV}\ and a KK gluon is excluded for $0.50< m_{\gkk} < 1.13$ \TeV, both at 95\% credibility level (CL). 
With a boosted reconstruction technique, a leptophobic \Zprime\ is excluded for $0.60<m_{\Zprime} < 1.15$~\TeV\ and a KK gluon is ruled out for $0.60< m_{\gkk} < 1.5$ \TeV\ at 95\% CL. 
Using both resolved and boosted reconstruction techniques on an integrated luminosity of 5~\ifb\ of lepton plus jets events at \mbox{$\sqrt{s}=7$~\TeV}, the CMS experiment excludes a narrow leptophobic topcolor \Zprime\ resonance in the mass range 0.50--1.49~\TeV\ 
and a KK gluon in the mass range 1.00--1.82~\TeV~\cite{CMS:2012cx}. 
A CMS study conducted on the same dataset, but using dilepton plus jets final states, sets slightly less stringent limits on the narrow \Zprime\ resonance, $0.75 < m_{\Zprime} < 1.3$~\TeV~\cite{cms:2012rq}. 
The ATLAS and CMS experiments also performed searches where the top quark pair decays hadronically, using 4.7~\ifb\ and 5.0~\ifb\ of integrated luminosity respectively at \mbox{$\sqrt{s}=7$~\TeV}.  
ATLAS excludes a narrow \Zprime\ resonance in the mass ranges 0.70--1.00~\TeV\ and 1.28--1.32~\TeV~\cite{ATLAS:2012qa} as well as a broad KK gluon with mass 0.7--1.62~\TeV. 
The CMS collaboration excludes a narrow \Zprime\ particle in the mass range $1.3$--$1.5$\, \TeV~\cite{Chatrchyan:2012ku}. 
The reach of the LHC searches now extends to far higher resonance masses than the Tevatron results~\cite{Aaltonen:2011ts,Aaltonen:2011vi,Abazov:2011gv,Aaltonen:2012af}.  

Using \totlumi\ of integrated luminosity, this analysis improves the previous ATLAS lepton plus jets analyses in that it uses a large-radius $R=1.0$ jet trigger, applies $b$-tagging in the boosted selection, utilizes an optimized method for charged lepton isolation and combines the resolved and boosted reconstruction analyses.

\section{The ATLAS detector}
\label{sec:atlas}

The ATLAS detector~\cite{Aad:2008zzm} is designed to measure the properties of a wide range of \TeV-scale physics processes that may occur in $pp$ interactions. 
It has a cylindrical geometry and close to 4$\pi$ solid-angle coverage. 

ATLAS uses a right-handed coordinate system with its origin at the nominal interaction point (IP) in the center of the detector and the $z$-axis along the beam pipe. 
The $x$-axis points from the IP to the center of the LHC ring, and the $y$-axis points upward.  
Cylindrical coordinates $(r,\phi)$ are used in the transverse plane, $\phi$ being the azimuthal angle around the beam pipe, measured from the $x$-axis. 
The pseudorapidity is defined in terms of the polar angle $\theta$ as $\eta=-\ln\tan(\theta/2)$. 
Transverse momentum and energy are defined in the $x$-$y$-plane, as $\pt=p\cdot \sin(\theta)$ and $\et=E\cdot \sin(\theta)$. 

The inner detector (ID) covers the pseudorapidity range $|\eta | <2.5$ and consists of multiple layers of silicon pixel and microstrip detectors as well as a straw-tube transition radiation tracker ($|\eta | <2.0$), which also provides electron identification information. 
The ID is surrounded by a superconducting solenoid that provides a 2~T axial magnetic field. 

The calorimeter system surrounds the ID and the solenoid and covers the pseudorapidity range \mbox{$|\eta|< 4.9$}. 
It consists of high-granularity lead/liquid-argon (LAr) electromagnetic calorimeters, a steel/scintillator-tile hadronic calorimeter within $|\eta| < 1.7$ and two copper/LAr hadronic endcap calorimeters covering $1.5 < |\eta| < 3.2$. 
The solid-angle coverage is completed out to \mbox{$|\eta|=4.9$} with forward copper/LAr and tungsten/LAr calorimeter modules. 

The muon spectrometer (MS) surrounds the calorimeters, incorporating multiple layers of trigger and tracking chambers in an azimuthal magnetic field produced by an air-core toroid magnet, which enables an independent, precise measurement of muon track momentum for $|\eta| < 2.7$. 
The muon trigger covers the region $|\eta| < 2.4$. 

A three-level trigger system~\cite{Aad:2012xs} 
is employed for the ATLAS detector. 
The first-level trigger is implemented in hardware, using a subset of detector information to reduce the event rate to a design value of 75 kHz. 
This is followed by two software-based trigger levels, which together reduced the event rate to about 300 Hz in 2011.

\section{Data and Monte Carlo samples}
\label{sec:samples}

The data used in this search were collected with the ATLAS detector at the LHC in 2011.  
The data are used only if they were recorded under stable beam conditions and with all relevant subdetector systems operational. 
The data sample used for resolved reconstruction was collected with a single-muon trigger with a transverse momentum threshold of 18~\GeV\ or a single-electron trigger with a transverse momentum threshold of 20~\GeV, which was raised to 22~\GeV\ later in the year.  
The data sample used for boosted reconstruction was collected with a single large-radius ($R= 1.0$) jet trigger with a transverse momentum threshold of 240~\GeV. $R$ is the radius parameter of the \akt\ jet algorithm, which is discussed in Sec.~\ref{sec:objectselection}. 
The jet trigger thresholds are measured at the electromagnetic (EM) energy scale, which, at threshold, is on average 80\% of the true energy scale, increasing with \pt. 
The integrated luminosity for the data sample is \totlumiwitherr. 

Simulated samples are used to predict the contributions from various Standard Model processes to the expected background and to model possible \ttbar\ resonance signals. 
After Monte Carlo event generation, all samples are passed through a {\sc GEANT}4-based~\cite{Agostinelli:2002hh} simulation~\cite{Aad:2010ah} of the ATLAS detector and reconstructed using the same software as for data. 
The effect of multiple $pp$ interactions is included in the simulated samples, and the simulated events are weighted such that the distribution of the average number of $pp$ interactions per bunch crossing agrees with data. 

The primary irreducible background is Standard Model \ttbar\ production, characterized by a smoothly falling invariant mass spectrum. 
It is modeled using the {\sc MC@NLO}~v4.01~\cite{Frixione:2002jk,Frixione:2003ei,Frixione:2010wd} generator with {\sc Herwig} v6.520~\cite{HerwigGC,Corcella:2002jc} for parton showering and hadronization and {\sc Jimmy}~\cite{Butterworth:1996zw} for modeling the multiple parton interactions. 
The CT10~\cite{Lai:2010vv} parton distribution functions (PDFs) are used and the top quark mass is set to 172.5~\GeV. 
Only events in which at least one of the $W$ bosons decays leptonically (including decays to $\tau$ leptons) are produced.  
This corresponds to an effective cross section times branching ratio at approximate NNLO (next-to-next-to-leading-order) of 90.5~pb~\cite{Beneke:2009ye,Aliev:2010zk}, obtained using the calculation described in Sec.~\ref{sec:systematics}. 
Additional \ttbar\ samples, generated with \Powheg~\cite{Frixione:2007vw} interfaced with \Pythia\ or \Herwig, are used to evaluate the model uncertainty in the parton showering and fragmentation, as described in Sec.~\ref{sec:systematics}. 

Single top quark production is modeled using multiple generators. 
Production in the $s$-channel and production with an associated $W$ boson ($Wt$) are modeled with {\sc MC@NLO}/{\sc Herwig}/{\sc Jimmy}~\cite{Frixione:2005vw,Frixione:2008yi} as above.  
Production in the $t$-channel is modeled using the {\sc AcerMC} v3.8~\cite{SAMPLES-ACER} generator and {\sc Pythia} v6.421~\cite{Sjostrand:2006za} for parton showering and hadronization. 
For the $s$- and $t$-channels, events are generated in which the $W$ boson is required to decay leptonically while for the $Wt$ process there is no such requirement. 
The cross section times branching ratios used are based on approximate NNLO calculations: $20.9$~pb ($t$-channel)~\cite{Kidonakis:2011wy}, $15.7$~pb ($Wt$ process)~\cite{Kidonakis:2010ux} and $1.5$~pb ($s$-channel)~\cite{Kidonakis:2010tc}. 

The Standard Model production of $W$ and $Z$ bosons that decay leptonically, accompanied by jets, is an important background. 
This includes decays to $\tau$ leptons. 
These samples are generated with {\sc Alpgen} v2.13~\cite{Mangano:2002ea} with up to five extra partons in the matrix element. 
Modeling of parton showering, hadronization and the underlying event uses {\sc Herwig} and {\sc Jimmy} as for the \ttbar\ samples, and the matching of the matrix element to the parton shower is done using the MLM method~\cite{Alwall:2007fs}. 
The PDFs used are CTEQ6L1~\cite{Pumplin:2002vw}. 
Specific $W$ boson plus heavy-flavor processes ($Wb\bar{b}$, $Wc\bar{c}$ and $Wc$) are generated separately with {\sc Alpgen} and double counting of the heavy-flavor contributions is removed from the $W$ plus light-quark jets samples. 
The $W$+jets samples are normalized to the inclusive NNLO cross sections~\cite{Hamberg:1990np,Gavin:2012sy} and then corrected using data as described in Sec.~\ref{sec:datadrivenbackground}. 
The $Z$+jets production, modeled using {\sc Alpgen}, includes contributions from the interference between photon and $Z$ boson exchanges, and events are required to have a dilepton invariant mass $40 < m_{\ell\ell} < 2000$~\GeV. 
The $Z$$b\bar{b}$ process is generated separately with {\sc Alpgen} and heavy-flavor contribution overlap removal is done as in the $W$+jets case. 

The diboson background is modeled using {\sc Herwig} and {\sc Jimmy} with MRST2007LO$^*$ PDFs~\cite{Sherstnev:2007nd}.  
A filter at generator level requiring the presence of at least one lepton with $\pt\ > 10$ \GeV\ and $|\eta| < 2.8$ is used.  
The NLO cross sections used for the samples before filtering are 17.0~pb for $WW$ production, 5.5~pb for $WZ$ production, and 1.3~pb for $ZZ$ production, estimated with the {\sc MCFM}~\cite{Campbell:1999ah} generator. 
The $WZ$ and $ZZ$ samples also include the off-shell photon contribution decaying to dilepton pairs~\cite{Aad:2010ey}. 

Signal samples of \Zprime\ events are modeled using {\sc Pythia} with CTEQ6L1 PDFs. 
This Monte Carlo sample, where the resonance width is 3\%, is used to interpret the data in the topcolor \Zprime\ model (where the width is 1.2\%) since in both cases the width is negligible compared to the detector resolution of ~10\%. 
The leptophobic topcolor \Zprime\ boson has a branching fraction  to \ttbar\ of 33\% for masses above 700\,\GeV, approaching exactly $1/3$ for very large masses. 
It is marginally smaller at lower masses with the smallest value being 31\% at a mass of 500\,\GeV~\cite{topcolor2,Harris:2011ez}. 
A $K$-factor of 1.3~\cite{Gao:2010bb} is applied to account for NLO effects~\footnote{A recent full NLO calculation~\cite{Caola:2012rs} gives smaller $K$-factors, which can partly be attributed to the use of different parameters than those in Ref.~\cite{Gao:2010bb}. The parameters used for signal generation in this paper corresponds more closely to Ref.~\cite{Gao:2010bb}.}. 
Signal samples of Randall--Sundrum KK gluons are generated via {\sc Madgraph}~\cite{Alwall:2007st} and then showered and hadronized using {\sc Pythia}. 
The width of the KK gluon is 15.3\% of its mass and its branching fraction to \ttbar\ is 92.5\%~\cite{Lillie:2007yh}. 
The production cross section times branching fractions for the two signals can be found in Table~\ref{tab:signalxsec}. 

\begin{table}[tbh]
\begin{center}
\caption{The production cross section times branching fraction (BR) for the resonant signal processes $pp \rightarrow \Zprime{} \rightarrow \ttbar$ in the topcolor model and $pp \rightarrow \gkk \rightarrow \ttbar$ for the KK gluon in a Randall--Sundrum model with warped extra dimensions. 
A $K$-factor of 1.3 has been applied to the \Zprime\ cross section to account for NLO effects. 
}
\vspace{2mm}
\begin{tabular}{rr@{.}lr@{.}l}
\hline \hline
Mass  & \multicolumn{2}{l}{~~~$\Zprime \to \ttbar$} & \multicolumn{2}{l}{~~~$\gkk \to \ttbar$} \\ 
$[$\GeV$]$ & \multicolumn{2}{l}{~~~$\sigma\times$ BR~$\times 1.3$ [pb]} & \multicolumn{2}{l}{~~~$\sigma\times$ BR [pb]} \\ 
\hline
  500  & \hspace{2em} 19&6   &  \hspace{1em} 81&3 \\
 1000  &  1&2   &   4&1 \\
 1500  &  0&13  &   0&50  \\
 2000  &  0&019 &   0&095 \\
 2500  &  0&0030 &   0&026 \\
 3000  &  0&00097&   0&0097 \\
\hline \hline 
\end{tabular}
\label{tab:signalxsec}
\end{center}
\end{table}

\section{Physics object selection}
\label{sec:objectselection}

The physics object selection criteria closely follow those in Ref.~\cite{Aad:2012qf}, the main exceptions being the treatment of charged lepton isolation and the use of large-radius jets. 

Jets are reconstructed using the \akt\ algorithm~\cite{Cacciari:2008gp,Cacciari:2011ma} applied to  topological clusters~\cite{Lampl:2008zz} of calorimeter cells with significant energy above the noise threshold.  
Jets with radius parameters of $R=0.4$ and $R=1.0$ are used.  
For the small-radius $R=0.4$ jets, topological clusters at the EM energy scale are used to form the jets~\cite{Aad:2011he}, while for the large-radius $R=1.0$ jets, locally calibrated topological clusters are used~\cite{ISS-0501,Barillari:2009zza,Aad:2011he}. 
The usage of locally calibrated clusters ensures a more correct description of the energy distribution inside the large-radius jet, which is needed when using jet substructure variables, as described in Sec.~\ref{sec:eventselection}. 
Both the small-radius and large-radius jets have their final transverse momentum and pseudorapidity adjusted with energy- and $\eta$-dependent correction factors.  
These are derived from simulation~\cite{Aad:2011he,ATLAS-CONF-2012-065} and verified using data~\cite{Aad:2011he}.  
The small-radius jets are required to have $\pt > 25$~\GeV\ and $|\eta| < 2.5$, while large-radius jets must have $\pt > 350$~\GeV\ and $|\eta| < 2.0$. 
Above this jet \pt\ value, the large-radius jet trigger is more than 99\% efficient. 
A $\Delta R = \sqrt{(\Delta\eta)^{2}+(\Delta\phi)^{2}}$ separation requirement between small-radius and large-radius jets in the boosted event selection below ensures no double-counting of topological cluster energy.

Small-radius jets are tagged as $b$-jets using a neural-network-based $b$-tagging algorithm that uses as input the results of impact parameter, secondary vertex, and decay topology algorithms~\cite{ATLAS-CONF-2012-043}.  
The operating point chosen for the resolved selection corresponds to an average $b$-tagging efficiency in simulated \ttbar\ events of 70\% and a light-quark rejection factor of 140 for $\pt > 20$~\GeV. 
For events passing the boosted selection criteria, the $b$-tagging efficiency estimated from simulated \ttbar\ events using small-radius jets with $\pt > 25$~\GeV\ is 75\% and the light-quark jet rejection factor is 85. 
The $b$-tag requirement for both the resolved and boosted event selections refers only to small-radius jets.  

Electrons are identified by the shape of the shower in the EM calorimeter and  must have a matching track in the inner detector~\cite{Aad:2011mk}.  
The cluster in the EM calorimeter must satisfy $|\eta| < 2.47$ with the transition region $1.37 < |\eta| < 1.52$ between EM calorimeter sections excluded.  
Electrons are required to be isolated as described below and their longitudinal impact distance ($|z_0|$) from the primary event vertex must be smaller than $2$~mm.  
The primary event vertex is defined as the vertex with the highest sum of the squared \pt\ values of the associated tracks ($\sum p_{\rm T,track}^2$) in the event. 
Electrons within a cone of $\Delta R = 0.4$ with respect to any small-radius jet are removed, which suppresses the background from multi-jet events with non-prompt electrons and removes events where the same calorimeter energy deposits would be counted within two physics objects. 
The electron transverse momentum, \pt, is calculated using the cluster energy and track direction, and must be greater than 25~\GeV\ to ensure a fully efficient trigger. 

Muon candidates are formed by matching reconstructed ID tracks with tracks reconstructed in the MS. 
Only muons with  $|\eta| < 2.5$ are used. 
Muons are required to be isolated as described below and to have $|z_0|<2$~mm. 
For the resolved reconstruction, muons are required to have a separation in $\Delta R$ of at least 0.1 from any small-radius jet.  
The muon momentum is calculated using both the MS and the ID tracks, taking the energy loss in the calorimeter into account. 
The transverse momentum of the muon must be greater than 25~\GeV, well above the trigger threshold, chosen to reduce the multi-jet background without impacting the signal. 

The isolation of charged leptons is typically defined using the transverse energy found in a fixed cone around the lepton~\cite{Aad:2010ey}. 
Because the angle between the charged lepton and the $b$-quark decreases as the top quark is more boosted, a better measure of isolation, named \emph{mini-isolation}, is used~\cite{Rehermann:2010vq,ATL-PHYS-PUB-2010-008}. 
The use of mini-isolation improves the lepton signal efficiency and background rejection with respect to the fixed-cone algorithm. 
For this analysis, mini-isolation is defined as  
\begin{equation}
I^{\ell}_{\mathrm{mini}} = \sum\limits_{\mathrm{tracks}} {\pt^{\mathrm{track}}}~,~~~\Delta R(\ell,\mathrm{track})<K_{\mathrm{T}}/\pt^{\ell},
\label{eq:miniiso}
\end{equation}
where the scalar sum runs over all tracks (except the matched lepton track) that have $\pt^{\mathrm{track}} > 1$~\GeV, 
pass quality selection criteria, and fulfill the $\Delta R(\ell,\mathrm{track})$ requirement shown in Eq.~(\ref{eq:miniiso}). 
Here $\pt^{\ell}$ is the lepton transverse momentum and $K_{\mathrm{T}}$ is an empirical scale parameter set to 10~\GeV, chosen so that the size of the isolation cone is optimal both at high and low \pt.   
The requirement $I_{\mathrm{mini}}^{\ell} / p^{\ell}_{\mathrm{T}}<0.05$ is used, corresponding to 95\% (98\%) efficiency for a muon (electron) in the \pt\ region used for this analysis. 
At low \pt, this mini-isolation criterion corresponds to a tighter isolation requirement than is used in other top quark analyses at ATLAS, and at high \pt\ it is looser. 

The missing transverse momentum, \met, is calculated~\cite{Aad:2012re} from the vector sum of the transverse energy in calorimeter cells associated with topological clusters.  
The direction of the energy deposits is given by the line joining the cell to the interaction point. 
Calorimeter cells are first uniquely associated with a physics object (e.g. electron, jet or muon). 
The transverse energy of each cell is then calibrated according to the type of object to which it belongs, and the vector sum of these is calculated. 
All muon transverse momenta are added and the associated calorimeter cell energies are subtracted. 
Finally, topological clusters not associated with any reconstructed object are calibrated at the EM energy scale and added to the transverse energy sum vector. 

\section{Event selection}
\label{sec:eventselection}

After passing a single-lepton trigger for the resolved reconstruction or a large-radius jet trigger for the boosted reconstruction, events are required to have exactly one isolated lepton with $\pt > 25$~\GeV\ and $I^{\ell}_{\mathrm{mini}}/\pt < 0.05$.  
The reconstructed primary event vertex is required to have at least five tracks with $\pt > 0.4$~\GeV. 
In the electron channel, \met\  must be larger than 30~\GeV\ and the transverse mass larger than 30~\GeV.  
The transverse mass is defined as \mbox{$\mt = \sqrt{2 \pt \met ( 1 - \cos{\Delta \phi})}$}, where \pt\ is the transverse momentum of the charged lepton and $\Delta \phi$ is the azimuthal angle between the charged lepton and the missing transverse momentum, which is assumed to be due to the neutrino. 
In the muon channel, the selection is $\met > 20$~\GeV\ and $\met + \mt >60$~\GeV. 
These selection criteria are chosen to suppress the multi-jet background. 
  
The selection requirements for jets differ for the cases of resolved or boosted reconstruction.  
For the resolved reconstruction, events are required to have at least three small-radius jets with $\pt > 25$~\GeV\ and $|\eta| < 2.5$. 
To reduce the effects of multiple $pp$ interactions in the same bunch crossing, at least 75\% of the scalar sum of the \pt\ of the tracks in each jet (called ``jet vertex fraction'') is required to be associated with the primary vertex. 
If one of the jets has a mass~\cite{Aad:2012ef} above 60~\GeV, it is assumed to contain the two quarks from the hadronic $W$ decay, or one of these quarks and the $b$-quark from the top-quark decay.  
If no jet has a mass above 60~\GeV\ then at least four jets are required, one of which must be tagged as a $b$-jet. 

For the boosted reconstruction, the three partons from the hadronic top-quark decay are expected to have merged into one large-radius jet.  
Thus, at least one large-radius jet with $\pt > 350$~\GeV\ and a mass larger than $100$~\GeV\ is required.  
The constituents of the large-radius jets are then reclustered with the exclusive \kt\ algorithm~\cite{Catani:1993hr,Ellis:1993tq} using {\sc FastJet}~\cite{Cacciari:2011ma} and $R=1.0$.  
The last step in the reclustering of subjets within a jet has the splitting scale $\sqrt{d_{12}}$. 
Generally $d_{ij}=\mathrm{min}(p_{\mathrm{T},i}^2,p_{\mathrm{T},j}^2) \Delta R_{ij}^2 / R^2$, where $i$ and $j$ refer to the two last proto-jets to be merged, $\Delta R_{ij}$ is a measure of the angle between them and $R=1.0$ is the fixed radius parameter~\cite{Catani:1993hr,Ellis:1993tq}. 
The value of $\sqrt{d_{12}}$ is expected to be larger for jets that contain a top-quark decay than for jets from the non-top-quark backgrounds. 
The splitting scale $\sqrt{d_{12}}$ is required to be larger than $40$~\GeV, a criterion that rejects about 40\% of the non-top-quark background, but only 10\% of the \ttbar\ sample. 

The jet formed by the $b$-quark from the semileptonically decaying top quark is selected as a small-radius jet that fulfills the same \pt, \eta\ and jet vertex fraction criteria as used in the resolved reconstruction and has a $\Delta R$ separation smaller than 1.5 from the lepton. 
If more than one jet fulfills these criteria, the one closest to the lepton is chosen. 
Two other requirements are applied to the event:  
the decay products from the two top quarks are required to be well separated through the criteria $\Delta \phi(\ell,j_{1.0}) > 2.3$ and $\Delta R(j_{0.4}, j_{1.0}) > 1.5$, where $j_{0.4}$ and $j_{1.0}$ denote the jets with $R=0.4$ and $R=1.0$, respectively. 
The $\Delta R(j_{0.4}, j_{1.0})$ requirement guarantees that there is no energy overlap between the two jets~\cite{Arce:1431929}. 
The highest-\pt\ large-radius jet passing these criteria is taken as the hadronically decaying top quark candidate.  

Finally, in both selections at least one small-radius $b$-tagged jet is required. 
In the boosted analysis this requirement is independent of any large-radius jet in the event, 
i.e. the $b$-tagged jet may originate from the hadronic top-quark decay and overlap with the large-radius jet or it may originate from the leptonic top-quark decay and thus be near the lepton.

\section{Event reconstruction}

The \ttbar\ candidate invariant mass, $m_{\ttbar}$, is computed from the four-momenta of the two reconstructed top quarks.  
For the semileptonically decaying top quark, in both the resolved and boosted selections, the longitudinal component of the neutrino momentum, $p_{z}$, is computed by imposing a $W$ boson mass constraint on the lepton plus \met\ system \cite{Aaltonen:2010jr,Aad:2012ux}. 
If only one real solution for $p_z$ exists, this is used. 
If two real solutions exist, the solution with the smallest $|p_z|$ is chosen or both are tested, depending on the reconstruction algorithm. 
In events where no real solution is found, the \MET{} vector is rescaled and rotated, applying the minimum variation necessary to find exactly one real solution. 
This procedure is justified since mismeasurement of the \MET{} vector is the likeliest explanation for the lack of a solution to the $p_z$ equation, assuming that the lepton indeed comes from a $W$ boson decay. 

For the resolved reconstruction, a $\chi^2$ minimization algorithm is used to select the best assignment of jets to the hadronically and semileptonically decaying top quarks.
The $\chi^2$ minimization uses the reconstructed top quark and $W$ boson masses as constraints: 
\begin{eqnarray}
\chi^2 &=&  \left[ \frac{m_{jj}-m_W}{\sigma_W} \right]^2 \nonumber \\
       &+&  \left[ \frac{m_{jjb}-m_{jj}-m_{t_{h}-W}}{\sigma_{t_{h}-W}} \right]^2 
       + \left[ \frac{m_{j\ell\nu}-m_{t_{\ell}}}{\sigma_{t_{\ell}}} \right]^2 \nonumber \\ 
       &+&  \left[ \frac{(p_{\mathrm T,jjb}-p_{\mathrm T,j\ell\nu}) - 
            (p_{\mathrm T,t_{h}}-p_{\mathrm T,t_{\ell}})}{\sigma_{\mathrm{diff}\pt}}  \right]^2,
\label{eq:chi2}
\end{eqnarray}
where $t_{h}$ and $t_{\ell}$ denote the hadronically and semileptonically decaying top quarks respectively, and $j$ and $b$ denote the jets originated by the light quarks and $b$-quarks. 
The first term constrains the hadronically decaying $W$ boson.  
The second term corresponds to the invariant mass of the hadronically decaying top quark, but since the invariant mass of the jets from the $W$ candidate ($m_{jj}$) is heavily correlated with the mass of the three jets from the hadronic top quark candidate ($m_{jjb}$), the mass of the hadronically decaying $W$ boson is subtracted to decouple this term from the first one. 
The third term represents the semileptonically decaying top quark, and the last term constrains the transverse momenta of the two top quarks to be similar, as expected for a resonance decay.  
The parameters of Eq.~(\ref{eq:chi2}) ($m_W$,  $m_{t_{h}-W}$, $m_{t_{\ell}}$, $\sigma_W$, $\sigma_{t_{h}-W}$, $\sigma_{t_{\ell}}$, $p_{\mathrm T,t_{h}}-p_{\mathrm T,t_{\ell}}$ and $\sigma_{\mathrm{diff}\pt}$) are determined from Monte Carlo simulation studies comparing partons from the top-quark decay with reconstructed objects~\footnote{The values used are 
$m_W = 83.2$~\GeV,  $m_{t_{h}-W} = 90.9$~\GeV, $m_{t_{\ell}} = 167.6$~\GeV, $\sigma_W = 10.7$~\GeV, $\sigma_{t_{h}-W} = 12.8$~\GeV, $\sigma_{t_{\ell}} = 20.5$~\GeV, $p_{\mathrm T,t_{h}}-p_{\mathrm T,t_{\ell}} =-7.4$~\GeV\ and $\sigma_{\mathrm{diff}\pt} =64.0$~\GeV.}.  
All small-radius jets satisfying the physics object selection requirements of Sec.~\ref{sec:objectselection} and $\pt>20$~\GeV\ are tried and the permutation with the lowest $\chi^2$ is used to calculate $m_{\ttbar}$. 
The correct assignment of the jets to the partons of the \ttbar\ decay ($q$, $\bar{q}'$, $b$, $\bar{b}$) 
is achieved in approximately 65\% of the \ttbar\ events for which all the decay products of the top quarks are in the acceptance of the detector and can be matched to reconstructed objects. 
If one of the jets has a mass larger than $60$~\GeV, the $\chi^2$ is slightly modified to allow the heavy jet to contain either the two light quarks from the $W$ boson decay or one quark from the $W$ boson and the $b$-quark from the top-quark decay. 
The reconstructed \ttbar\ invariant mass, \mttbarreco, in simulated events is shown in Fig.~\ref{fig:resolved_mass_reconstruction} for a selection of \Zprime\ and \gkk\ mass values.

\begin{figure}[tbp]
\centering
\subfigure[~Resolved reconstruction. \label{fig:resolved_mass_reconstruction}]
{
\includegraphics[width=0.45\textwidth]{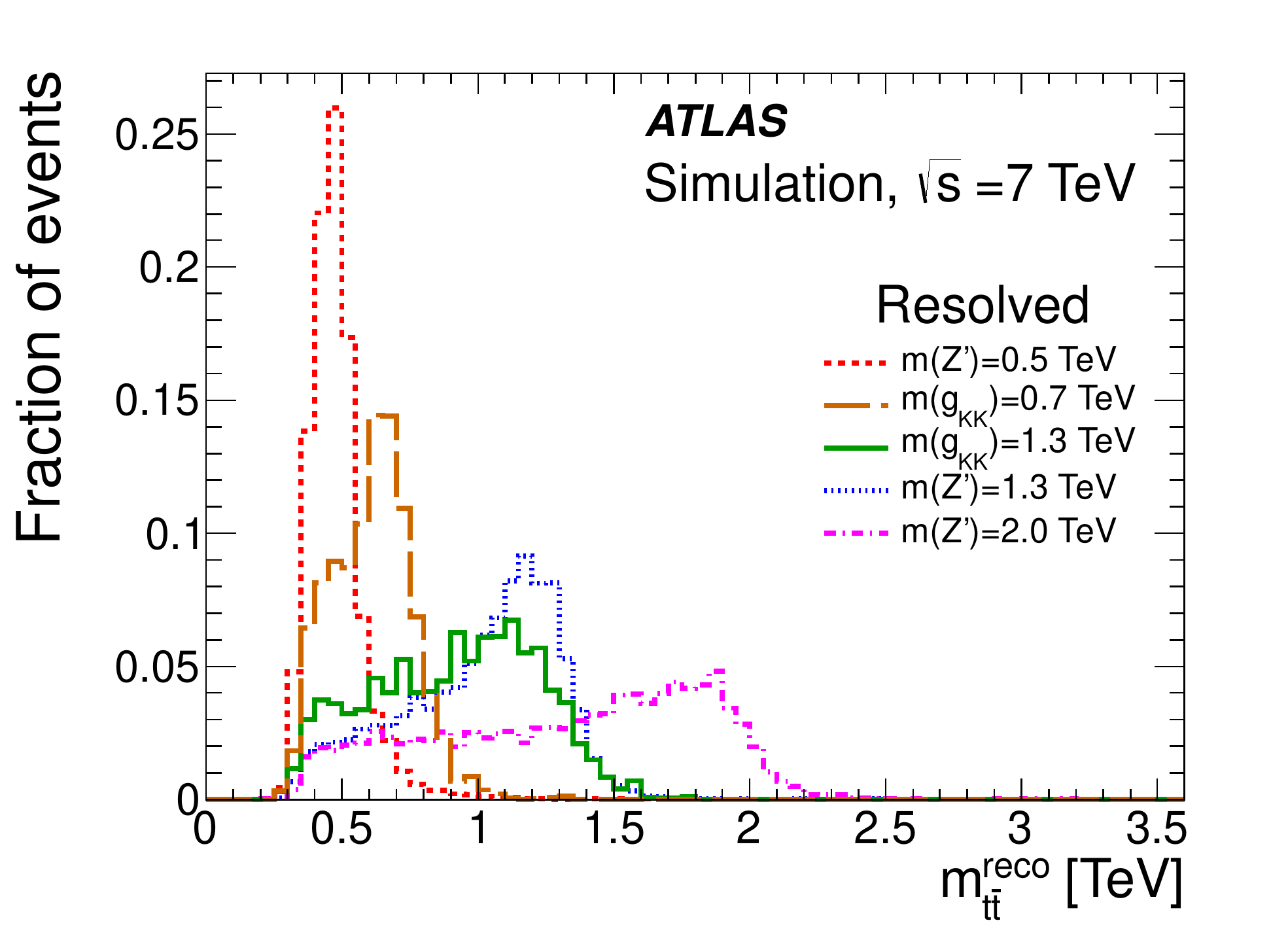} }
\subfigure[~Boosted reconstruction.\label{fig:boosted_mass_reconstruction}]
{
\includegraphics[width=0.45\textwidth]{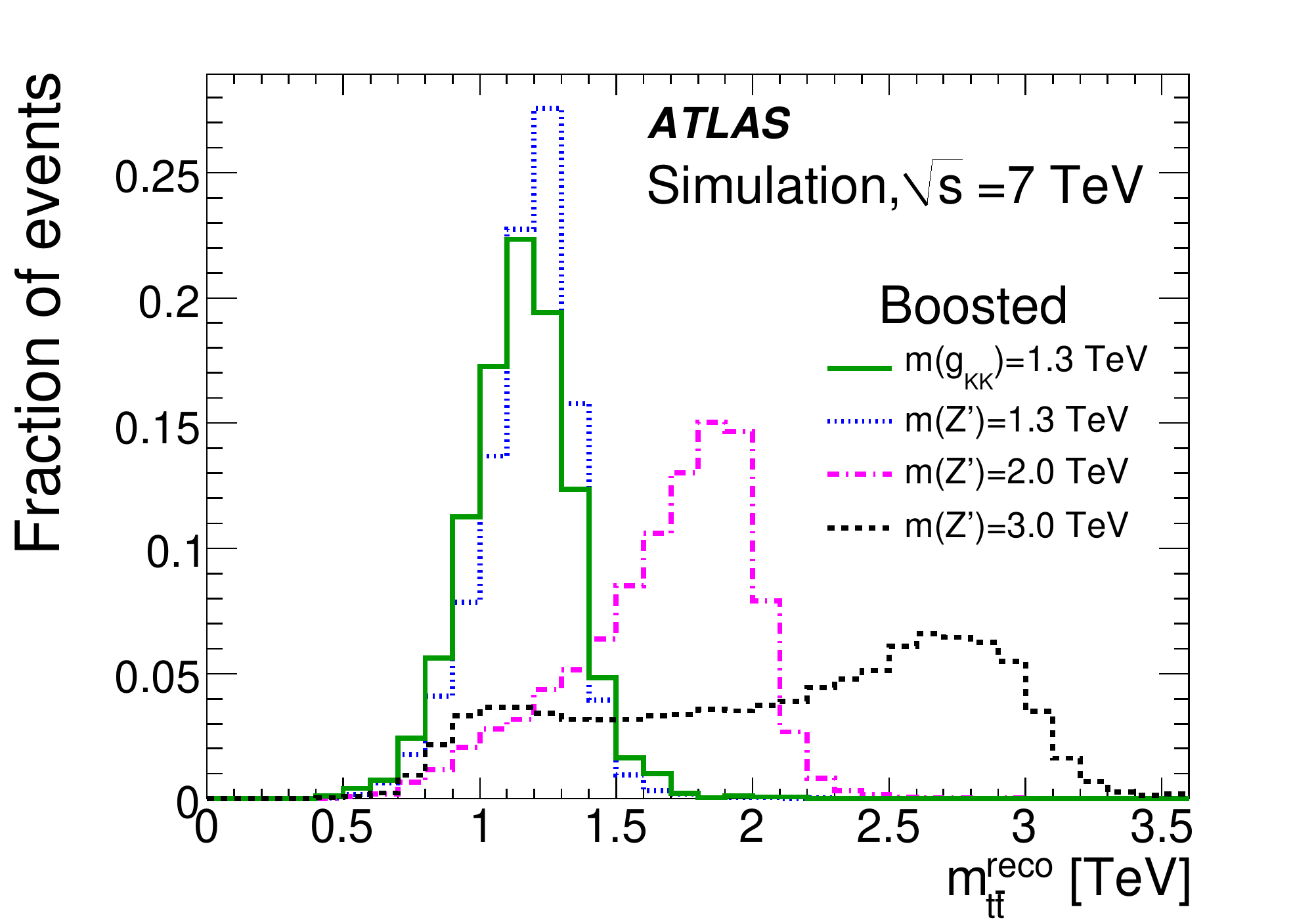} }
\caption{The reconstructed \ttbar\ invariant mass, \mttbarreco, using the (a) resolved and (b) boosted selection, for a variety of simulated \Zprime\ masses ($m(\Zprime)$). The broad Kaluza--Klein gluon resonance at masses 0.7~\TeV\ and 1.3~\TeV\ are also shown for comparison.  
\label{fig:mass_reconstruction}}
\end{figure}

For the boosted reconstruction, the hadronically decaying top-quark four-momentum is taken to be that of the large-radius jet, while the semileptonically decaying top-quark four-momentum is formed from the neutrino solution from the $W$ boson mass constraint, the high-\pt\ lepton and the nearest small-radius jet.  
In this case there is no ambiguity in the assignment of the objects to the original top quarks. 
The reconstructed \ttbar\ invariant mass for a selection of simulated \Zprime\ and \gkk\ mass points is shown in Fig.~\ref{fig:boosted_mass_reconstruction}.

The extended tails at low masses for high-mass resonances in Fig.~\ref{fig:mass_reconstruction} are caused mainly by the convolution of the \Zprime\ line shape and the steeply falling parton distribution functions. 
The $\chi^2$ method sometimes also reconstructs a slightly lower \mttbar\ value in the case of hard final-state radiation, since it tends to select the soft jets from the light quarks in the top-quark decay, rather than hard jets from final-state radiation.

Four independent $m_{\ttbar}$ invariant mass spectra are used to search for \ttbar\ resonances.  
For each of the $e$+jets and $\mu$+jets decay channels, two orthogonal data samples are created. 
The first sample contains all events that pass the boosted event selection. 
For these events, $m_{\ttbar}$ is estimated using the boosted reconstruction. 
This first sample includes events that also pass the resolved event selection.  
For these events the boosted reconstruction is used because of its better reconstructed mass resolution. 
The second sample, referred to as the \emph{resolved selection} in the remainder
of the paper, contains all events that pass the resolved event selection but do not pass the boosted event selection.  

\begin{figure}[htbp]
\centering
\includegraphics[width=0.45\textwidth]{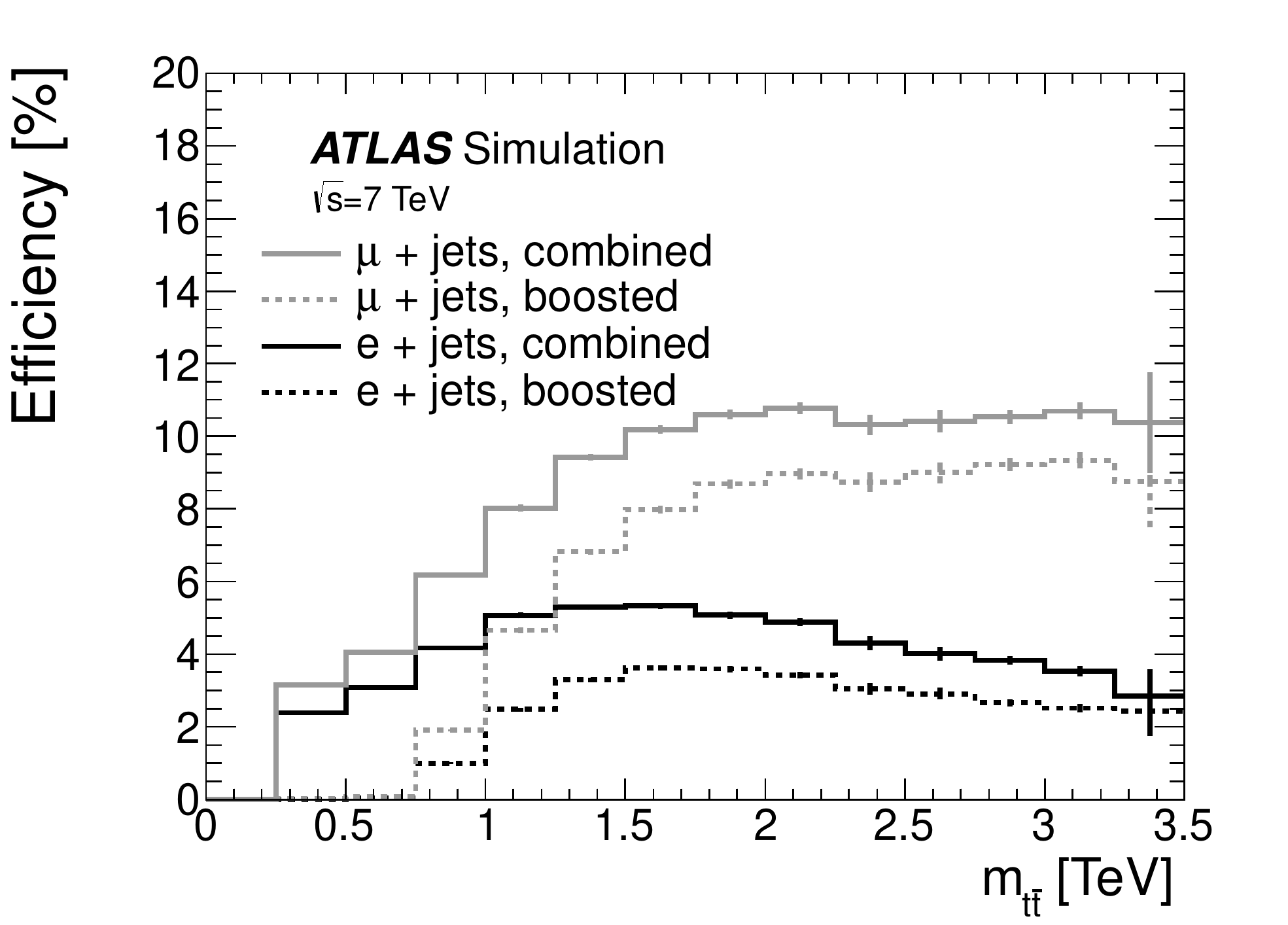}
\caption{The selection efficiency as a function of the true $m_{\ttbar}$ for simulated \Zprime\ resonances at various mass points. The $\mu$+jets channel is shown with gray lines and the $e$+jets channel with black lines. Dashed lines show the boosted selection and solid lines the total selection efficiency. 
\label{fig:SelEffCombinedPaper}}
\end{figure}

The efficiency of the boosted event selection for selecting $\Zprime \rightarrow \ttbar$ events (including all possible \ttbar\ decay channels) as a function of the true invariant mass of the \ttbar\ system is shown in Fig.~\ref{fig:SelEffCombinedPaper}, together with the selection efficiency for all events passing either selection method. 
These efficiencies are given with respect to the full set of $\Zprime \rightarrow \ttbar$ events and they include both the fraction of events within the fiducial acceptance and the fraction of those events that pass the criteria for reconstructed objects, as well as the branching fraction to the various final states. 
At masses below 1~\TeV, the resolved selection is the most efficient, whereas the boosted selection gains in importance at high masses. 
The $e$+jets efficiency drops at high masses, due to the $\Delta R(j,e)>0.4$ requirement, which removes highly collimated top-quark decays. 
The overall selection efficiency is larger for the $\mu$+jets channel because of an inherent larger selection efficiency of muons compared with electrons, and also because of the differences in the requirements on the missing transverse energy and transverse mass.

\section{Backgrounds determined from data}
\label{sec:datadrivenbackground}

The $W$+jets and multi-jet backgrounds and their uncertainties are largely determined from data.
The expected shape of the \mttbarreco\ distribution of the $W$+jets background is estimated using {\sc Alpgen} simulation samples, but the overall normalization and flavor fractions are determined from data. 

The total number of $W$+jets events passing selection criteria in the data, 
$N_{W^+} + N_{W^-}$, is estimated from the observed charge asymmetry in data~\cite{ATLAS:2012an,Aad:2012hg} and the predicted charge asymmetry in $W$+jets events from Monte Carlo simulation:
\begin{equation}
N_{W^+} + N_{W^-} = \left(\frac{r_{\mathrm{MC}} + 1}{r_{\mathrm{MC}} - 1}\right)(D_{\mathrm{corr}+} - D_{\mathrm{corr}-}), 
\label{eq:Wchargeasymm}
\end{equation}
where $r_{\mathrm{MC}}$ is the predicted ratio in Monte Carlo simulation of the $W^+$ to $W^-$ boson cross sections after event selection criteria are applied (but without $b$-tagging) and $D_{\mathrm{corr}+(-)}$ is the number of observed events with a positively (negatively) charged lepton.  
Charge-symmetric contributions from \ttbar\ and $Z$+jets processes cancel in the difference and the contributions from the remaining, slightly charge-asymmetric processes are accounted for by  Monte Carlo simulation. 
To increase the sample size for the boosted selection, the jet mass and $\sqrt{d_{12}}$ requirements are not applied and the \pt\ requirement on the large-radius jet is relaxed to be $>300$~\GeV. 
From stability tests performed by varying the \pt\ requirement, it is concluded that no additional uncertainty for the extrapolation to the signal region is needed. 
The resulting corrections for the $W$+jets yields from Monte Carlo simulation to agree with data are unity within their uncertainties (10\%--20\%) for both the boosted and resolved selections.

Data are also used to determine scale factors for the relative fraction of $W$+jets events with heavy-flavor jets. 
A system of three equations is solved to determine the fractions of events containing two $b$-quarks, two $c$-quarks, one $c$-quark or only light quarks, for each jet multiplicity $i$ of the events. 
The ratio of events containing two $b$-quarks to events with two $c$-quarks is taken from Monte Carlo simulation. The sum of all flavor fractions is constrained to unity.  
By comparing the number of events with $i$ jets before and after $b$-tagging (separately for positively and negatively charged leptons) between data and Monte Carlo simulation, correction factors for the flavor fractions for each jet bin $i$ are determined~\cite{Aad:2012qf,Aad:2011kp,Aad:2012hg}. 

The normalization and shape of the multi-jet backgrounds are determined directly from data using a matrix method~\cite{Aad:2012qf} for both the resolved and boosted selections. 
The multi-jet backgrounds include all background sources from processes with non-prompt leptons or jets misreconstructed as leptons, including the fully hadronic decays of $W$ and $Z$ bosons and \ttbar\ pairs (both from Standard Model production and from possible signal). 
The matrix method uses efficiencies, measured in data, associated with prompt leptons (from $W$ and $Z$ boson decays) and non-prompt leptons (from multi-jet events) passing the required isolation criterion. 
An alternative method, called the jet--electron method~\cite{Aad:2012ux} is used to estimate the systematic uncertainty of the normalization and shape of the invariant mass spectrum associated with this background. 
Consistency checks comparing the matrix method with the jet--electron method show that the normalization uncertainty is 60\%, for both the resolved and boosted selections, and that the impact of the shape uncertainty is negligible. 

For both selection methods, the modeling of the multi-jet contribution is validated using a multi-jet-enriched control region with $\met<50$~\GeV\ and $\mt < 50$~\GeV. 
For muons, both selection methods require a transverse impact parameter significance $|d_0|/\sigma(d_0)>4$  to enhance the fraction of heavy-flavor jets in the sample, which is the dominant source of multi-jet events after $b$-tagging. 
For the boosted selection, at least one large-radius jet with $\pt>150$~\GeV\ is also required and the jet mass and \kt\ splitting scale requirements are inverted to $m_{\mathrm{jet}}<100$~\GeV\ and $\dsplit<40$~\GeV. 
The control region for the boosted selection is disjoint from the signal region, while 14\% (6\%) of events from the control region for the resolved selection also pass the signal region criteria for the $e$+jets ($\mu$+jets) channel.
Within the quoted systematic uncertainties, the modeling of the multi-jet background agrees with the data in the control regions, as shown in Fig.~\ref{fig:MMmtt}. 
\begin{figure*}[tbp]
\begin{center}
\subfigure[~Resolved selection, $e$+jets channel.]{
\label{F:MMmtteRes}
\includegraphics[width=0.45\textwidth]{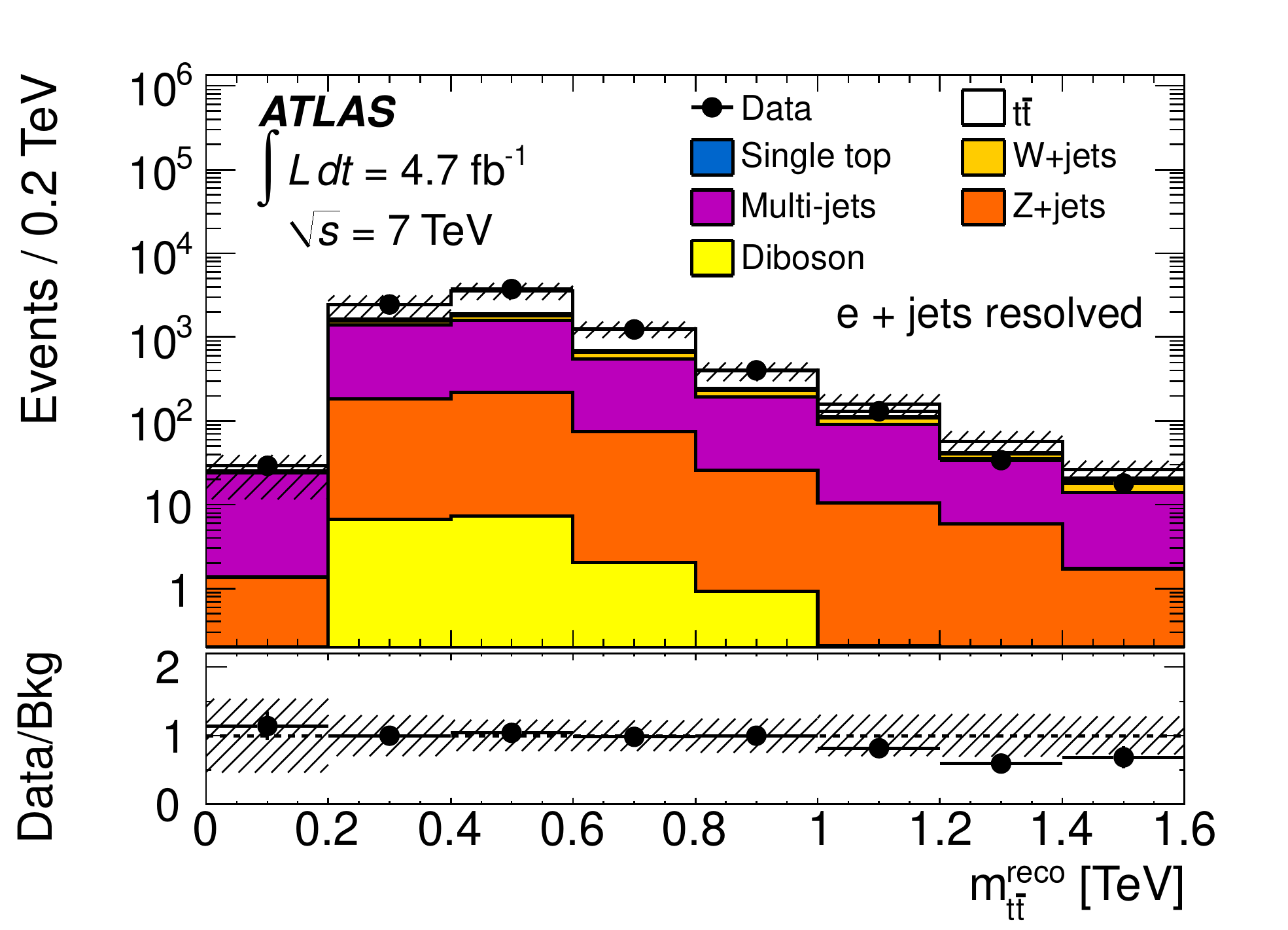} }
\subfigure[~Resolved selection, $\mu$+jets channel.]{
\label{F:MMmttmuRes}
\includegraphics[width=0.45\textwidth]{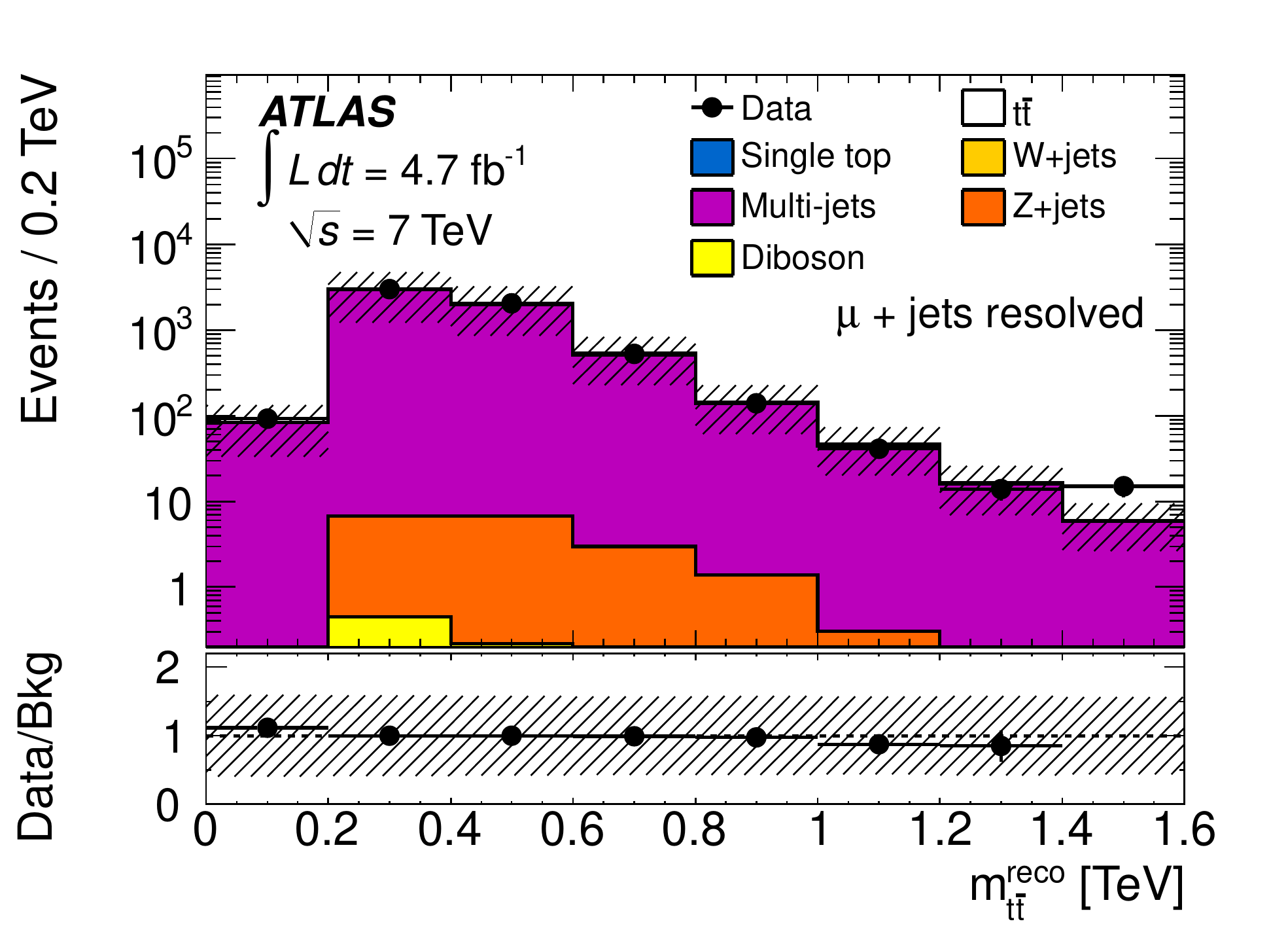} }
\subfigure[~Boosted selection, $e$+jets channel.]{
\label{F:MMmtteBoo}
\includegraphics[width=0.45\textwidth]{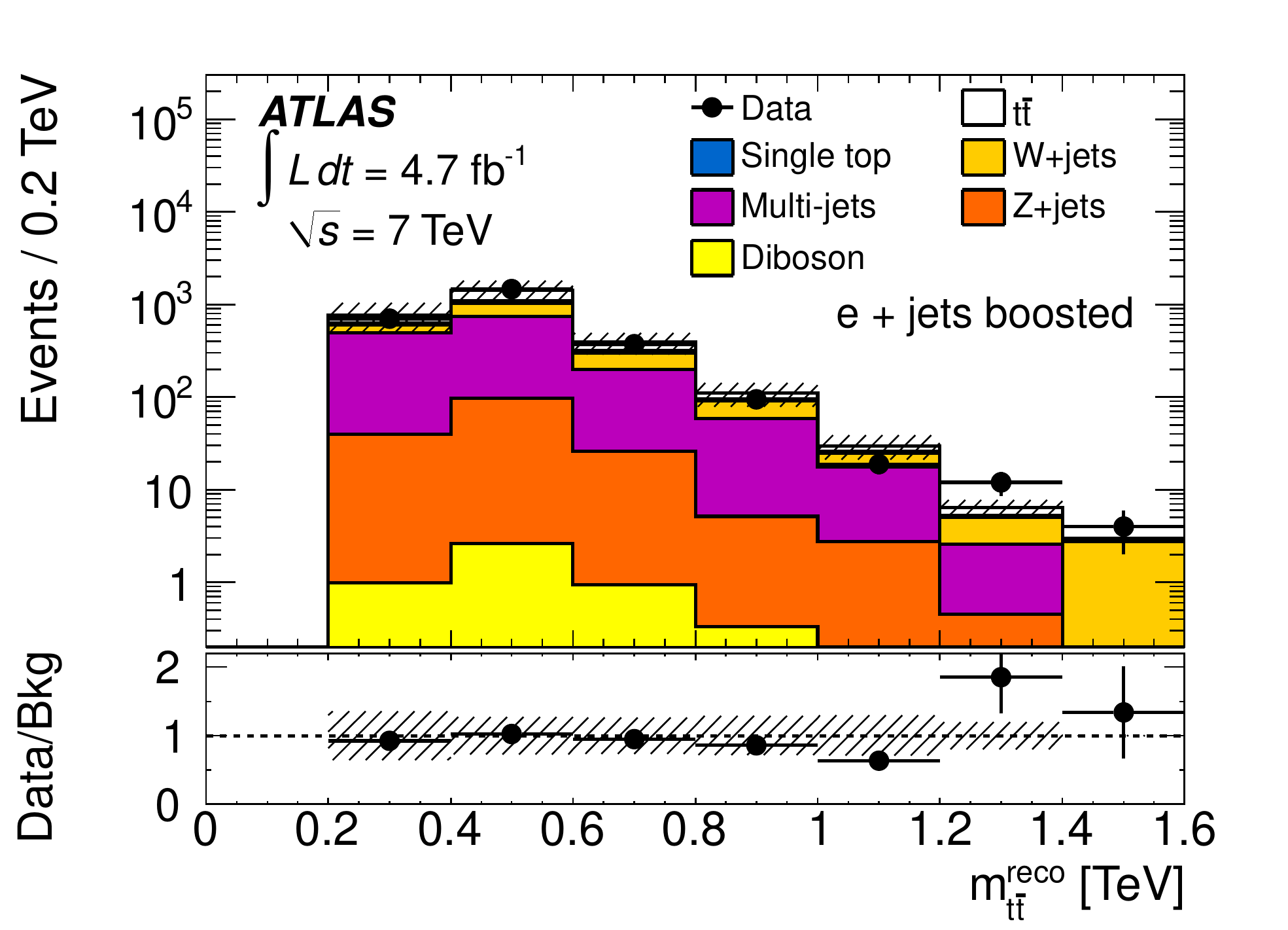} }
\subfigure[~Boosted selection, $\mu$+jets channel.]{
\label{F:MMmttmuBoo}
\includegraphics[width=0.45\textwidth]{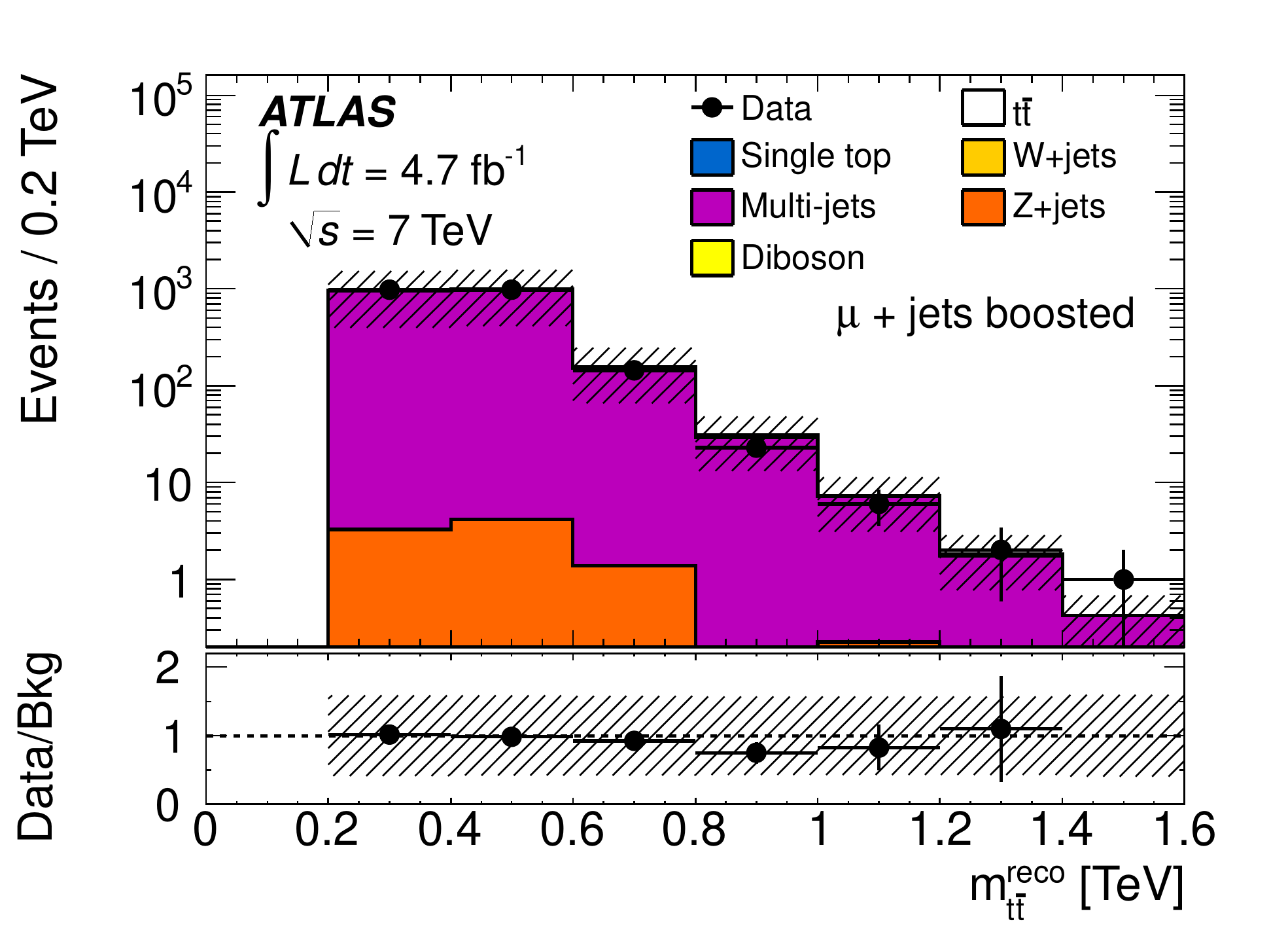} }
\caption{The reconstructed \ttbar\ invariant mass, \mttbarreco, in the multi-jet control regions for the resolved (a, b) and boosted (c, d) selections. The 60\% uncertainty on the multi-jet contribution is indicated as the shaded area. The multi-jet fraction is significantly larger for the $\mu$+jets channel than for the $e$+jet channel because of the impact parameter requirement on the muons, which suppresses prompt muons. 
\label{fig:MMmtt}}
\end{center}
\end{figure*}

\section{Systematic uncertainties}
\label{sec:systematics}

The final observables are the four \ttbar\ invariant mass spectra (two selections and two decay channels). 
The uncertainties can be broadly divided into two categories: uncertainties that affect reconstructed physics objects (such as jets) and uncertainties that affect the modeling of certain backgrounds or signals. 
Some of the uncertainties affect both the shape and the normalization of the spectrum, while others affect the normalization only. 

The dominant normalization uncertainty on the total background is the Standard Model \ttbar\ cross-section uncertainty of 11\%.  
The uncertainty has been calculated at approximate NNLO in QCD~\cite{Beneke:2009ye} with {\sc Hathor} 1.2~\cite{Aliev:2010zk} using the MSTW2008 90\% confidence-level NNLO PDF sets~\cite{Martin:2009iq} and PDF+$\alpha_S$ uncertainties according to the MSTW prescription~\cite{Martin:2009bu}.  
These uncertainties are then added in quadrature to the normalization and factorization scale uncertainty and found to give results consistent with the NLO+NNLL calculation of Ref.~\cite{Cacciari:2011hy} as implemented in Top++ 1.0~\cite{Czakon:2011xx}. 
A cross section uncertainty from varying the top-quark mass by $\pm$1~\GeV, added in quadrature to the scale and PDF+$\alpha_S$ uncertainties, is also included. 

The $W$+jets normalization is varied within the uncertainty, dominated by statistics, of the data-driven determination, corresponding to 12\% (10\%) for the resolved selection in the $e$+jets ($\mu$+jets) channel and 19\% (18\%) for the boosted selection in the $e$+jets ($\mu$+jets) channel. 
Four variations of the flavor composition are considered, including the statistical uncertainty on their data-driven determination, the uncertainty on the extrapolation to different jet multiplicities, and the correlations between different flavor fractions, giving a change in the $W$+jets event yield of about 10\% per variation. 
The normalization uncertainty on the multi-jet background is $60$\%, coming from the difference between the matrix and jet--electron methods. 
The single top quark background uncertainty~\cite{Kidonakis:2011wy,Kidonakis:2010tc,Kidonakis:2010ux} is $7.7$\%. 
The normalization uncertainty of the $Z$+jets sample is $48$\%, estimated using Berends--Giele scaling~\cite{Aad:2010ey}. 
The diboson normalization uncertainty is 34\%, which is a combination of the PDF uncertainty and additional uncertainties from each extra selected jet. 

The preliminary estimate of the 2011 luminosity uncertainty of $3.9$\% is used, based on the techniques explained in Ref.~\cite{Aad:2013ucp}, and is applied to the signal samples and all backgrounds except multi-jets and $W$+jets, which are estimated from data. 

The variation in the shape of the \ttbar\ mass spectrum due to the next-to-leading-order scale variation is accounted for as a mass-dependent scaling, obtained by varying the renormalization and factorization scales up and down by a factor of two in MC@NLO, and normalizing to the nominal cross section (described in Sec.~\ref{sec:samples}). 
The resulting uncertainties range from 10\% of the \ttbar\ background at low $\mttbar$ to 20\% at masses above 1~\TeV. 
The PDF uncertainty on all Monte Carlo samples is estimated by taking the envelope of the MSTW2008NLO, NNPDF2.3~\cite{Ball:2012cx} and CT10 PDF set uncertainties at 68\% confidence-level~\footnote{The CT10 PDF uncertainties are scaled down by a factor 1.6645 to reach the 68\% confidence-level.} following the PDF4LHC recommendation~\cite{Botje:2011sn} and normalizing to the nominal cross section. 
The PDF uncertainty has a much larger effect on the \ttbar\ mass spectrum in the boosted sample than in the resolved sample, 
with variations in the number of \ttbar\ events increasing from 5\% at 1~\TeV\ to over 50\% above 2~\TeV, due primarily to the larger relative PDF uncertainties in the higher-mass (higher partonic $x$) regime. 
The effect on the total background from the PDF variations is 4.7\% (7.3\%) after the resolved (boosted) selection. 

One of the dominant uncertainties affecting reconstructed physics objects is the jet energy scale (JES) uncertainty, especially for large-radius jets~\cite{Aad:2012ef,ATLAS-CONF-2012-065}, which has an effect of 17\% on the background yield in the boosted selection. 
This uncertainty also includes variations in the jet mass scale (JMS) and the \kt\ splitting scales within their uncertainties~\cite{ATLAS-CONF-2012-065}. 
The uncertainty is smaller for the resolved selection, since the large-radius jets are only used indirectly there, in the veto of events that pass the boosted selection. 
For small-radius jets, the uncertainties in the JES, the jet reconstruction efficiency and the jet energy resolution (JER) are considered~\cite{Aad:2011he}. 
The $b$-tagging uncertainty is modeled through simultaneous variations of the uncertainties on the efficiency and rejection scale factors~\cite{ATLAS-CONF-2012-043,ATLAS-CONF-2012-040}. 
An additional $b$-tagging uncertainty is applied for high-momentum jets ($\pt > 200$~\GeV) to account for uncertainties in the modeling of the track reconstruction in dense environments with high track multiplicities~\footnote{The additional $b$-tagging uncertainty is an extrapolation of uncertainty from regions of lower \pt, and it is approximately 12\% for $b$-jets and 17\% for $c$-jets, added in quadrature with the jet efficiency correction factor for the 140--200 \GeV\ region. 
}.  
The effect of uncertainties associated with the jet vertex fraction is also considered. 

The uncertainty on the Standard Model \ttbar\ background due to uncertainties in the modeling of QCD initial- and final-state radiation (ISR/FSR) is estimated using \AcerMC~\cite{SAMPLES-ACER} plus \Pythia\ Monte Carlo samples by varying the \Pythia\ ISR and FSR parameters while retaining consistency with a previous ATLAS measurement of \ttbar\ production with a veto on additional central jet activity~\cite{ATLAS:2012al}. 
The magnitude of the variations comes from a measurement of extra radiation in top quark events. 
Higher-order electroweak virtual corrections to the \ttbar\ mass spectrum have been estimated in Ref.~\cite{Manohar:2012rs} and are used as an estimate of the systematic uncertainty of the \ttbar\ Monte Carlo sample normalization. 
The parton showering and fragmentation uncertainty on the \ttbar\ background is estimated by comparing the result from samples generated with \Powheg\ interfaced with \Pythia\ or \Herwig\ for the parton showering and hadronization. 

For the $W$+jets background, the uncertainty on the shape of the mass distribution is estimated by varying the parameterization of the renormalization and factorization scales~\cite{Mangano:2002ea}. 

The shape uncertainty of the multi-jet background is estimated by comparing the matrix method and the jet--electron method, and its impact on the expected upper cross-section limit of the signal models (discussed in Sec.~\ref{sec:Results}) is found to be negligible.  

For the leptons, the uncertainties on the mini-isolation efficiency, the single-lepton trigger and the reconstruction efficiency are estimated using $Z \rightarrow ee$ and $Z \rightarrow \mu\mu$ events. 
The difference between $Z$ boson and \ttbar\ events is part of the mini-isolation uncertainty. 
Uncertainties on the \MET\ reconstruction, as well as on the energy scale and energy resolution of the leptons are also considered, and generally have a smaller impact on the yield and the expected limits than the uncertainties mentioned above.

In Table~\ref{tab:systimpact}, an overview of the effects of the dominant systematic uncertainties on the background and signal yields is given. 
Only the impact on the overall normalization is shown in the table.  
Some of the systematic uncertainties also have a significant dependence on the reconstructed \ttbar\ mass and this is fully taken into account in the analysis.

\begin{table}[tbh]
\begin{center}
\caption{Average uncertainty from the dominant systematic effects on the total background yield and on the estimated yield of a \Zprime\ with $m=1.6$~\TeV. 
The $e$+jets and $\mu$+jets spectra are added. 
The shift is given in percent of the nominal value. 
The error on the yield from all systematic effects is estimated as the quadratic sum of all systematic uncertainties. 
Certain systematic effects are not relevant for the \Zprime\ samples, which is indicated with a bar ($-$) in the table. 
}
\vspace{2mm}

\footnotesize{
\begin{tabular}{lrrrr}
\hline \hline
                  & \multicolumn{2}{c}{Resolved selection} & \multicolumn{2}{c}{Boosted selection} \\
                  & \multicolumn{2}{c}{uncertainty [\%]} & \multicolumn{2}{c}{uncertainty [\%]}  \\
Systematic effect & tot.~bkg. & \Zprime& tot.~bkg. & \Zprime\\
\hline
Luminosity                    & $3.3$& $3.9$& $3.5$& $3.9$\\
PDF                           & $4.7$& $3.2$& $7.3$& $1.5$\\
ISR/FSR                       & $0.5$& $-$  & $0.9$& $-$  \\
Parton shower and fragm.      & $0.1$& $-$  & $7.4$& $-$  \\
\ttbar\ normalization         & $8.2$& $-$  & $9.0$& $-$  \\
\ttbar\ EW virtual correction & $1.9$& $-$  & $4.2$& $-$  \\
\ttbar\ NLO scale variation   & $1.2$& $-$  & $8.9$& $-$  \\
$W$+jets $bb$+$cc$+$c$ vs.~light&$1.7$&$-$  & $1.1$& $-$  \\
$W$+jets $bb$ variation       & $1.3$& $-$  & $1.1$& $-$  \\
$W$+jets $c$ variation        & $0.8$& $-$  & $0.1$& $-$  \\
$W$+jets normalization        & $1.3$& $-$  & $1.5$& $-$  \\
Multi-jets norm, $e$+jets     & $1.7$& $-$  & $0.4$& $-$  \\
Multi-jets norm, $\mu$+jets   & $1.0$& $-$  & $1.1$& $-$  \\
JES, small-radius jets        & $7.9$& $3.1$& $0.6$& $0.4$\\
JES+JMS, large-radius jets    & $0.2$& $4.7$&$17.3$& $2.8$\\  
Jet energy resolution         & $1.3$& $0.7$& $0.5$& $0.2$\\
Jet vertex fraction           & $1.4$& $1.8$& $1.9$& $1.9$\\
$b$-tag efficiency            & $3.8$& $7.9$& $6.1$& $3.7$\\
$c$-tag efficiency            & $1.2$& $0.6$& $0.1$& $2.6$\\
Mistag rate                   & $1.0$& $0.3$& $0.6$& $0.1$\\
Electron efficiency           & $0.6$& $0.7$& $0.5$& $0.5$\\
Muon efficiency               & $0.9$& $0.9$& $0.6$& $0.6$\\
\hline 
All systematic effects        &$14.1$&$11.2$&$25.4$& $7.1$\\
\hline \hline
\end{tabular}
}
\label{tab:systimpact}
\end{center}
\end{table}

\section{Comparison of data and the Standard Model prediction}
\label{sec:comparison}

After all event selection criteria are applied, \NdataResolvedTot\ resolved and \NdataBoostedTot\ boosted events remain.  
A total of \NdataOverlap\ events pass both sets of selection criteria, and in the analysis they are treated as boosted events. 
The event yields from data and from the expected backgrounds for \totlumi\ 
are listed in Table~\ref{tab:datamcyield}, along with the total systematic uncertainties, described in Sec.~\ref{sec:systematics}. 

\begin{table}[tbh!]
\begin{center}
\caption{Data and expected background event yields after the resolved and boosted selections. The total systematic uncertainty of the expected background yields is listed. 
}
\vspace{2mm}
\begin{tabular}{lr@{$~\pm~$}lr@{$~\pm~$}l}
\hline \hline
Type          & \multicolumn{2}{c}{Resolved selection} & \multicolumn{2}{c}{Boosted selection} \\  
\hline
\ttbar           & 44200 & 7000 & 940 & 260 \\ 
Single top       &  3200 &  500 &  50 &  10 \\ 
Multi-jets $e$   &  1600 & 1000 &   8 &   5 \\ 
Multi-jets $\mu$ &  1000 &  600 &  19 &  11 \\ 
$W$+jets         &  7000 & 2200 &  90 &  30 \\ 
$Z$+jets         &   800 &  500 &  11 &   6 \\ 
Dibosons         &   120 &   50 &  0.9&   0.6 \\
\hline
Total            & 58000 & 8000 &1120 & 280 \\ 
\hline
Data & \multicolumn{2}{c}{\NdataResolvedTot{}} &\multicolumn{2}{c}{\NdataBoostedTot{}} \\
\hline \hline
\end{tabular}
\label{tab:datamcyield}
\end{center}
\end{table} 

Figures~\ref{fig:jet0pt_resolved} and~\ref{fig:top_mass_pt_boosted} show the transverse momentum of the leading (small-radius) jet after the resolved selection and the transverse momentum of the selected large-radius jet after the boosted selection, respectively. 
\begin{figure}[tbp]
\begin{center}
\subfigure[~$e$+jets channel.]{
\label{F:jet0pt_resolved_ele}
\includegraphics[width=0.45\textwidth]{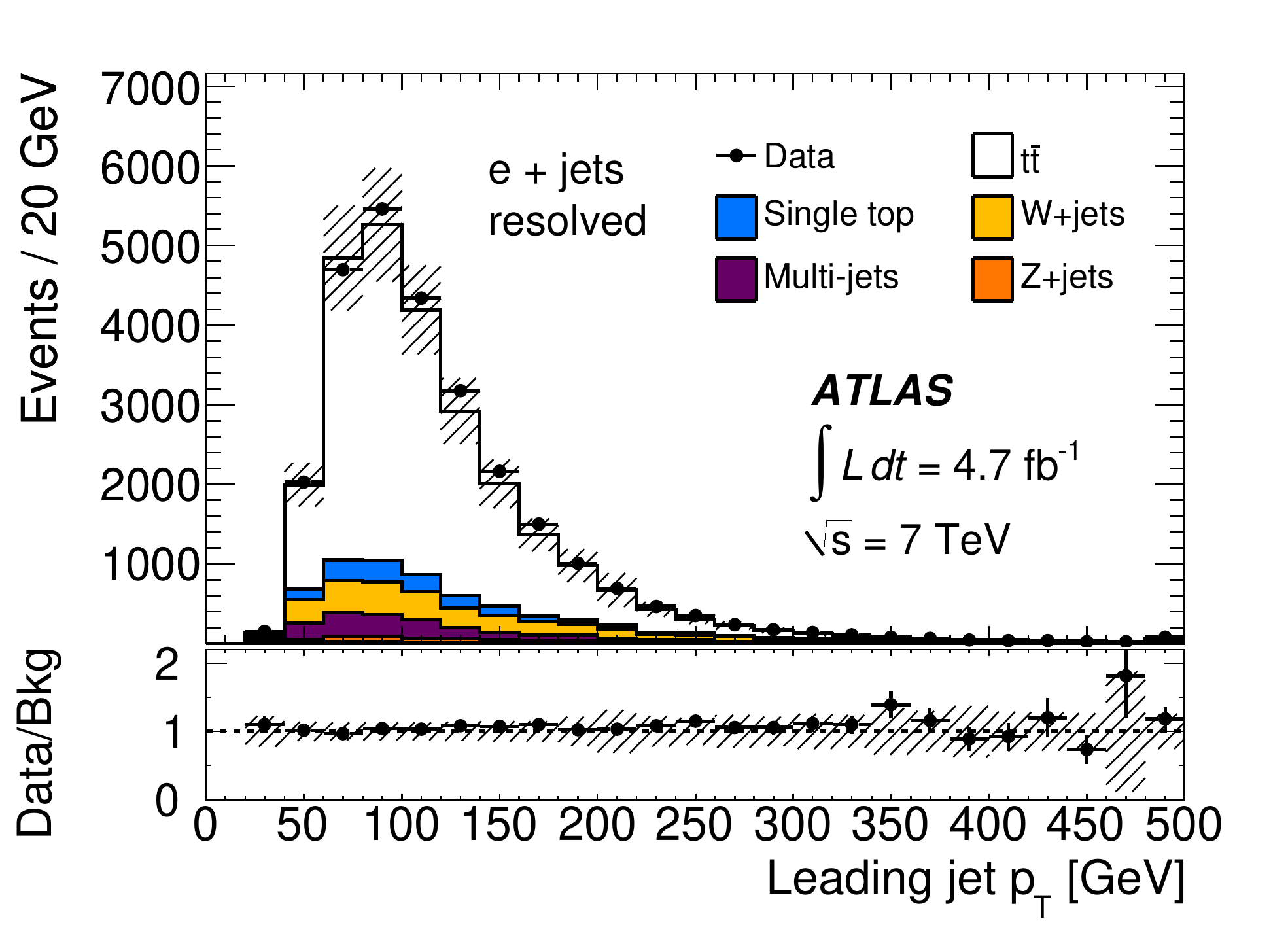} }
\subfigure[~$\mu$+jets channel.]{
\label{F:jet0pt_resolved_muo}
\includegraphics[width=0.45\textwidth]{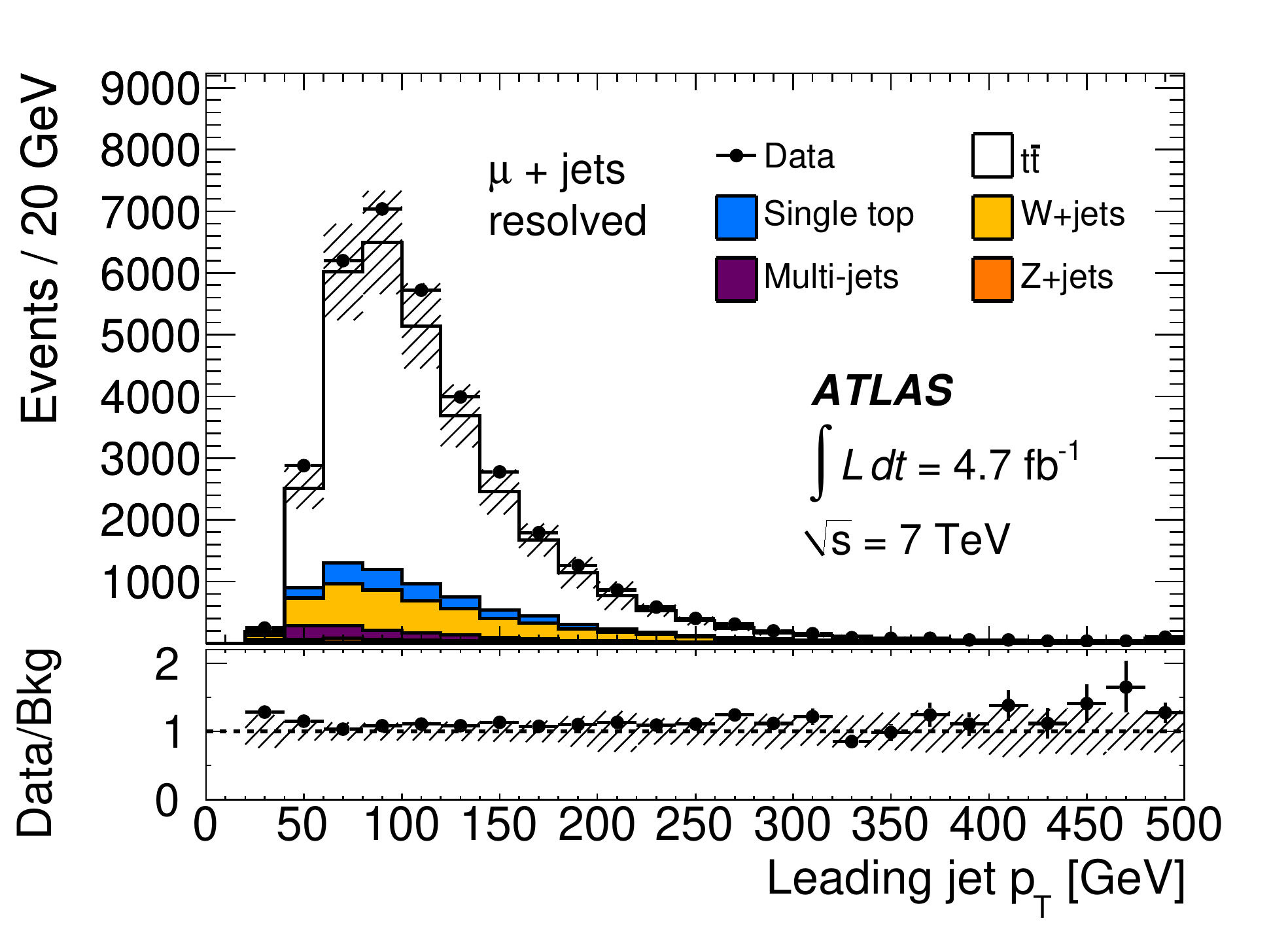} }
\caption{The transverse momentum of the leading jet in (a) the $e$+jets and (b) the $\mu$+jets channels, after the resolved selection. The shaded area indicates the total systematic uncertainties. 
\label{fig:jet0pt_resolved} }
\end{center}
\end{figure}
\begin{figure}[tbp]
\begin{center}
\subfigure[~$e$+jets channel.]{
\label{F:top_pt_el}
\includegraphics[width=0.45\textwidth]{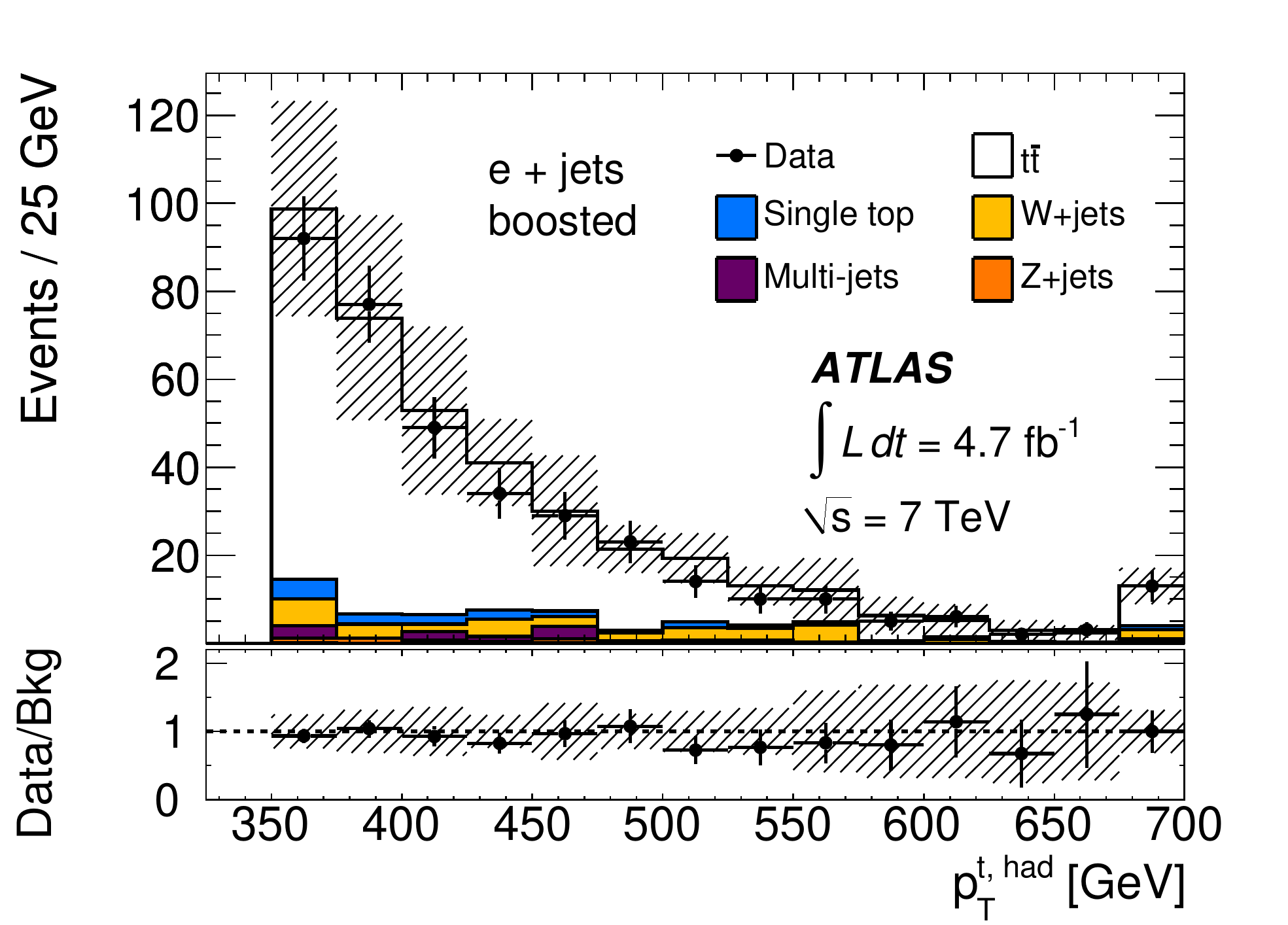} }
\subfigure[~$\mu$+jets channel.]{
\label{F:top_pt_mu}
\includegraphics[width=0.45\textwidth]{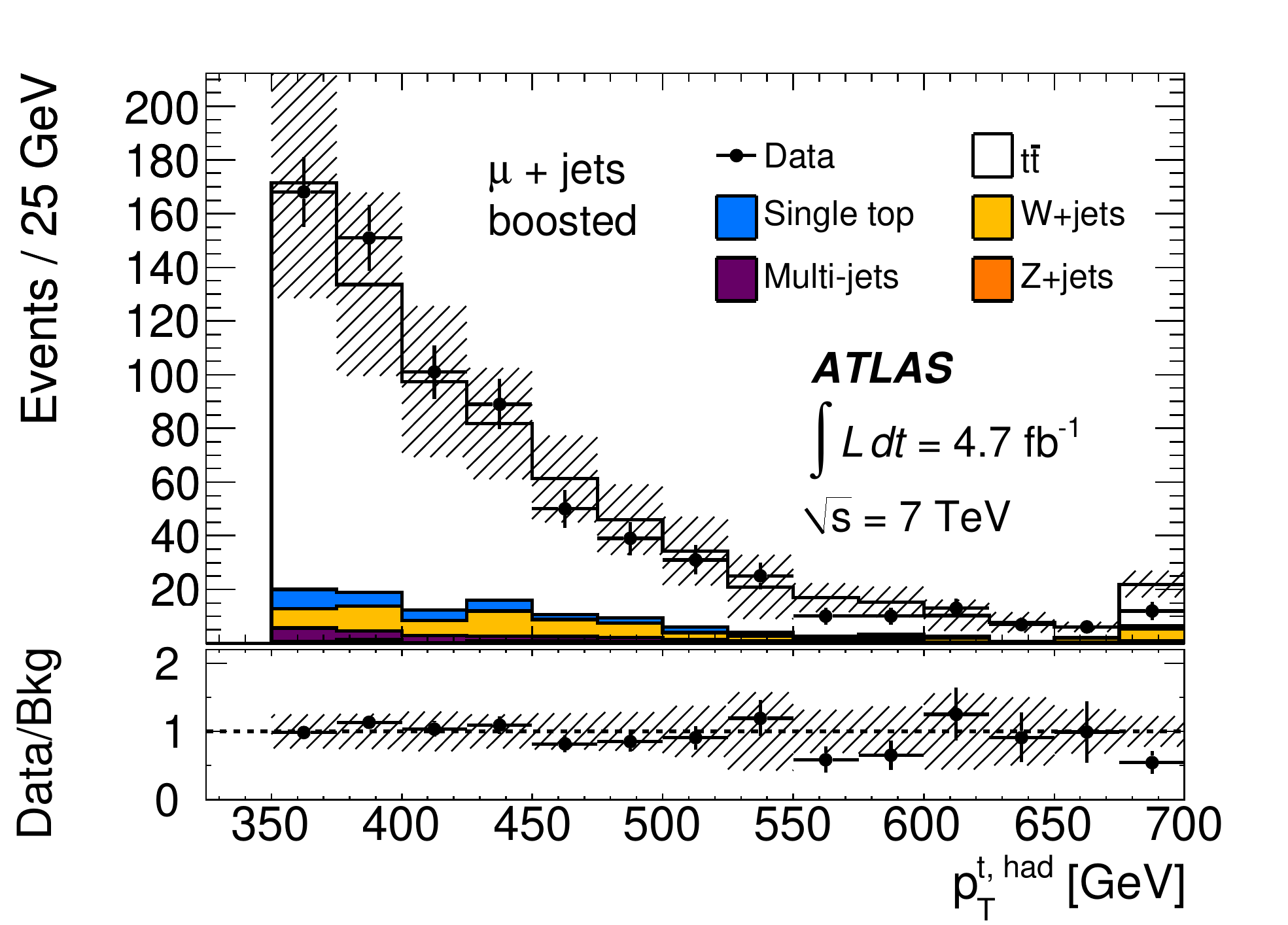} }
\caption{The transverse momentum of the hadronically decaying top quark candidate in (a) the $e$+jets and (b) the $\mu$+jets channels, after the boosted selection. The shaded area indicates the total systematic uncertainties. The last bin contains histogram limit overflows. 
\label{fig:top_mass_pt_boosted} }
\end{center}
\end{figure}
In Figs.~\ref{fig:leptop_mass_boosted} and \ref{fig:hadtopmass_boosted}, the reconstructed mass distributions of the semileptonically and hadronically decaying top quark candidates are shown, using the boosted event selection. Figure~\ref{fig:hadtopsplit_boosted} shows the distribution of the first \kt\ splitting scale of the selected large-radius jet. 
\begin{figure}[tbp]
\begin{center}
\subfigure[~$e$+jets channel.]{
\label{F:leptop_mass_el}
\includegraphics[width=0.45\textwidth]{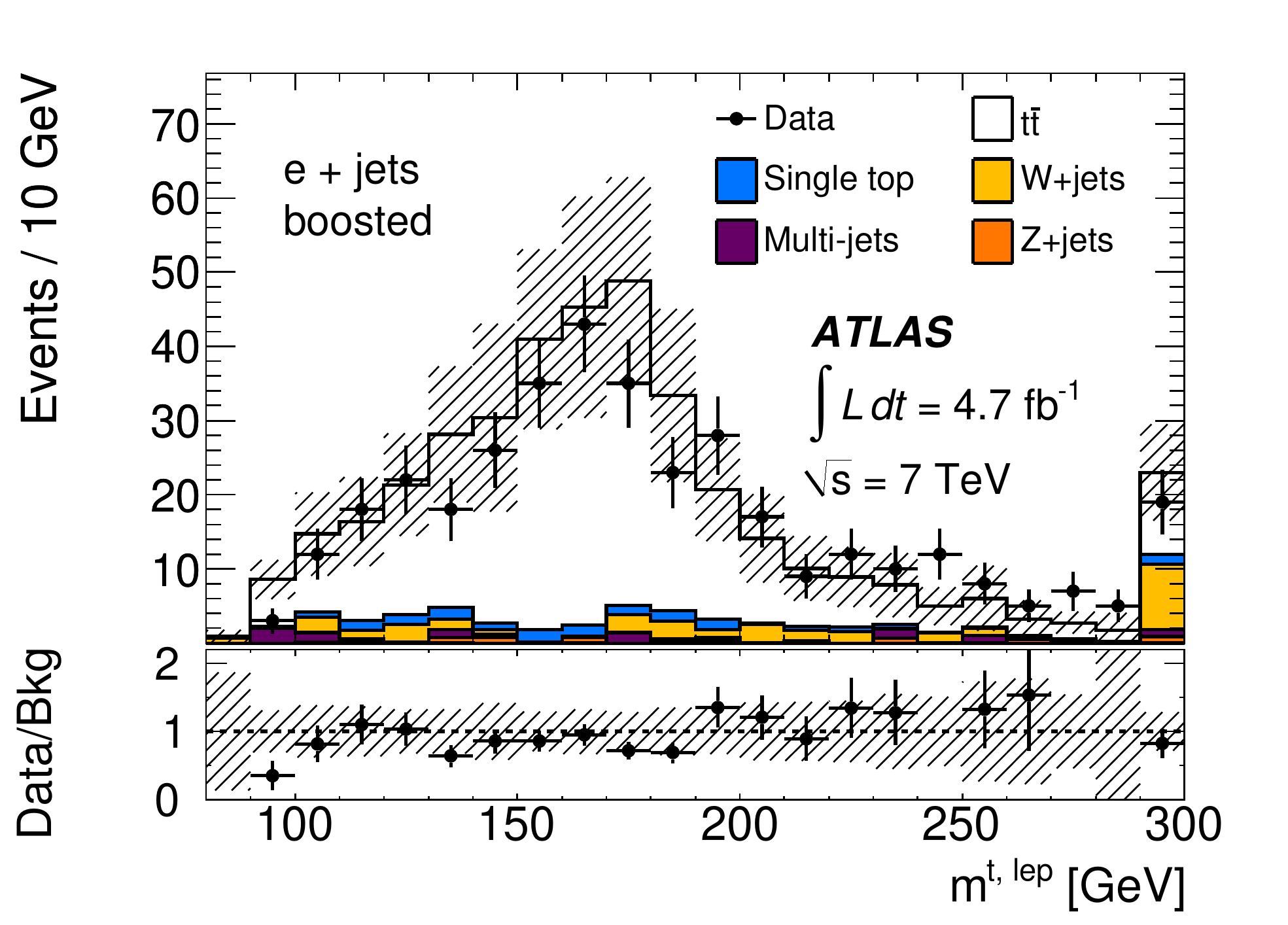} }
\subfigure[~$\mu$+jets channel.]{
\label{F:leptop_mass_mu}
\includegraphics[width=0.45\textwidth]{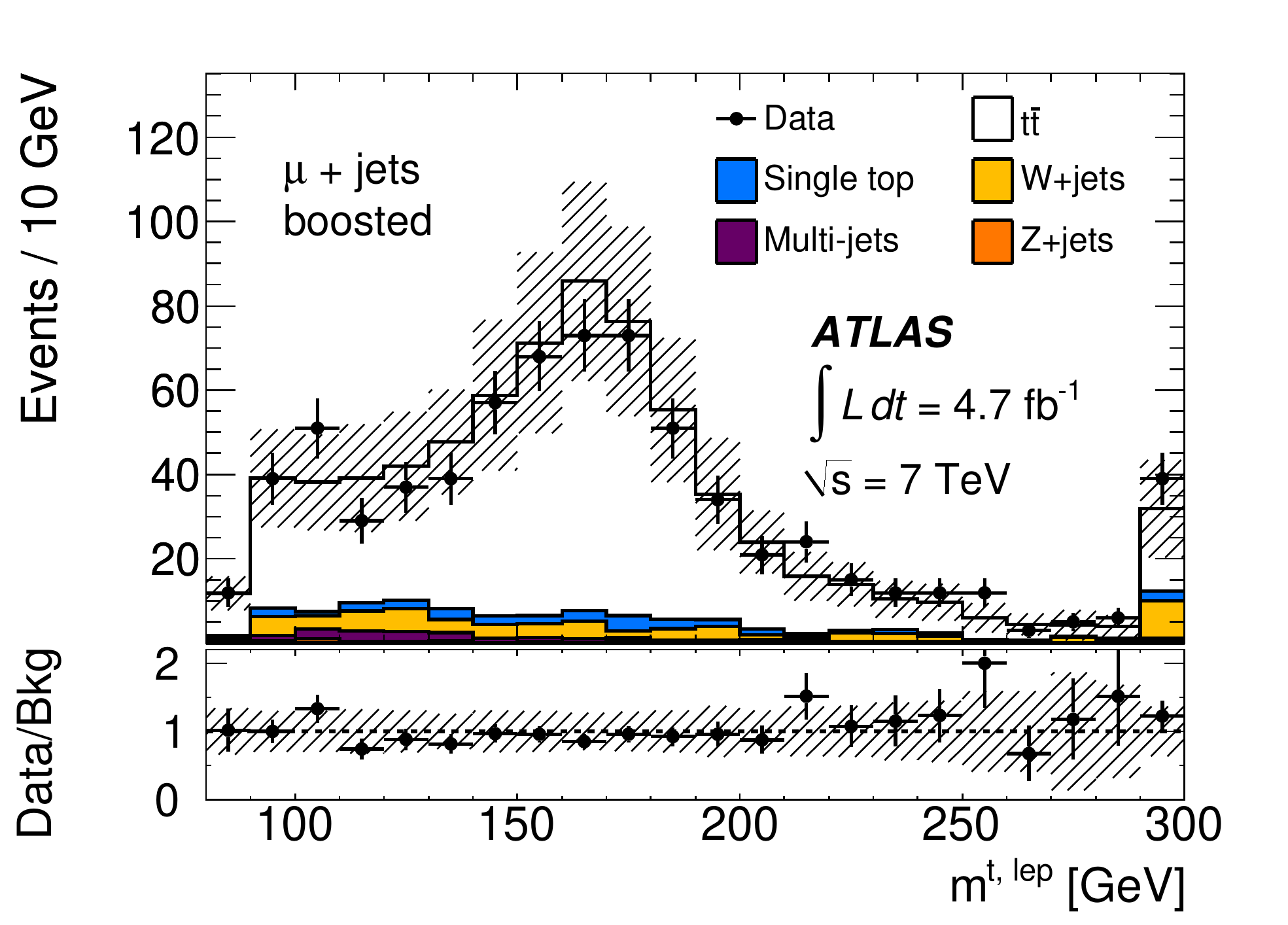} }
\caption{The invariant mass of the semileptonically decaying top quark candidate, \mtlep, in (a) the $e$+jets and (b) the $\mu$+jets channels, after the boosted selection. The mass has been reconstructed from the small-radius jet, the charged lepton and the missing transverse momentum, using a $W$ mass constraint to obtain the longitudinal momentum of the neutrino. The shaded area indicates the total systematic uncertainties. The last bin contains histogram limit overflows.
\label{fig:leptop_mass_boosted} }
\end{center}
\end{figure}
\begin{figure}[tbp]
\begin{center}
\subfigure[~$e$+jets channel.]{
\label{F:hadtop_mass_ele}
\includegraphics[width=0.45\textwidth]{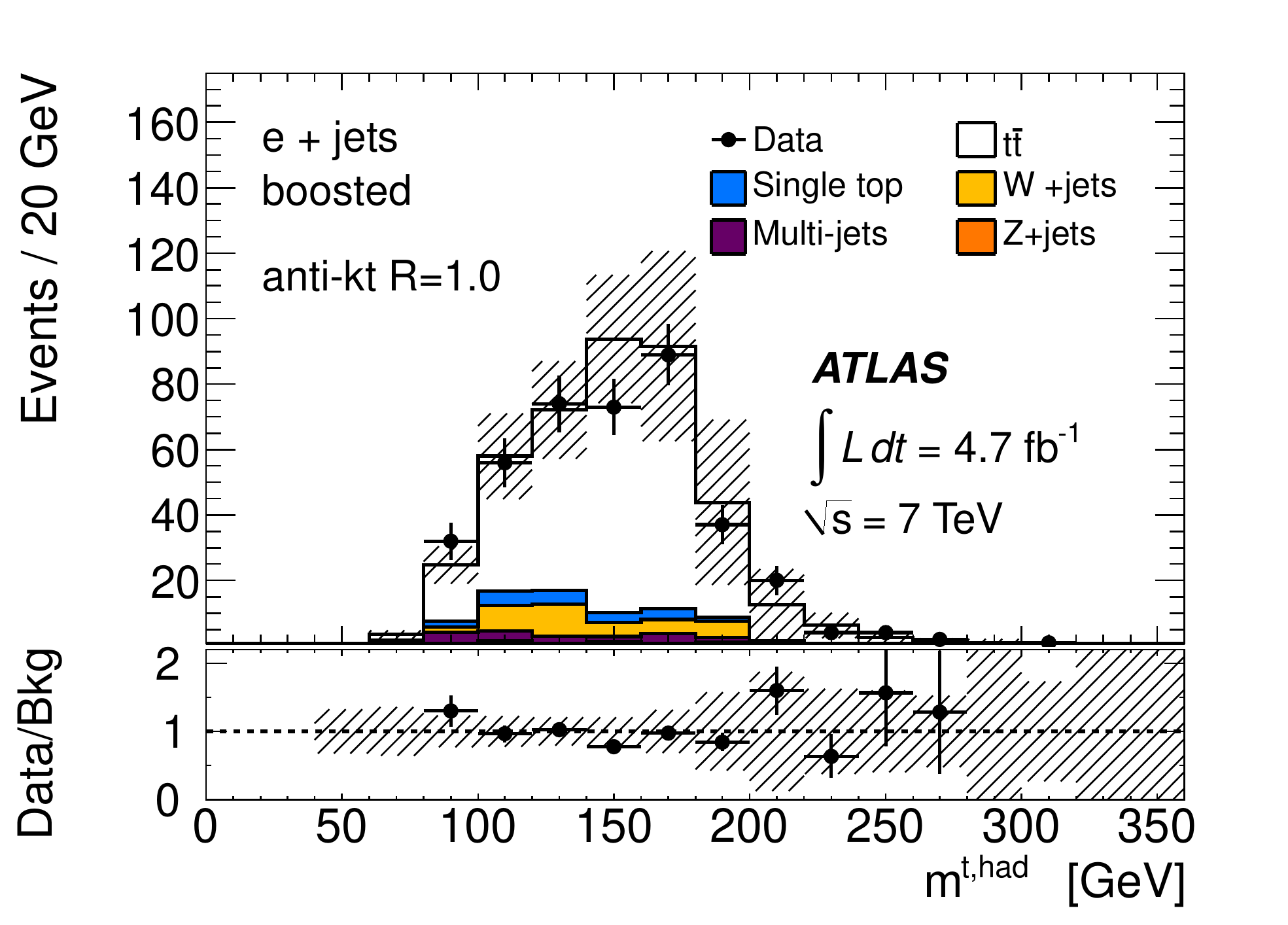} }
\subfigure[~$\mu$+jets channel.]{
\label{F:hadtop_mass_muo}
\includegraphics[width=0.45\textwidth]{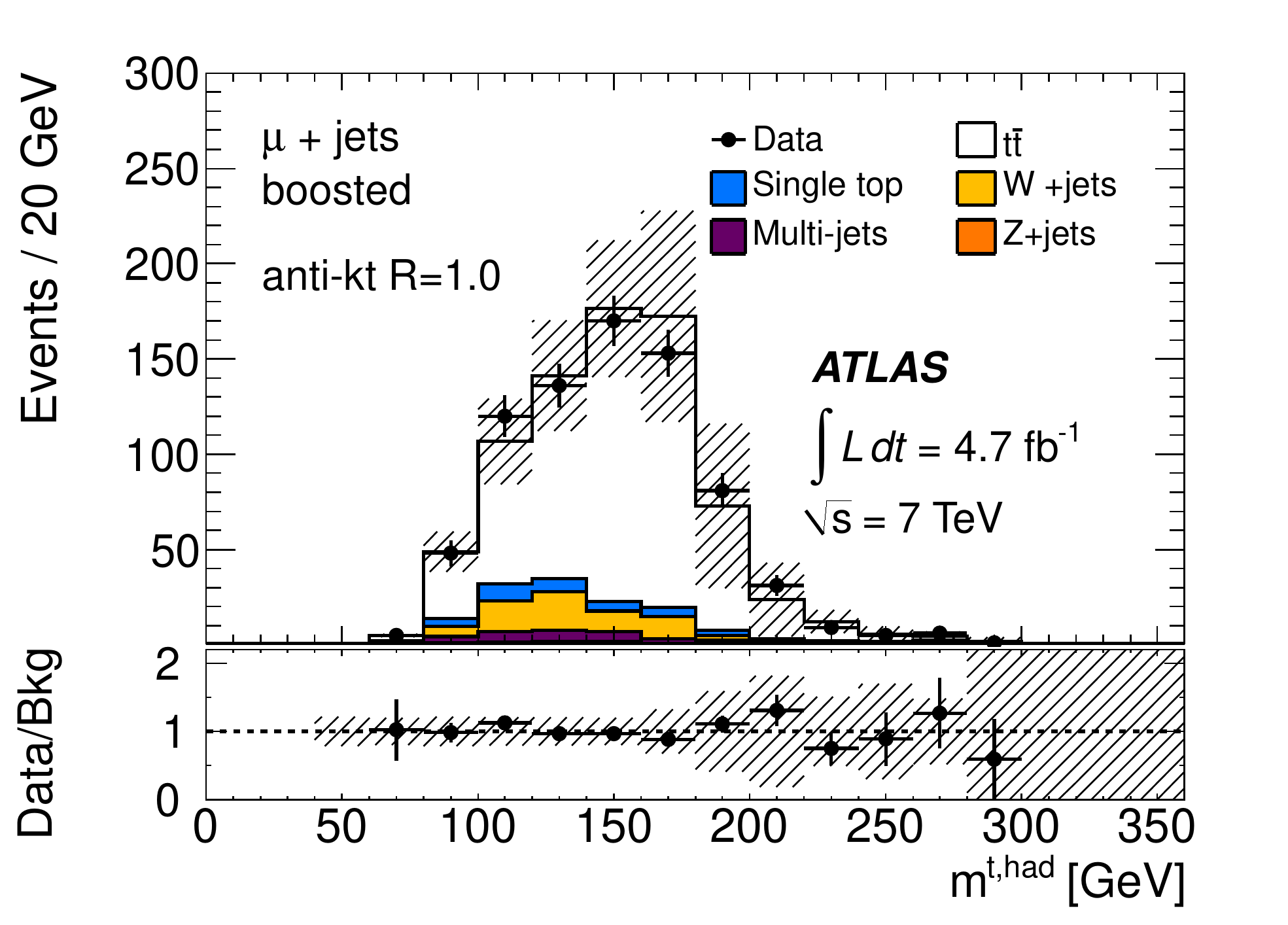} }
\caption{The mass of the large-radius jet from the hadronically decaying top quark, \mthad, in (a) the $e$+jets and (b) the $\mu$+jets channels, after the boosted selection, except the requirement $\mthad >100$~\GeV. The shaded area indicates the total systematic uncertainties. 
\label{fig:hadtopmass_boosted} }
\end{center}
\end{figure}
In these figures, the diboson background is too small to be visible. 
Good agreement is observed between the data and the expected background. 

\begin{figure}[tbp]
\begin{center}
\subfigure[~$e$+jets channel.]{
\label{F:hadtop_split_ele}
\includegraphics[width=0.45\textwidth]{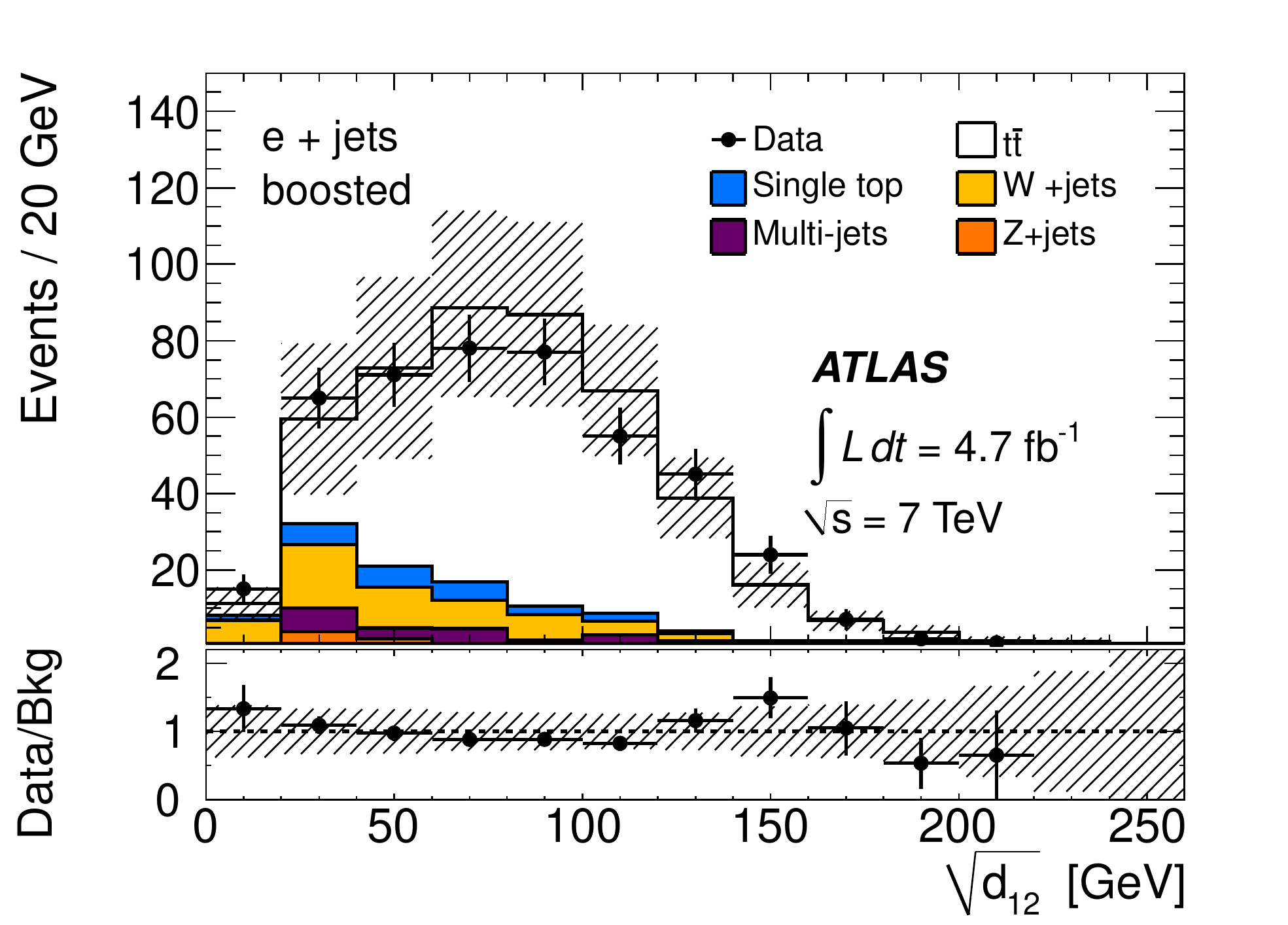} }
\subfigure[~$\mu$+jets channel.]{
\label{F:hadtop_split_muo}
\includegraphics[width=0.45\textwidth]{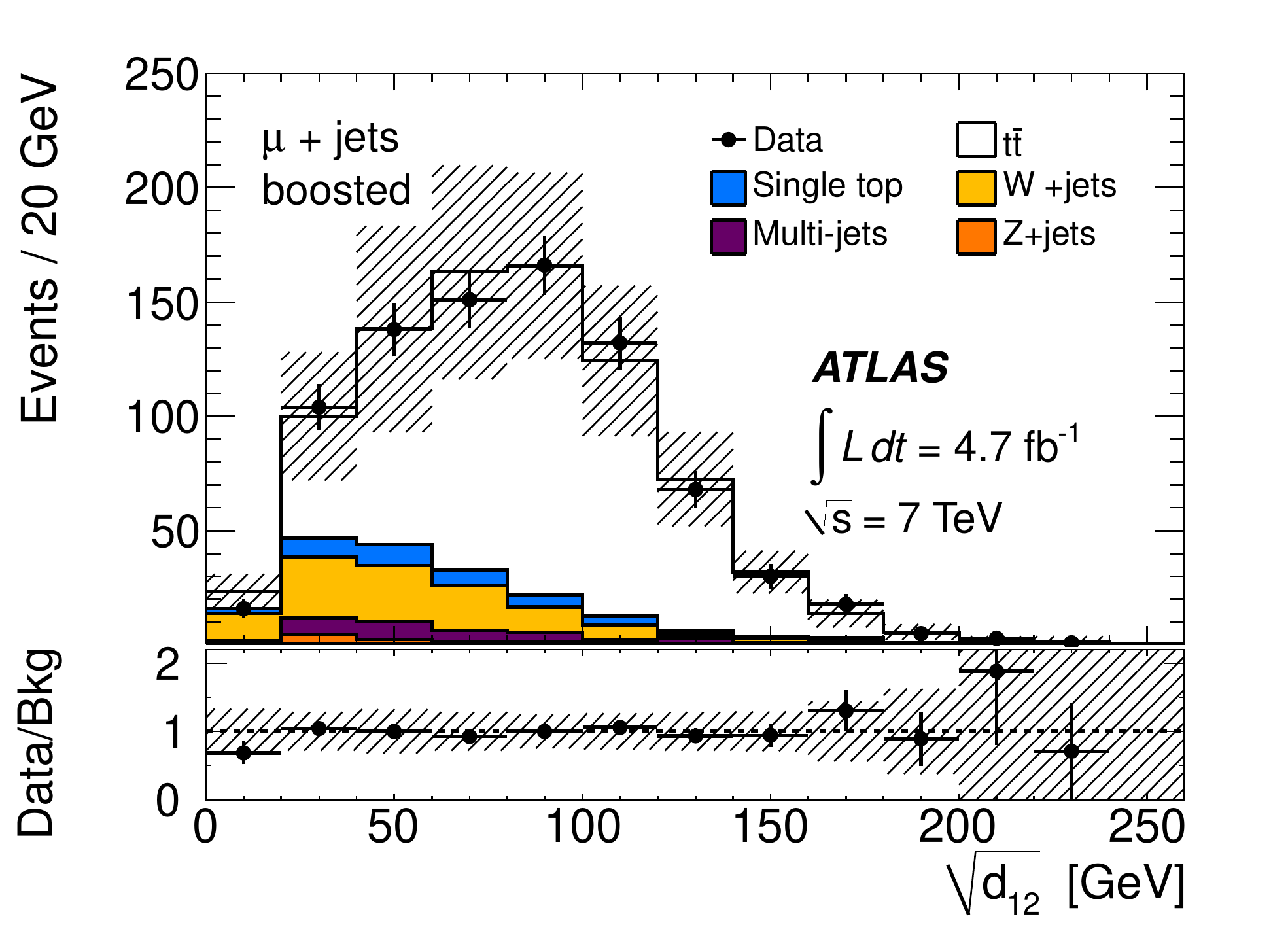} }
\caption{The first \kt\ splitting scale, \dsplit, of the large-radius jet from the hadronically decaying top quark in (a) the $e$+jets and (b) the $\mu$+jets channels, after the boosted selection, except the requirement $\dsplit>40$~\GeV. The shaded area indicates the total systematic uncertainties. 
\label{fig:hadtopsplit_boosted} }
\end{center}
\end{figure}

The \ttbar\ invariant mass spectra for the resolved and the boosted selections in the $e$+jets and $\mu$+jets decay channels are shown in Fig.~\ref{fig:mttlog}. Figure~\ref{fig:mttadded} shows the \ttbar\ invariant mass spectrum for all channels added together. 
The data agree with the Standard Model prediction within the uncertainties. 
The slight shape mismatch between data and the Standard Model prediction seen in Fig.~\ref{fig:mttlog}, especially for the resolved selection, is fully covered by the uncertainties. 
Systematic uncertainties that tilt the shape in this way include the \ttbar\ generator uncertainty, the small-radius jet energy scale and resolution uncertainties, and the ISR/FSR modeling.

\begin{figure*}[tbp]
\begin{center}
\subfigure[~$e$+jets channel, resolved selection.]{
\label{F:mtt_resolved_el}
\includegraphics[width=0.45\textwidth]{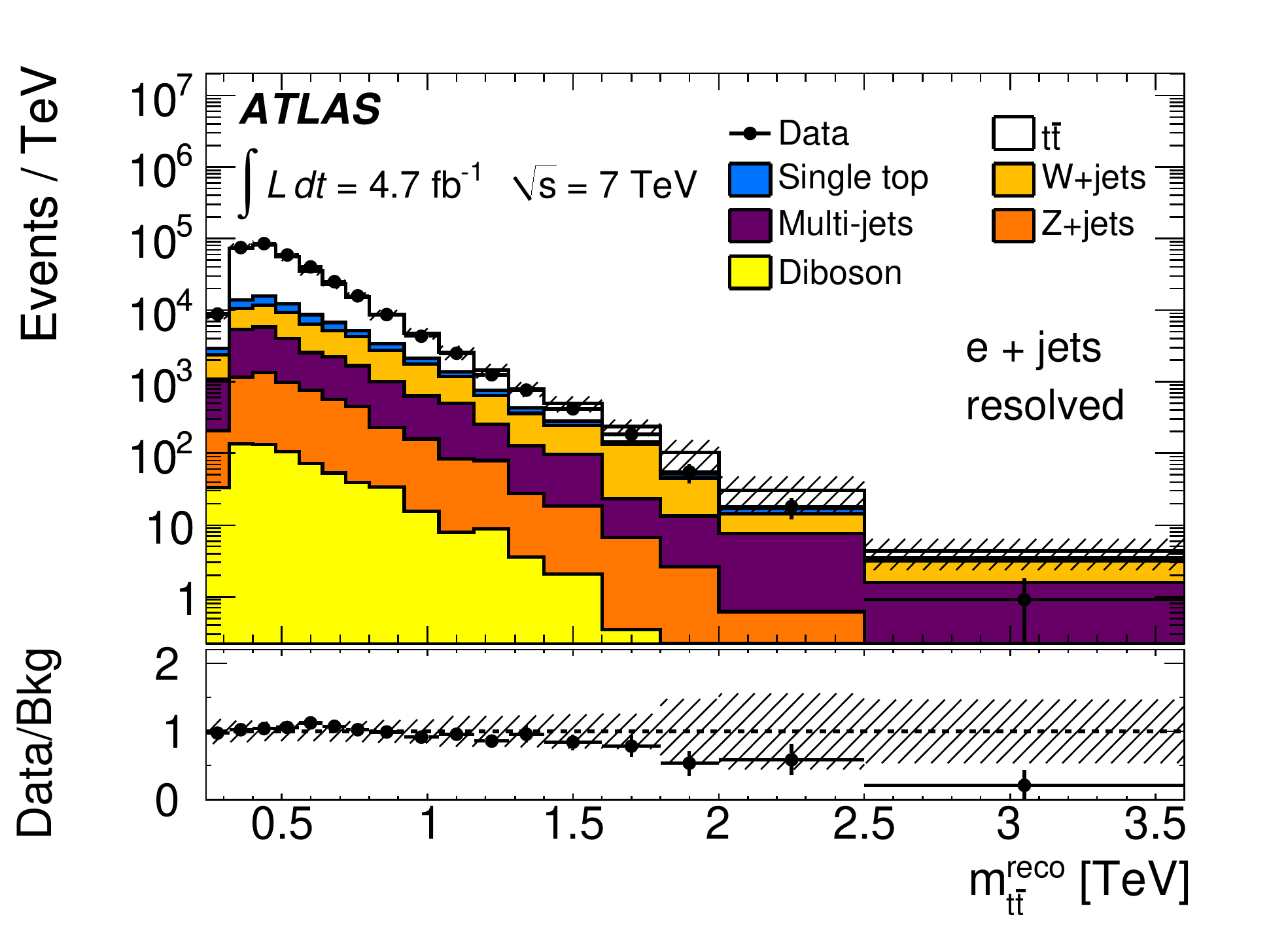} }
\subfigure[~$\mu$+jets channel, resolved selection.]{
\label{F:mtt_resolved_mu}
\includegraphics[width=0.45\textwidth]{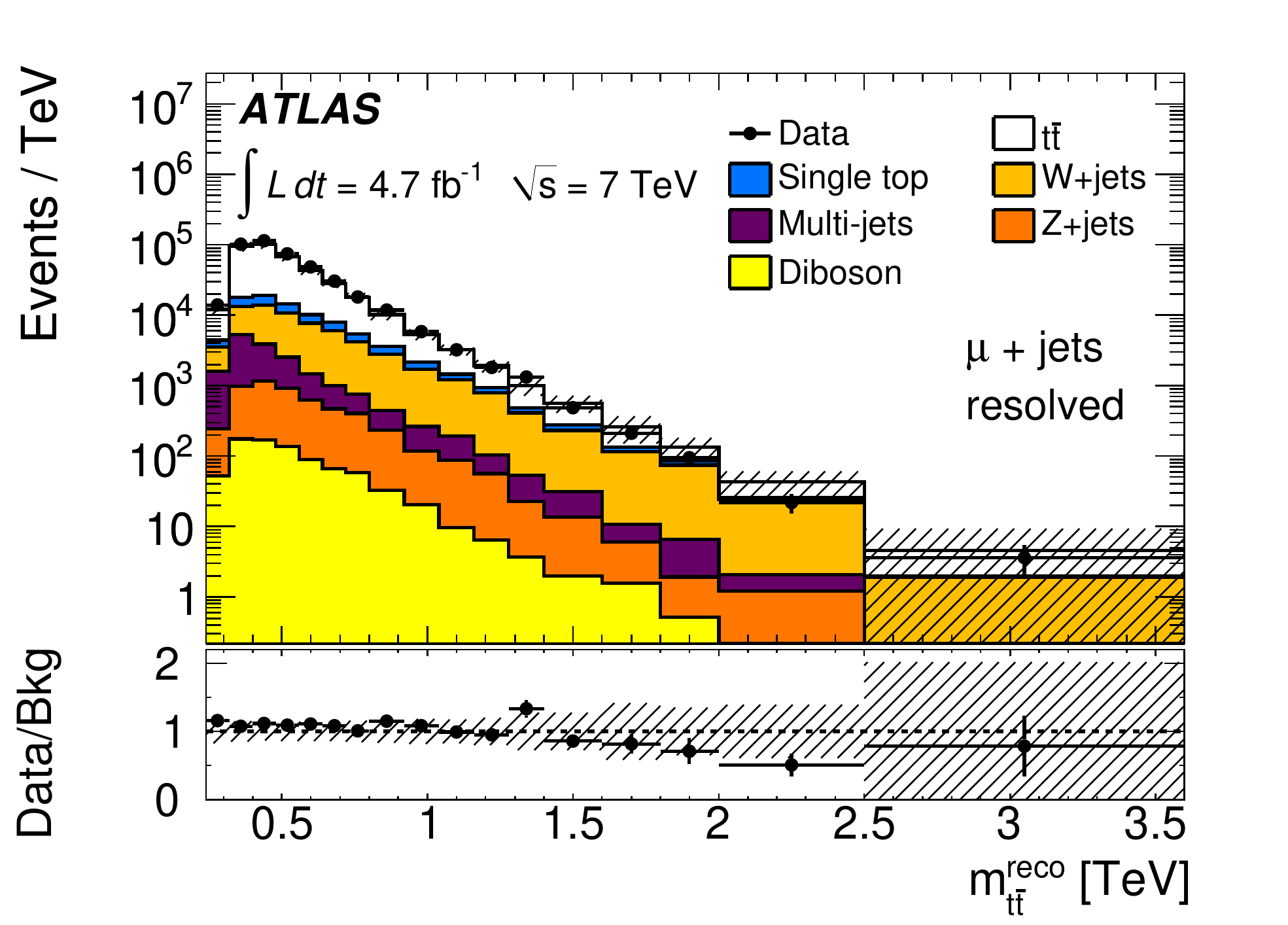} }
\subfigure[~$e$+jets channel, boosted selection.]{
\label{F:mtt_boosted_el}
\includegraphics[width=0.45\textwidth]{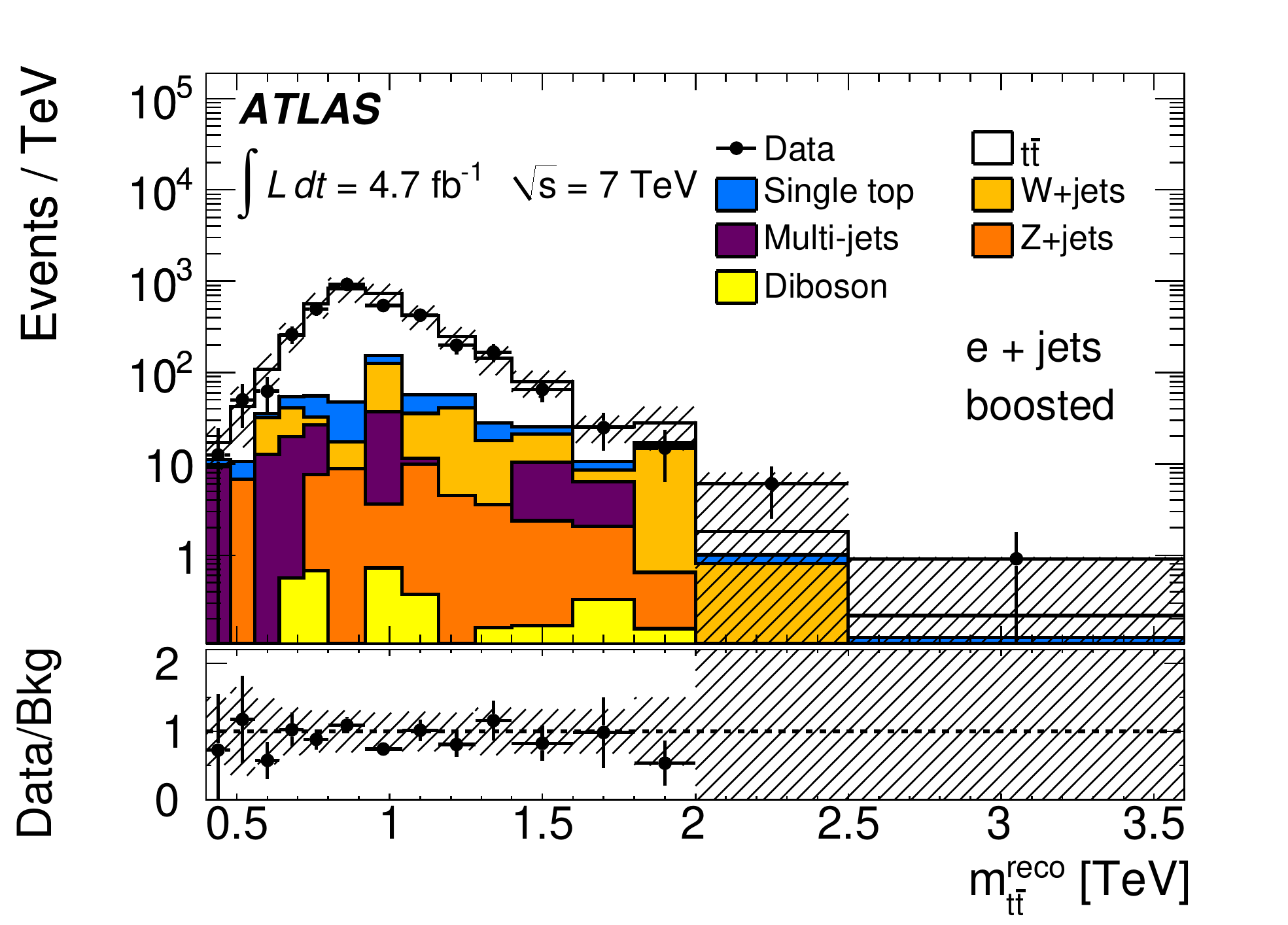} }
\subfigure[~$\mu$+jets channel, boosted selection.]{
\label{F:mtt_boosted_mu}
\includegraphics[width=0.45\textwidth]{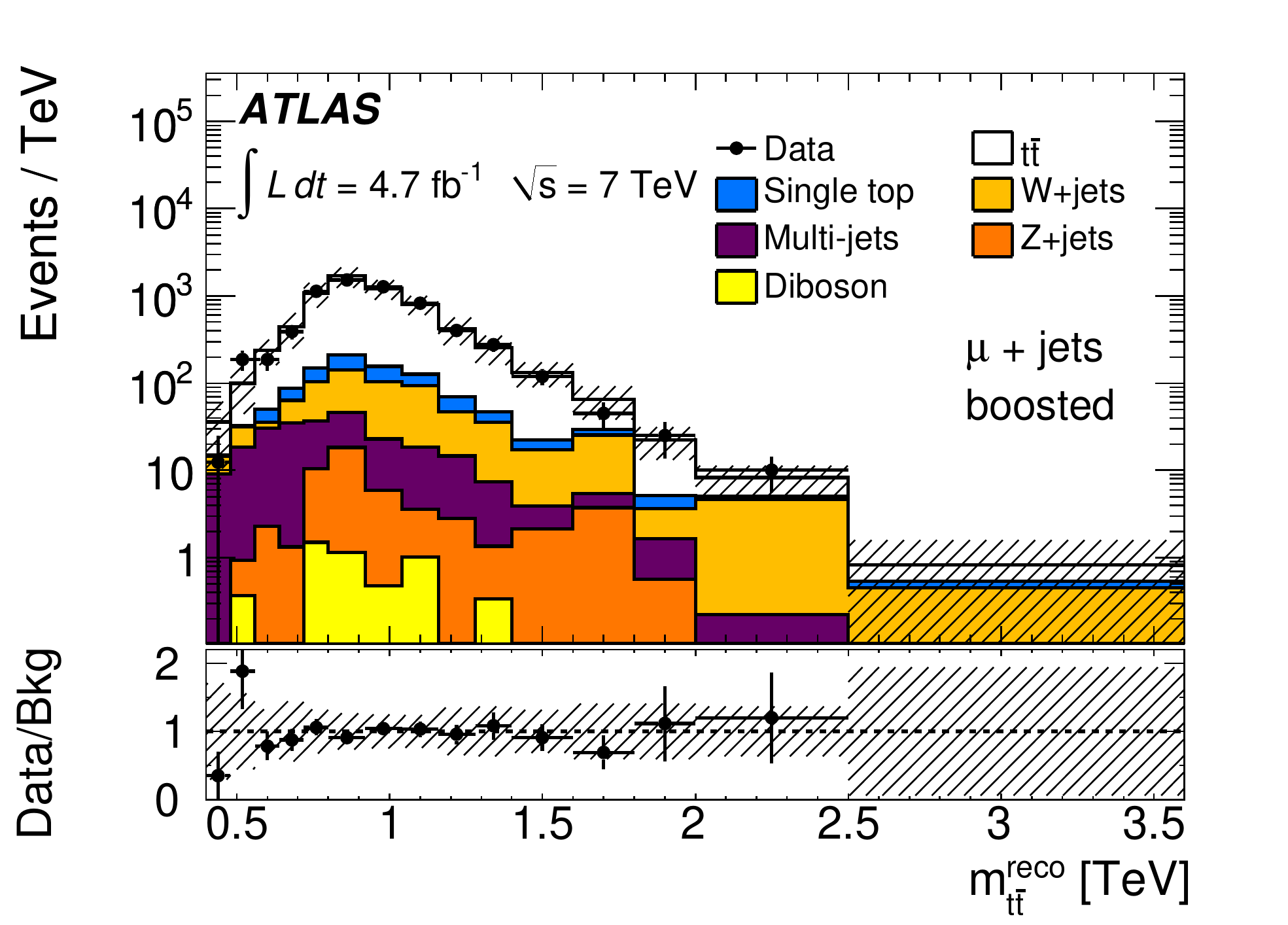} }
\caption{The \ttbar\ invariant mass spectra for the two channels and the two selection methods. The shaded area indicates the total systematic uncertainties. 
\label{fig:mttlog} }
\end{center}
\end{figure*}

\begin{figure}[tbp]
\begin{center}
\includegraphics[width=0.45\textwidth]{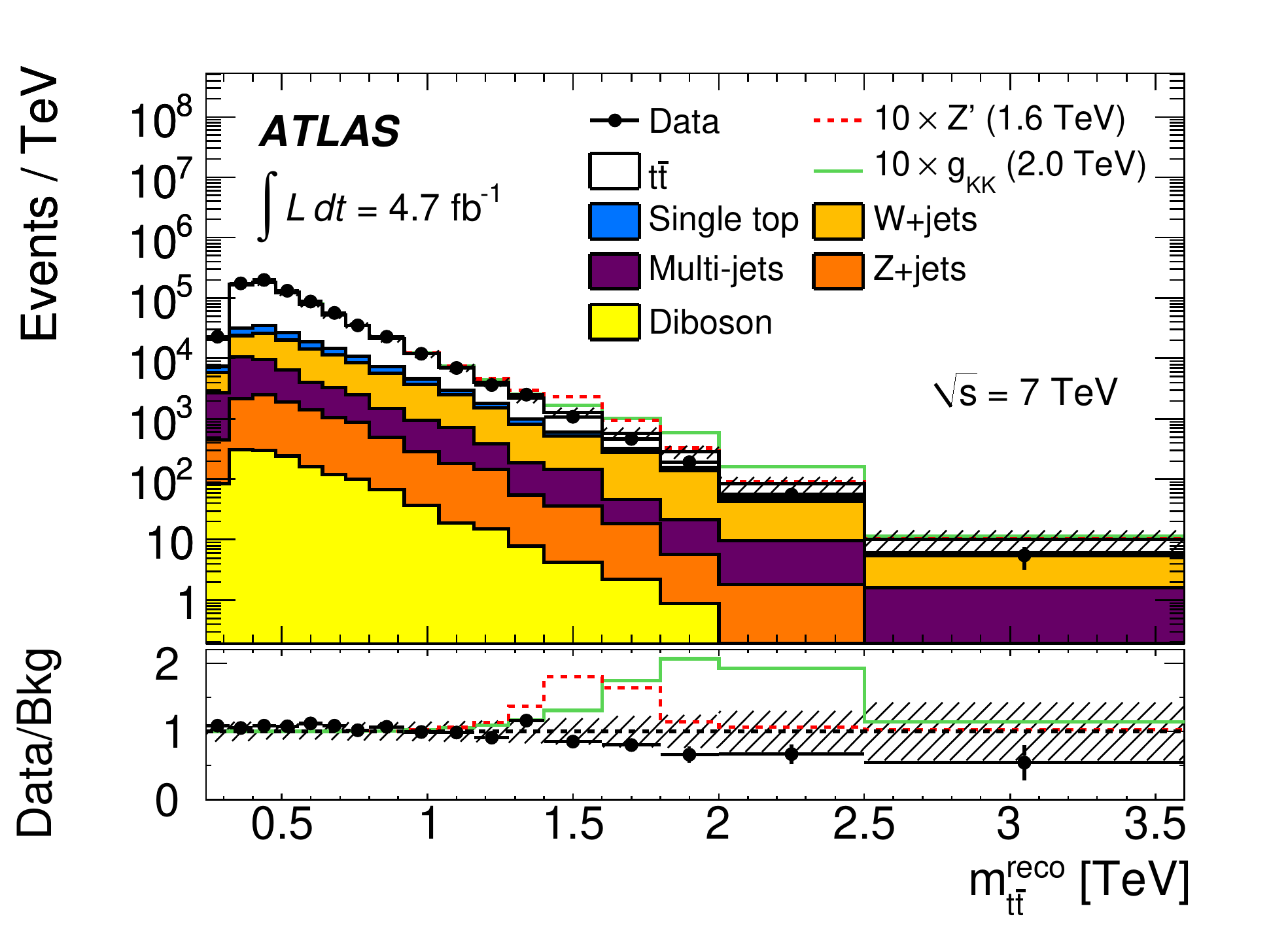}
\caption{The \ttbar\ invariant mass spectrum, adding the spectra from the two channels and both selection methods. The shaded area indicates the total systematic uncertainties. Two benchmark signals are indicated on top of the background, a \Zprime\ resonance with $m=1.6$~\TeV\ and a KK gluon with $m=2.0$~\TeV. The assumed cross sections of the signals in this figure are ten times larger than the theoretical predictions in Table~\ref{tab:signalxsec}. 
\label{fig:mttadded} }
\end{center}
\end{figure}

\section{Results}
\label{sec:Results}

After the reconstruction of the \ttbar\ mass spectra, the data and expected background distributions are compared to search for hints of phenomena associated with new physics using {\sc BumpHunter}~\cite{Choudalakis:2011qn}. 
This is a hypothesis-testing tool that uses pseudo-experiments to search for local excesses or deficits in the data compared to the Standard Model prediction in binned histograms, taking the look-elsewhere effect into account over the full mass spectrum. 
The Standard Model prediction is allowed to float within the systematic uncertainties. 
After accounting for the systematic uncertainties, no significant deviation from the expected background is found. 
Upper limits are set on the cross section times branching ratio of the \Zprime\ and KK gluon benchmark models using a Bayesian technique, implemented in a tool developed by the D0  collaboration~\cite{Bertram:2000br}. 
The prior is taken to be constant in the signal cross section, which in this case is an excellent approximation of the reference prior that maximizes the amount of missing information~\cite{Berger_prior:2009}, as given in Ref.~\cite{Casadei:2011hx}. 
The Bayesian limits are in good agreement with results obtained using the $CL_s$ method~\cite{Junk:1999kv,Read:2002hq}. 
For each of the models investigated, 95\% CL upper limits are set on the product of production cross section and branching ratio into \ttbar\ pairs.

\begin{figure}[tbp]
\begin{center}
\subfigure[~\Zprime{} upper cross-section limits. ]{
\label{F:limits_Zprime}
\includegraphics[width=0.45\textwidth]{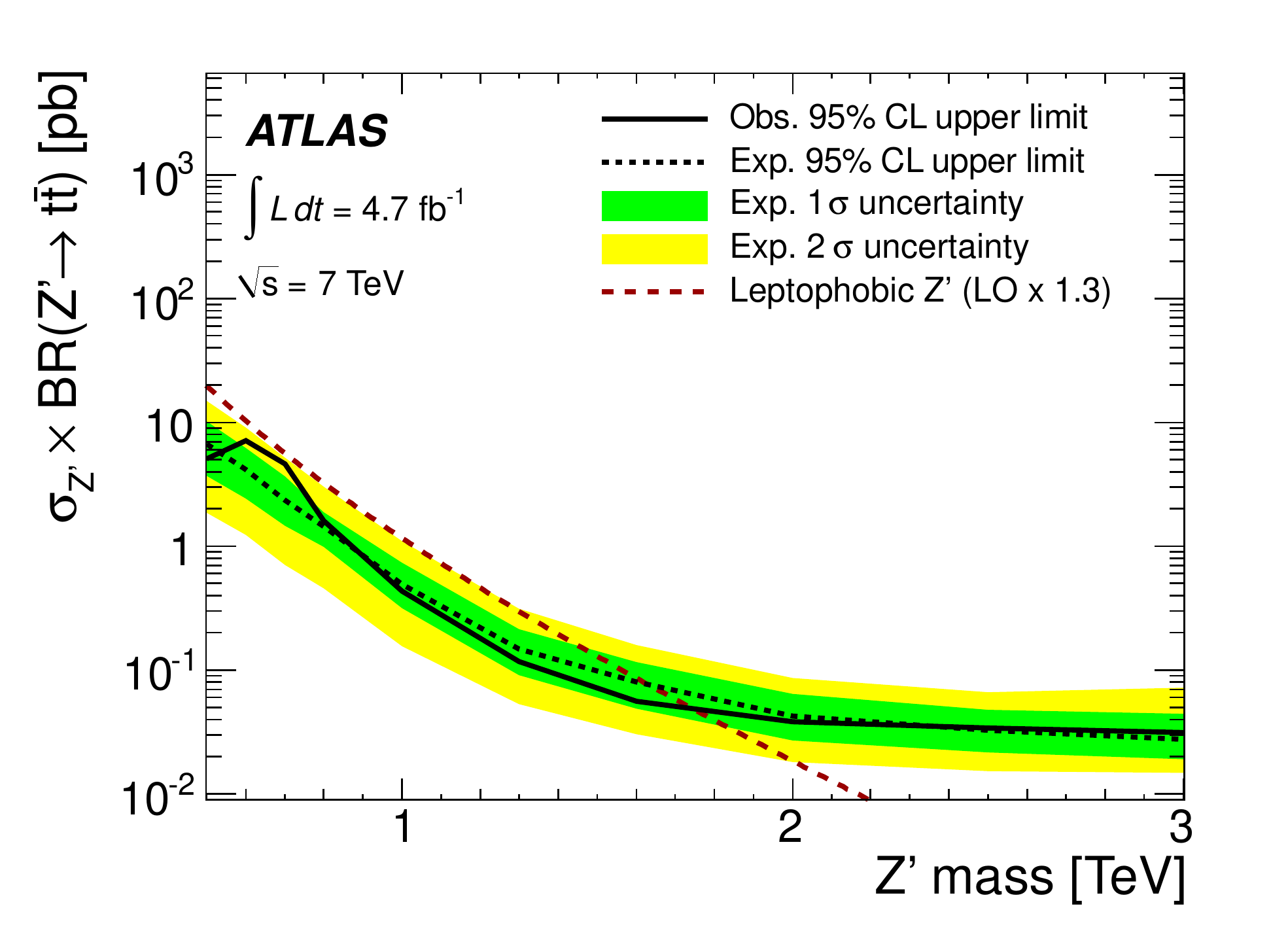} }
\subfigure[~\gkk\ upper cross-section limits.]{
\label{F:limits_kkg}
\includegraphics[width=0.45\textwidth]{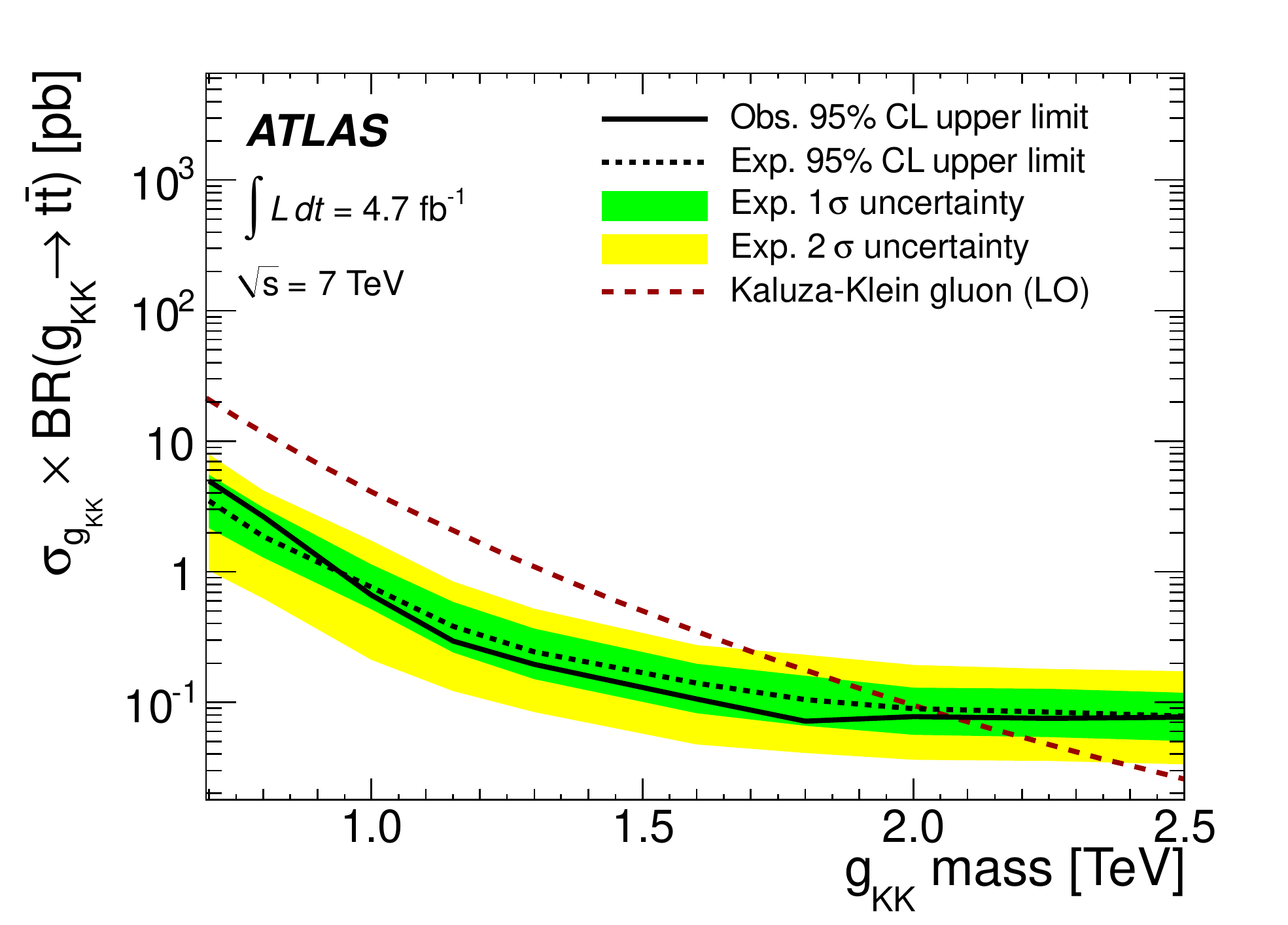} }\\
\caption{Observed and expected upper cross-section limits times the \ttbar{} branching ratio on (a) narrow \Zprime{} resonances and (b) Kaluza--Klein gluons. The resolved and the boosted selections have been combined in the estimation of the limits. Systematic and statistical uncertainties are included. 
\label{fig:limit_Zprime_kkg}
}
\end{center}
\end{figure}

Figure~\ref{fig:limit_Zprime_kkg} displays the upper limits on the cross section, with systematic and statistical uncertainties, obtained from the combination of the two selections for each of the benchmark models. 
The numerical values of the upper limits on the cross section are given in Table~\ref{tab:limits} (\Zprime) and Table~\ref{tab:limits_kkg} (\gkk). 
The expected limits and uncertainty band are obtained using pseudo-experiments based on Poisson distributions for the number of entries in each bin. 
In the combination, the four disjoint spectra are used, corresponding to boosted and resolved selections, as well as $e$+jets and $\mu$+jets decay channels.
Due to the improvements in the analysis, the upper limits on the cross section for a \Zprime\ resonance at 1.6~\TeV\ and a KK gluon resonance at 2.0~\TeV\ are less than half the values that would be obtained by a simple rescaling of the previous best high-mass ATLAS limits~\cite{Arce:1431929} to account for the larger integrated luminosity.  
Using the combined upper limits on the cross section, a leptophobic topcolor \Zprime{} boson (KK gluon) with mass between 0.5~\TeV\ and \ZpexcludedMass\ (0.7~\TeV\ and \kkgexcludedMass) is excluded at 95\% CL.

\begin{table}[tbhp]
\begin{center}
\caption{Upper limits on the cross-section times branching ratio, at 95\% CL, for a leptophobic topcolor \Zprime\ decaying to \ttbar, using the combination of all four samples. 
The observed and expected limits for each mass point are given, as well as the $\pm1\sigma$ variation of the expected limit. }
\vspace{2mm}
\begin{tabular}{cr@{.}lr@{.}lr@{.}lr@{.}l}
\hline \hline
Mass  &  \multicolumn{2}{c}{Obs. }  &  \multicolumn{2}{c}{Exp. }  &  \multicolumn{2}{c}{Exp.$-$1$\sigma$~~ }  &  \multicolumn{2}{c}{Exp.$+$1$\sigma$~~} \\
$[$\TeV$]$ &  \multicolumn{2}{c}{[pb]}  &  \multicolumn{2}{c}{[pb]}  &  \multicolumn{2}{c}{[pb]}  &  \multicolumn{2}{c}{[pb]} \\
\hline
 0.50&    \hspace{1em}5&1&     \hspace{1em}6&7&     \hspace{1em}3&7&    \hspace{1ex}10&2\\
 0.60&    7&1&     4&2&     2&4&     6&2\\
 0.70&    4&6&     2&3&     1&5&     3&7\\
 0.80&    1&61&    1&45&    0&98&    1&89\\
 1.00&    0&43&    0&49&    0&31&    0&74\\
 1.30&    0&117&   0&148&   0&090&   0&213\\
 1.60&    0&056&   0&080&   0&049&   0&115\\
 2.00&    0&038&   0&042&   0&027&   0&064\\
 2.50&    0&034&   0&033&   0&022&   0&048\\
 3.00~&   0&031~~& 0&028~~& 0&019~~& 0&044\\
\hline \hline
\end{tabular}
\label{tab:limits}
\end{center}
\end{table}

\begin{table}[tbhp]
\begin{center}
\caption{Upper limits on the cross-section times branching ratio, at 95\% CL, for a Kaluza--Klein gluon decaying to \ttbar, using the combination of all four samples. 
The observed and expected limits for each mass point are given, as well as the $\pm1\sigma$ variation of the expected limit. }
\vspace{2mm}
\begin{tabular}{cr@{.}lr@{.}lr@{.}lr@{.}l}
\hline \hline
Mass  &  \multicolumn{2}{c}{Obs. }  &  \multicolumn{2}{c}{Exp. }  &  \multicolumn{2}{c}{Exp.$-$1$\sigma$~~ }  &  \multicolumn{2}{c}{Exp.$+$1$\sigma$~~} \\
$[$\TeV$]$ &  \multicolumn{2}{c}{[pb]}  &  \multicolumn{2}{c}{[pb]}  &  \multicolumn{2}{c}{[pb]}  &  \multicolumn{2}{c}{[pb]} \\
\hline
 0.70& \hspace{1em}5&0& \hspace{1em}3&5& \hspace{1em}2&2& \hspace{1em}5&5\\
 0.80&    2&6&     1&86&    1&29&    3&1\\
 1.00&    0&66&    0&76&    0&51&    1&14\\
 1.15&    0&29&    0&38&    0&24&    0&58\\
 1.30&    0&20&    0&24&    0&15&    0&37\\
 1.60&    0&106&   0&140&   0&082&   0&198\\
 1.80&    0&072&   0&105&   0&066&   0&159\\
 2.00&    0&077&   0&089&   0&056&   0&129\\
 2.25&    0&075&   0&084&   0&054&   0&126\\
 2.50&    0&077&   0&078&   0&050&   0&119\\
\hline
\end{tabular}
\label{tab:limits_kkg}
\end{center}
\end{table}

\section{Summary}
\label{sec:Summary}

A search for \ttbar\ resonances in the lepton plus jets decay channel has been carried out with the ATLAS experiment at the LHC.  
The search uses a data sample corresponding to an integrated luminosity of \totlumi\ of proton--proton collisions at a center-of-mass energy of 7~\TeV.  
The \ttbar\ system is reconstructed in two different ways. 
For the resolved selection, the hadronic top-quark decay is reconstructed as two or three $R = 0.4$ jets, and for the boosted selection, it is reconstructed as one $R = 1.0$ jet.  
No excess of events beyond the Standard Model predictions is observed in the \ttbar\ invariant mass spectrum.  
Upper limits on the cross section times branching ratio are set for two benchmark models: a narrow \Zprime\ resonance from Ref.~\cite{Hill:1994hp} and a broad Randall--Sundrum Kaluza--Klein gluon from Ref.~\cite{Lillie:2007yh}. 
The 95\% credibility upper limits on the cross section times branching ratio for the narrow resonance range from \Zpxseclow\ at a resonance mass of 0.5~\TeV\  to \Zpxsechigh\ at 3~\TeV. 
The upper limits on the cross section determined for the broad resonance are higher, \kkgxseclow\ (\kkgxsechigh) at 0.7~(2.0)~\TeV. 
Based on these results, the existence of a narrow leptophobic topcolor \Zprime\ boson with mass $0.5$--\ZpexcludedMass\ is excluded at 95\% CL. 
A broad Kaluza--Klein gluon in the mass range 0.7--\kkgexcludedMass\ is also excluded at 95\% CL.

\section*{Acknowledgments}
\label{sec:Acknowledgements}


We thank CERN for the very successful operation of the LHC, as well as the 
support staff from our institutions without whom ATLAS could not be 
operated efficiently. 

We acknowledge the support of ANPCyT, Argentina; YerPhI, Armenia; ARC, 
Australia; BMWF and FWF, Austria; ANAS, Azerbaijan; SSTC, Belarus; CNPq and FAPESP, 
Brazil; NSERC, NRC and CFI, Canada; CERN; CONICYT, Chile; CAS, MOST and NSFC, 
China; COLCIENCIAS, Colombia; MSMT CR, MPO CR and VSC CR, Czech Republic; 
DNRF, DNSRC and Lundbeck Foundation, Denmark; EPLANET, ERC and NSRF, European Union; 
IN2P3-CNRS, CEA-DSM/IRFU, France; GNSF, Georgia; BMBF, DFG, HGF, MPG and AvH 
Foundation, Germany; GSRT and NSRF, Greece; ISF, MINERVA, GIF, DIP and Benoziyo Center, 
Israel; INFN, Italy; MEXT and JSPS, Japan; CNRST, Morocco; FOM and NWO, 
Netherlands; BRF and RCN, Norway; MNiSW, Poland; GRICES and FCT, Portugal; MERYS 
(MECTS), Romania; MES of Russia and ROSATOM, Russian Federation; JINR; MSTD, 
Serbia; MSSR, Slovakia; ARRS and MIZ\v{S}, Slovenia; DST/NRF, South Africa; 
MICINN, Spain; SRC and Wallenberg Foundation, Sweden; SER, SNSF and Cantons of 
Bern and Geneva, Switzerland; NSC, Taiwan; TAEK, Turkey; STFC, the Royal 
Society and Leverhulme Trust, United Kingdom; DOE and NSF, United States of 
America. 

The crucial computing support from all WLCG partners is acknowledged 
gratefully, in particular from CERN and the ATLAS Tier-1 facilities at 
TRIUMF (Canada), NDGF (Denmark, Norway, Sweden), CC-IN2P3 (France), 
KIT/GridKA (Germany), INFN-CNAF (Italy), NL-T1 (Netherlands), PIC (Spain), 
ASGC (Taiwan), RAL (UK) and BNL (USA) and in the Tier-2 facilities 
worldwide. 

\clearpage

\bibliography{prdv2.bib}

\begin{thebibliography}{10}%
\makeatletter
\providecommand \@ifxundefined [1]{%
 \ifx #1\undefined \expandafter \@firstoftwo
 \else \expandafter \@secondoftwo
\fi
}%
\providecommand \@ifnum [1]{%
 \ifnum #1\expandafter \@firstoftwo
 \else \expandafter \@secondoftwo
\fi
}%
\providecommand \enquote [1]{``#1''}%
\providecommand \bibnamefont  [1]{#1}%
\providecommand \bibfnamefont [1]{#1}%
\providecommand \citenamefont [1]{#1}%
\providecommand\href[0]{\@sanitize\@href}%
\providecommand\@href[1]{\endgroup\@@startlink{#1}\endgroup\@@href}%
\providecommand\@@href[1]{#1\@@endlink}%
\providecommand \@sanitize [0]{\begingroup\catcode`\&12\catcode`\#12\relax}%
\@ifxundefined \pdfoutput {\@firstoftwo}{%
 \@ifnum{\z@=\pdfoutput}{\@firstoftwo}{\@secondoftwo}%
}{%
 \providecommand\@@startlink[1]{\leavevmode\special{html:<a href="#1">}}%
 \providecommand\@@endlink[0]{\special{html:</a>}}%
}{%
 \providecommand\@@startlink[1]{%
  \leavevmode
  \pdfstartlink
   attr{/Border[0 0 1 ]/H/I/C[0 1 1]}%
   user{/Subtype/Link/A<</Type/Action/S/URI/URI(#1)>>}%
  \relax
 }%
 \providecommand\@@endlink[0]{\pdfendlink}%
}%
\providecommand \url  [0]{\begingroup\@sanitize \@url }%
\providecommand \@url [1]{\endgroup\@href {#1}{\urlprefix}}%
\providecommand \urlprefix [0]{URL }%
\providecommand \Eprint[0]{\href }%
\@ifxundefined \urlstyle {%
  \providecommand \doi [1]{doi:\discretionary{}{}{}#1}%
}{%
  \providecommand \doi [0]{doi:\discretionary{}{}{}\begingroup
  \urlstyle{rm}\Url }%
}%
\providecommand \doibase [0]{http://dx.doi.org/}%
\providecommand \Doi[1]{\href{\doibase#1}}%
\providecommand \bibAnnote [3]{%
  \BibitemShut{#1}%
  \begin{quotation}\noindent
    \textsc{Key:}\ #2\\\textsc{Annotation:}\ #3%
  \end{quotation}%
}%
\providecommand \bibAnnoteFile [2]{%
  \IfFileExists{#2}{\bibAnnote {#1} {#2} {\input{#2}}}{}%
}%
\providecommand \typeout [0]{\immediate \write \m@ne }%
\providecommand \selectlanguage [0]{\@gobble}%
\providecommand \bibinfo [0]{\@secondoftwo}%
\providecommand \bibfield [0]{\@secondoftwo}%
\providecommand \translation [1]{[#1]}%
\providecommand \BibitemOpen[0]{}%
\providecommand \bibitemStop [0]{}%
\providecommand \bibitemNoStop [0]{.\EOS\space}%
\providecommand \EOS [0]{\spacefactor3000\relax}%
\providecommand \BibitemShut [1]{\csname bibitem#1\endcsname}%
\bibitem{Hill:1994hp}%
  \BibitemOpen
  \bibfield{author}{%
  \bibinfo {author} {\bibfnamefont{C.~T.}\ \bibnamefont{Hill}},\ }%
  \bibfield{journal}{%
  \Doi{10.1016/0370-2693(94)01660-5}{\bibinfo {journal} {Phys. Lett. B}}\ }%
  \textbf{\bibinfo {volume} {345}},\ \bibinfo {pages} {483} (\bibinfo {year}
  {1995}),\ \Eprint{http://arxiv.org/abs/hep-ph/9411426}{arXiv:hep-ph/9411426}%
  \bibAnnoteFile{NoStop}{Hill:1994hp}%
\bibitem{topcolor2}%
  \BibitemOpen
  \bibfield{author}{%
  \bibinfo {author} {\bibnamefont{{R.~M.~Harris, C.~T.~Hill, and
  S.~J.~Parke,}}}\ }%
  \Eprint{http://arxiv.org/abs/hep-ph/9911288}{arXiv:hep-ph/9911288}%
  \bibAnnoteFile{NoStop}{topcolor2}%
\bibitem{Harris:2011ez}%
  \BibitemOpen
  \bibfield{author}{%
  \bibinfo {author} {\bibfnamefont{R.~M.}\ \bibnamefont{Harris}}\ and\ \bibinfo
  {author} {\bibfnamefont{S.}~\bibnamefont{Jain}},\ }%
  \bibfield{journal}{%
  \Doi{10.1140/epjc/s10052-012-2072-4}{\bibinfo {journal} {Eur. Phys. J. C}}\
  }%
  \textbf{\bibinfo {volume} {72}},\ \bibinfo {pages} {2072} (\bibinfo {year}
  {2012}),\ \Eprint{http://arxiv.org/abs/1112.4928}{arXiv:1112.4928}%
  \bibAnnoteFile{NoStop}{Harris:2011ez}%
\bibitem{Note1}%
  \BibitemOpen
  \bibinfo {note} {In common with other experimental searches, the specific
  model used is the leptophobic scenario, model IV in Ref.~\cite {topcolor2}
  with $f_1 =1$ and $f_2=0$. The corrections to the Lagrangian discussed in
  Ref.~\cite {Harris:2011ez} are included.}%
  \bibAnnoteFile{Stop}{Note1}%
\bibitem{Lillie:2007yh}%
  \BibitemOpen
  \bibfield{author}{%
  \bibinfo {author} {\bibfnamefont{B.}~\bibnamefont{Lillie}}, \bibinfo {author}
  {\bibfnamefont{L.}~\bibnamefont{Randall}},\ and\ \bibinfo {author}
  {\bibfnamefont{L.-T.}\ \bibnamefont{Wang}},\ }%
  \bibfield{journal}{%
  \Doi{10.1088/1126-6708/2007/09/074}{\bibinfo {journal} {J. High Energy
  Phys.}}\ }%
  \textbf{\bibinfo {volume} {0709}},\ \bibinfo {pages} {074} (\bibinfo {year}
  {2007}),\ \Eprint{http://arxiv.org/abs/hep-ph/0701166}{arXiv:hep-ph/0701166}%
  \bibAnnoteFile{NoStop}{Lillie:2007yh}%
\bibitem{Lillie:2007ve}%
  \BibitemOpen
  \bibfield{author}{%
  \bibinfo {author} {\bibfnamefont{B.}~\bibnamefont{Lillie}}, \bibinfo {author}
  {\bibfnamefont{J.}~\bibnamefont{Shu}},\ and\ \bibinfo {author}
  {\bibfnamefont{T.~M.~P.}\ \bibnamefont{Tait}},\ }%
  \bibfield{journal}{%
  \Doi{10.1103/PhysRevD.76.115016}{\bibinfo {journal} {Phys. Rev. D}}\ }%
  \textbf{\bibinfo {volume} {76}},\ \bibinfo {pages} {115016} (\bibinfo {year}
  {2007}),\ \Eprint{http://arxiv.org/abs/0706.3960}{arXiv:0706.3960}%
  \bibAnnoteFile{NoStop}{Lillie:2007ve}%
\bibitem{Agashe:2006hk}%
  \BibitemOpen
  \bibfield{author}{%
  \bibinfo {author} {\bibfnamefont{K.}~\bibnamefont{Agashe}} \emph{et~al.},\ }%
  \bibfield{journal}{%
  \Doi{10.1103/PhysRevD.77.015003}{\bibinfo {journal} {Phys. Rev. D}}\ }%
  \textbf{\bibinfo {volume} {77}},\ \bibinfo {pages} {015003} (\bibinfo {year}
  {2008}),\ \Eprint{http://arxiv.org/abs/hep-ph/0612015}{arXiv:hep-ph/0612015}%
  \bibAnnoteFile{NoStop}{Agashe:2006hk}%
\bibitem{Djouadi:2007eg}%
  \BibitemOpen
  \bibfield{author}{%
  \bibinfo {author} {\bibfnamefont{A.}~\bibnamefont{Djouadi}}, \bibinfo
  {author} {\bibfnamefont{G.}~\bibnamefont{Moreau}},\ and\ \bibinfo {author}
  {\bibfnamefont{R.~K.}\ \bibnamefont{Singh}},\ }%
  \bibfield{journal}{%
  \Doi{10.1016/j.nuclphysb.2007.12.024}{\bibinfo {journal} {Nucl. Phys. B}}\ }%
  \textbf{\bibinfo {volume} {797}},\ \bibinfo {pages} {1} (\bibinfo {year}
  {2008}),\ \Eprint{http://arxiv.org/abs/0706.4191}{arXiv:0706.4191}%
  \bibAnnoteFile{NoStop}{Djouadi:2007eg}%
\bibitem{Agashe:2007zd}%
  \BibitemOpen
  \bibfield{author}{%
  \bibinfo {author} {\bibfnamefont{K.}~\bibnamefont{Agashe}}, \bibinfo {author}
  {\bibfnamefont{H.}~\bibnamefont{Davoudiasl}}, \bibinfo {author}
  {\bibfnamefont{G.}~\bibnamefont{Perez}},\ and\ \bibinfo {author}
  {\bibfnamefont{A.}~\bibnamefont{Soni}},\ }%
  \bibfield{journal}{%
  \Doi{10.1103/PhysRevD.76.036006}{\bibinfo {journal} {Phys. Rev. D}}\ }%
  \textbf{\bibinfo {volume} {76}},\ \bibinfo {pages} {036006} (\bibinfo {year}
  {2007}),\ \Eprint{http://arxiv.org/abs/hep-ph/0701186}{arXiv:hep-ph/0701186}%
  \bibAnnoteFile{NoStop}{Agashe:2007zd}%
\bibitem{Aad:2012wm}%
  \BibitemOpen
  \bibfield{author}{%
  \bibinfo {author} {\bibnamefont{\mbox{ATLAS Collaboration}}},\ }%
  \bibfield{journal}{%
  \bibinfo {journal} {Eur. Phys. J. C}\ }%
  \textbf{\bibinfo {volume} {72}},\ \bibinfo {pages} {2083} (\bibinfo {year}
  {2012}),\ \Eprint{http://arxiv.org/abs/1205.5371}{arXiv:1205.5371}%
  \bibAnnoteFile{NoStop}{Aad:2012wm}%
\bibitem{Arce:1431929}%
  \BibitemOpen
  \bibfield{author}{%
  \bibinfo {author} {\bibnamefont{\mbox{ATLAS Collaboration}}},\ }%
  \bibfield{journal}{%
  \Doi{10.1007/JHEP09(2012)041}{\bibinfo {journal} {J. High Energy Phys.}}\ }%
  \textbf{\bibinfo {volume} {1209}},\ \bibinfo {pages} {041} (\bibinfo {year}
  {2012}),\ \Eprint{http://arxiv.org/abs/1207.2409}{arXiv:1207.2409}%
  \bibAnnoteFile{NoStop}{Arce:1431929}%
\bibitem{CMS:2012cx}%
  \BibitemOpen
  \bibfield{author}{%
  \bibinfo {author} {\bibnamefont{\mbox{CMS Collaboration}}},\ }%
  \bibfield{journal}{%
  \Doi{10.1007/JHEP12(2012)015}{\bibinfo {journal} {J. High Energy Phys.}}\ }%
  \textbf{\bibinfo {volume} {1212}},\ \bibinfo {pages} {015} (\bibinfo {year}
  {2012}),\ \Eprint{http://arxiv.org/abs/1209.4397}{arXiv:1209.4397}%
  \bibAnnoteFile{NoStop}{CMS:2012cx}%
\bibitem{cms:2012rq}%
  \BibitemOpen
  \bibfield{author}{%
  \bibinfo {author} {\bibnamefont{\mbox{CMS Collaboration}}},\ }%
  \bibfield{journal}{%
  \Doi{10.1103/PhysRevD.87.072002}{\bibinfo {journal} {Phys. Rev. D}}\ }%
  \textbf{\bibinfo {volume} {87}},\ \bibinfo {pages} {072002} (\bibinfo {year}
  {2013}),\ \Eprint{http://arxiv.org/abs/1211.3338}{arXiv:1211.3338}%
  \bibAnnoteFile{NoStop}{cms:2012rq}%
\bibitem{ATLAS:2012qa}%
  \BibitemOpen
  \bibfield{author}{%
  \bibinfo {author} {\bibnamefont{\mbox{ATLAS Collaboration}}},\ }%
  \bibfield{journal}{%
  \Doi{10.1007/JHEP01(2013)116}{\bibinfo {journal} {J. High Energy Phys.}}\ }%
  \textbf{\bibinfo {volume} {1301}},\ \bibinfo {pages} {116} (\bibinfo {year}
  {2012}),\ \Eprint{http://arxiv.org/abs/1211.2202}{arXiv:1211.2202}%
  \bibAnnoteFile{NoStop}{ATLAS:2012qa}%
\bibitem{Chatrchyan:2012ku}%
  \BibitemOpen
  \bibfield{author}{%
  \bibinfo {author} {\bibnamefont{\mbox{CMS Collaboration}}},\ }%
  \bibfield{journal}{%
  \Doi{10.1007/JHEP09(2012)029}{\bibinfo {journal} {J. High Energy Phys.}}\ }%
  \textbf{\bibinfo {volume} {1209}},\ \bibinfo {pages} {029} (\bibinfo {year}
  {2012}),\ \Eprint{http://arxiv.org/abs/1204.2488}{arXiv:1204.2488}%
  \bibAnnoteFile{NoStop}{Chatrchyan:2012ku}%
\bibitem{Aaltonen:2011ts}%
  \BibitemOpen
  \bibfield{author}{%
  \bibinfo {author} {\bibnamefont{{CDF Collaboration, T.~Aaltonen
  \emph{et~al}.}}},\ }%
  \bibfield{journal}{%
  \Doi{10.1103/PhysRevD.84.072004}{\bibinfo {journal} {Phys. Rev. D}}\ }%
  \textbf{\bibinfo {volume} {84}},\ \bibinfo {pages} {072004} (\bibinfo {year}
  {2011}),\ \Eprint{http://arxiv.org/abs/1107.5063}{arXiv:1107.5063}%
  \bibAnnoteFile{NoStop}{Aaltonen:2011ts}%
\bibitem{Aaltonen:2011vi}%
  \BibitemOpen
  \bibfield{author}{%
  \bibinfo {author} {\bibnamefont{{CDF Collaboration, T.~Aaltonen
  \emph{et~al}.}}},\ }%
  \bibfield{journal}{%
  \bibinfo {journal} {Phys. Rev. D}\ }%
  \textbf{\bibinfo {volume} {84}},\ \bibinfo {pages} {072003} (\bibinfo {year}
  {2011}),\ \Eprint{http://arxiv.org/abs/1108.4755}{arXiv:1108.4755}%
  \bibAnnoteFile{NoStop}{Aaltonen:2011vi}%
\bibitem{Abazov:2011gv}%
  \BibitemOpen
  \bibfield{author}{%
  \bibinfo {author} {\bibnamefont{{D0 Collaboration, V.~Abazov
  \emph{et~al}.}}},\ }%
  \bibfield{journal}{%
  \bibinfo {journal} {Phys. Rev. D}\ }%
  \textbf{\bibinfo {volume} {85}},\ \bibinfo {pages} {051101} (\bibinfo {year}
  {2012}),\ \Eprint{http://arxiv.org/abs/1111.1271}{arXiv:1111.1271}%
  \bibAnnoteFile{NoStop}{Abazov:2011gv}%
\bibitem{Aaltonen:2012af}%
  \BibitemOpen
  \bibfield{author}{%
  \bibinfo {author} {\bibnamefont{{CDF Collaboration, T.~Aaltonen
  \emph{et~al}.}}}\ }%
  \Eprint{http://arxiv.org/abs/1211.5363}{arXiv:1211.5363}%
  \bibAnnoteFile{NoStop}{Aaltonen:2012af}%
\bibitem{Aad:2008zzm}%
  \BibitemOpen
  \bibfield{author}{%
  \bibinfo {author} {\bibnamefont{\mbox{ATLAS Collaboration}}},\ }%
  \bibfield{journal}{%
  \Doi{10.1088/1748-0221/3/08/S08003}{\bibinfo {journal} {JINST}}\ }%
  \textbf{\bibinfo {volume} {3}},\ \bibinfo {pages} {S08003} (\bibinfo {year}
  {2008})%
  \bibAnnoteFile{NoStop}{Aad:2008zzm}%
\bibitem{Aad:2012xs}%
  \BibitemOpen
  \bibfield{author}{%
  \bibinfo {author} {\bibnamefont{\mbox{ATLAS Collaboration}}},\ }%
  \bibfield{journal}{%
  \Doi{10.1140/epjc/s10052-011-1849-1}{\bibinfo {journal} {Eur. Phys. J. C}}\
  }%
  \textbf{\bibinfo {volume} {72}},\ \bibinfo {pages} {1849} (\bibinfo {year}
  {2012}),\ \Eprint{http://arxiv.org/abs/1110.1530}{arXiv:1110.1530}%
  \bibAnnoteFile{NoStop}{Aad:2012xs}%
\bibitem{Agostinelli:2002hh}%
  \BibitemOpen
  \bibfield{author}{%
  \bibinfo {author} {\bibfnamefont{S.}~\bibnamefont{Agostinelli}} \emph{et~al.}
  (\bibinfo {collaboration} {GEANT4}),\ }%
  \bibfield{journal}{%
  \Doi{10.1016/S0168-9002(03)01368-8}{\bibinfo {journal} {Nucl. Instrum.
  Methods Phys. Res., Sect. A}}\ }%
  \textbf{\bibinfo {volume} {506}},\ \bibinfo {pages} {250} (\bibinfo {year}
  {2003})%
  \bibAnnoteFile{NoStop}{Agostinelli:2002hh}%
\bibitem{Aad:2010ah}%
  \BibitemOpen
  \bibfield{author}{%
  \bibinfo {author} {\bibnamefont{\mbox{ATLAS Collaboration}}},\ }%
  \bibfield{journal}{%
  \Doi{10.1140/epjc/s10052-010-1429-9}{\bibinfo {journal} {Eur. Phys. J. C}}\
  }%
  \textbf{\bibinfo {volume} {70}},\ \bibinfo {pages} {823} (\bibinfo {year}
  {2010}),\ \Eprint{http://arxiv.org/abs/1005.4568}{arXiv:1005.4568}%
  \bibAnnoteFile{NoStop}{Aad:2010ah}%
\bibitem{Frixione:2002jk}%
  \BibitemOpen
  \bibfield{author}{%
  \bibinfo {author} {\bibfnamefont{S.}~\bibnamefont{Frixione}}\ and\ \bibinfo
  {author} {\bibfnamefont{B.~R.}\ \bibnamefont{Webber}},\ }%
  \bibfield{journal}{%
  \bibinfo {journal} {J. High Energy Phys.}\ }%
  \textbf{\bibinfo {volume} {0206}},\ \bibinfo {pages} {029} (\bibinfo {year}
  {2002}),\ \Eprint{http://arxiv.org/abs/hep-ph/0204244}{arXiv:hep-ph/0204244}%
  \bibAnnoteFile{NoStop}{Frixione:2002jk}%
\bibitem{Frixione:2003ei}%
  \BibitemOpen
  \bibfield{author}{%
  \bibinfo {author} {\bibfnamefont{S.}~\bibnamefont{Frixione}}, \bibinfo
  {author} {\bibfnamefont{P.}~\bibnamefont{Nason}},\ and\ \bibinfo {author}
  {\bibfnamefont{B.~R.}\ \bibnamefont{Webber}},\ }%
  \bibfield{journal}{%
  \bibinfo {journal} {J. High Energy Phys.}\ }%
  \textbf{\bibinfo {volume} {0308}},\ \bibinfo {pages} {007} (\bibinfo {year}
  {2003}),\ \Eprint{http://arxiv.org/abs/hep-ph/0305252}{arXiv:hep-ph/0305252}%
  \bibAnnoteFile{NoStop}{Frixione:2003ei}%
\bibitem{Frixione:2010wd}%
  \BibitemOpen
  \bibfield{author}{%
  \bibinfo {author} {\bibfnamefont{S.}~\bibnamefont{Frixione}}, \bibinfo
  {author} {\bibfnamefont{F.}~\bibnamefont{Stoeckli}}, \bibinfo {author}
  {\bibfnamefont{P.}~\bibnamefont{Torrielli}}, \bibinfo {author}
  {\bibfnamefont{B.~R.}\ \bibnamefont{Webber}},\ and\ \bibinfo {author}
  {\bibfnamefont{C.~D.}\ \bibnamefont{White}}}%
   (\bibinfo {year} {2010}),\
  \Eprint{http://arxiv.org/abs/1010.0819}{arXiv:1010.0819}%
  \bibAnnoteFile{NoStop}{Frixione:2010wd}%
\bibitem{HerwigGC}%
  \BibitemOpen
  \bibfield{author}{%
  \bibinfo {author} {\bibfnamefont{G.}~\bibnamefont{Corcella}} \emph{et~al.},\
  }%
  \bibfield{journal}{%
  \bibinfo {journal} {J. High Energy Phys.}\ }%
  \textbf{\bibinfo {volume} {0101}},\ \bibinfo {pages} {010} (\bibinfo {year}
  {2001}),\ \Eprint{http://arxiv.org/abs/hep-ph/0011363}{hep-ph/0011363}%
  \bibAnnoteFile{NoStop}{HerwigGC}%
\bibitem{Corcella:2002jc}%
  \BibitemOpen
  \bibfield{author}{%
  \bibinfo {author} {\bibnamefont{{G.~Corcella \emph{et al.},}}}\ }%
  \Eprint{http://arxiv.org/abs/hep-ph/0210213}{arXiv:hep-ph/0210213}%
  \bibAnnoteFile{NoStop}{Corcella:2002jc}%
\bibitem{Butterworth:1996zw}%
  \BibitemOpen
  \bibfield{author}{%
  \bibinfo {author} {\bibfnamefont{J.~M.}\ \bibnamefont{Butterworth}}, \bibinfo
  {author} {\bibfnamefont{J.~R.}\ \bibnamefont{Forshaw}},\ and\ \bibinfo
  {author} {\bibfnamefont{M.~H.}\ \bibnamefont{Seymour}},\ }%
  \bibfield{journal}{%
  \Doi{10.1007/s002880050286}{\bibinfo {journal} {Z. Phys. C}}\ }%
  \textbf{\bibinfo {volume} {72}},\ \bibinfo {pages} {637} (\bibinfo {year}
  {1996}),\ \Eprint{http://arxiv.org/abs/hep-ph/9601371}{arXiv:hep-ph/9601371}%
  \bibAnnoteFile{NoStop}{Butterworth:1996zw}%
\bibitem{Lai:2010vv}%
  \BibitemOpen
  \bibfield{author}{%
  \bibinfo {author} {\bibfnamefont{H.-L.}\ \bibnamefont{Lai}} \emph{et~al.},\
  }%
  \bibfield{journal}{%
  \Doi{10.1103/PhysRevD.82.074024}{\bibinfo {journal} {Phys. Rev. D}}\ }%
  \textbf{\bibinfo {volume} {82}},\ \bibinfo {pages} {074024} (\bibinfo {year}
  {2010}),\ \Eprint{http://arxiv.org/abs/1007.2241}{arXiv:1007.2241}%
  \bibAnnoteFile{NoStop}{Lai:2010vv}%
\bibitem{Beneke:2009ye}%
  \BibitemOpen
  \bibfield{author}{%
  \bibinfo {author} {\bibfnamefont{M.}~\bibnamefont{Beneke}} \emph{et~al.},\ }%
  \bibfield{journal}{%
  \Doi{10.1016/j.physletb.2010.05.038}{\bibinfo {journal} {Phys. Lett. B}}\ }%
  \textbf{\bibinfo {volume} {690}},\ \bibinfo {pages} {483} (\bibinfo {year}
  {2010}),\ \Eprint{http://arxiv.org/abs/0911.5166}{arXiv:0911.5166}%
  \bibAnnoteFile{NoStop}{Beneke:2009ye}%
\bibitem{Aliev:2010zk}%
  \BibitemOpen
  \bibfield{author}{%
  \bibinfo {author} {\bibfnamefont{M.}~\bibnamefont{Aliev}} \emph{et~al.},\ }%
  \bibfield{journal}{%
  \Doi{10.1016/j.cpc.2010.12.040}{\bibinfo {journal} {Comput. Phys. Commun.}}\
  }%
  \textbf{\bibinfo {volume} {182}},\ \bibinfo {pages} {1034} (\bibinfo {year}
  {2011}),\ \Eprint{http://arxiv.org/abs/1007.1327}{arXiv:1007.1327}%
  \bibAnnoteFile{NoStop}{Aliev:2010zk}%
\bibitem{Frixione:2007vw}%
  \BibitemOpen
  \bibfield{author}{%
  \bibinfo {author} {\bibfnamefont{S.}~\bibnamefont{Frixione}}, \bibinfo
  {author} {\bibfnamefont{P.}~\bibnamefont{Nason}},\ and\ \bibinfo {author}
  {\bibfnamefont{C.}~\bibnamefont{Oleari}},\ }%
  \bibfield{journal}{%
  \Doi{10.1088/1126-6708/2007/11/070}{\bibinfo {journal} {J. High Energy
  Phys.}}\ }%
  \textbf{\bibinfo {volume} {0711}},\ \bibinfo {pages} {070} (\bibinfo {year}
  {2007}),\ \Eprint{http://arxiv.org/abs/0709.2092}{arXiv:0709.2092}%
  \bibAnnoteFile{NoStop}{Frixione:2007vw}%
\bibitem{Frixione:2005vw}%
  \BibitemOpen
  \bibfield{author}{%
  \bibinfo {author} {\bibfnamefont{S.}~\bibnamefont{Frixione}}, \bibinfo
  {author} {\bibfnamefont{E.}~\bibnamefont{Laenen}}, \bibinfo {author}
  {\bibfnamefont{P.}~\bibnamefont{Motylinski}},\ and\ \bibinfo {author}
  {\bibfnamefont{B.~R.}\ \bibnamefont{Webber}},\ }%
  \bibfield{journal}{%
  \Doi{10.1088/1126-6708/2006/03/092}{\bibinfo {journal} {J. High Energy
  Phys.}}\ }%
  \textbf{\bibinfo {volume} {0603}},\ \bibinfo {pages} {092} (\bibinfo {year}
  {2006}),\ \Eprint{http://arxiv.org/abs/hep-ph/0512250}{arXiv:hep-ph/0512250}%
  \bibAnnoteFile{NoStop}{Frixione:2005vw}%
\bibitem{Frixione:2008yi}%
  \BibitemOpen
  \bibfield{author}{%
  \bibinfo {author} {\bibfnamefont{S.}~\bibnamefont{Frixione}} \emph{et~al.},\
  }%
  \bibfield{journal}{%
  \Doi{10.1088/1126-6708/2008/07/029}{\bibinfo {journal} {J. High Energy
  Phys.}}\ }%
  \textbf{\bibinfo {volume} {0807}},\ \bibinfo {pages} {029} (\bibinfo {year}
  {2008}),\ \Eprint{http://arxiv.org/abs/0805.3067}{arXiv:0805.3067}%
  \bibAnnoteFile{NoStop}{Frixione:2008yi}%
\bibitem{SAMPLES-ACER}%
  \BibitemOpen
  \bibfield{author}{%
  \bibinfo {author} {\bibnamefont{{B.~P.~Kersevan and E.~Richter-Was,}}}\ }%
  \Eprint{http://arxiv.org/abs/hep-ph/0405247}{arXiv:hep-ph/0405247}%
  \bibAnnoteFile{NoStop}{SAMPLES-ACER}%
\bibitem{Sjostrand:2006za}%
  \BibitemOpen
  \bibfield{author}{%
  \bibinfo {author} {\bibfnamefont{T.}~\bibnamefont{Sj{\"o}strand}}, \bibinfo
  {author} {\bibfnamefont{S.}~\bibnamefont{Mrenna}},\ and\ \bibinfo {author}
  {\bibfnamefont{P.~Z.}\ \bibnamefont{Skands}},\ }%
  \bibfield{journal}{%
  \Doi{10.1088/1126-6708/2006/05/026}{\bibinfo {journal} {J. High Energy
  Phys.}}\ }%
  \textbf{\bibinfo {volume} {0605}},\ \bibinfo {pages} {026} (\bibinfo {year}
  {2006}),\ \Eprint{http://arxiv.org/abs/hep-ph/0603175}{arXiv:hep-ph/0603175}%
  \bibAnnoteFile{NoStop}{Sjostrand:2006za}%
\bibitem{Kidonakis:2011wy}%
  \BibitemOpen
  \bibfield{author}{%
  \bibinfo {author} {\bibfnamefont{N.}~\bibnamefont{Kidonakis}},\ }%
  \bibfield{journal}{%
  \Doi{10.1103/PhysRevD.83.091503}{\bibinfo {journal} {Phys. Rev. D}}\ }%
  \textbf{\bibinfo {volume} {83}},\ \bibinfo {pages} {091503} (\bibinfo {year}
  {2011}),\ \Eprint{http://arxiv.org/abs/1103.2792}{arXiv:1103.2792}%
  \bibAnnoteFile{NoStop}{Kidonakis:2011wy}%
\bibitem{Kidonakis:2010ux}%
  \BibitemOpen
  \bibfield{author}{%
  \bibinfo {author} {\bibfnamefont{N.}~\bibnamefont{Kidonakis}},\ }%
  \bibfield{journal}{%
  \Doi{10.1103/PhysRevD.82.054018}{\bibinfo {journal} {Phys. Rev. D}}\ }%
  \textbf{\bibinfo {volume} {82}},\ \bibinfo {pages} {054018} (\bibinfo {year}
  {2010}),\ \Eprint{http://arxiv.org/abs/1005.4451}{arXiv:1005.4451}%
  \bibAnnoteFile{NoStop}{Kidonakis:2010ux}%
\bibitem{Kidonakis:2010tc}%
  \BibitemOpen
  \bibfield{author}{%
  \bibinfo {author} {\bibfnamefont{N.}~\bibnamefont{Kidonakis}},\ }%
  \bibfield{journal}{%
  \Doi{10.1103/PhysRevD.81.054028}{\bibinfo {journal} {Phys. Rev. D}}\ }%
  \textbf{\bibinfo {volume} {81}},\ \bibinfo {pages} {054028} (\bibinfo {year}
  {2010}),\ \Eprint{http://arxiv.org/abs/1001.5034}{arXiv:1001.5034}%
  \bibAnnoteFile{NoStop}{Kidonakis:2010tc}%
\bibitem{Mangano:2002ea}%
  \BibitemOpen
  \bibfield{author}{%
  \bibinfo {author} {\bibfnamefont{M.~L.}\ \bibnamefont{Mangano}}
  \emph{et~al.},\ }%
  \bibfield{journal}{%
  \bibinfo {journal} {J. High Energy Phys.}\ }%
  \textbf{\bibinfo {volume} {0307}},\ \bibinfo {pages} {001} (\bibinfo {year}
  {2003}),\ \Eprint{http://arxiv.org/abs/hep-ph/0206293}{arXiv:hep-ph/0206293}%
  \bibAnnoteFile{NoStop}{Mangano:2002ea}%
\bibitem{Alwall:2007fs}%
  \BibitemOpen
  \bibfield{author}{%
  \bibinfo {author} {\bibfnamefont{J.}~\bibnamefont{Alwall}} \emph{et~al.},\ }%
  \bibfield{journal}{%
  \Doi{10.1140/epjc/s10052-007-0490-5}{\bibinfo {journal} {Eur. Phys. J. C}}\
  }%
  \textbf{\bibinfo {volume} {53}},\ \bibinfo {pages} {473} (\bibinfo {year}
  {2008}),\ \Eprint{http://arxiv.org/abs/0706.2569}{arXiv:0706.2569}%
  \bibAnnoteFile{NoStop}{Alwall:2007fs}%
\bibitem{Pumplin:2002vw}%
  \BibitemOpen
  \bibfield{author}{%
  \bibinfo {author} {\bibfnamefont{J.}~\bibnamefont{Pumplin}} \emph{et~al.},\
  }%
  \bibfield{journal}{%
  \bibinfo {journal} {J. High Energy Phys.}\ }%
  \textbf{\bibinfo {volume} {0207}},\ \bibinfo {pages} {012} (\bibinfo {year}
  {2002}),\ \Eprint{http://arxiv.org/abs/hep-ph/0201195}{arXiv:hep-ph/0201195}%
  \bibAnnoteFile{NoStop}{Pumplin:2002vw}%
\bibitem{Hamberg:1990np}%
  \BibitemOpen
  \bibfield{author}{%
  \bibinfo {author} {\bibfnamefont{R.}~\bibnamefont{Hamberg}}, \bibinfo
  {author} {\bibfnamefont{W.}~\bibnamefont{van Neerven}},\ and\ \bibinfo
  {author} {\bibfnamefont{T.}~\bibnamefont{Matsuura}},\ }%
  \bibfield{journal}{%
  \Doi{10.1016/0550-3213(91)90064-5, 10.1016/0550-3213(91)90064-5}{\bibinfo
  {journal} {Nucl. Phys. B}}\ }%
  \textbf{\bibinfo {volume} {359}},\ \bibinfo {pages} {343} (\bibinfo {year}
  {1991})%
  \bibAnnoteFile{NoStop}{Hamberg:1990np}%
\bibitem{Gavin:2012sy}%
  \BibitemOpen
  \bibfield{author}{%
  \bibinfo {author} {\bibnamefont{{R.~Gavin, Y.~Li, F.~Petriello and
  S.~Quackenbush,}}}\ }%
  \Eprint{http://arxiv.org/abs/1201.5896}{arXiv:1201.5896}%
  \bibAnnoteFile{NoStop}{Gavin:2012sy}%
\bibitem{Sherstnev:2007nd}%
  \BibitemOpen
  \bibfield{author}{%
  \bibinfo {author} {\bibfnamefont{A.}~\bibnamefont{Sherstnev}}\ and\ \bibinfo
  {author} {\bibfnamefont{R.}~\bibnamefont{Thorne}},\ }%
  \bibfield{journal}{%
  \Doi{10.1140/epjc/s10052-008-0610-x}{\bibinfo {journal} {Eur. Phys. J. C}}\
  }%
  \textbf{\bibinfo {volume} {55}},\ \bibinfo {pages} {553} (\bibinfo {year}
  {2008}),\ \Eprint{http://arxiv.org/abs/0711.2473}{arXiv:0711.2473}%
  \bibAnnoteFile{NoStop}{Sherstnev:2007nd}%
\bibitem{Campbell:1999ah}%
  \BibitemOpen
  \bibfield{author}{%
  \bibinfo {author} {\bibfnamefont{J.~M.}\ \bibnamefont{Campbell}}\ and\
  \bibinfo {author} {\bibfnamefont{R.~K.}\ \bibnamefont{Ellis}},\ }%
  \bibfield{journal}{%
  \Doi{10.1103/PhysRevD.60.113006}{\bibinfo {journal} {Phys. Rev. D}}\ }%
  \textbf{\bibinfo {volume} {60}},\ \bibinfo {pages} {113006} (\bibinfo {year}
  {1999}),\ \Eprint{http://arxiv.org/abs/hep-ph/9905386}{arXiv:hep-ph/9905386}%
  \bibAnnoteFile{NoStop}{Campbell:1999ah}%
\bibitem{Aad:2010ey}%
  \BibitemOpen
  \bibfield{author}{%
  \bibinfo {author} {\bibnamefont{\mbox{ATLAS Collaboration}}},\ }%
  \bibfield{journal}{%
  \Doi{10.1140/epjc/s10052-011-1577-6}{\bibinfo {journal} {Eur. Phys. J. C}}\
  }%
  \textbf{\bibinfo {volume} {71}},\ \bibinfo {pages} {1577} (\bibinfo {year}
  {2011}),\ \Eprint{http://arxiv.org/abs/1012.1792}{arXiv:1012.1792}%
  \bibAnnoteFile{NoStop}{Aad:2010ey}%
\bibitem{Gao:2010bb}%
  \BibitemOpen
  \bibfield{author}{%
  \bibinfo {author} {\bibfnamefont{J.}~\bibnamefont{Gao}} \emph{et~al.},\ }%
  \bibfield{journal}{%
  \Doi{10.1103/PhysRevD.82.014020}{\bibinfo {journal} {Phys. Rev. D}}\ }%
  \textbf{\bibinfo {volume} {82}},\ \bibinfo {pages} {014020} (\bibinfo {year}
  {2010}),\ \Eprint{http://arxiv.org/abs/1004.0876}{arXiv:1004.0876}%
  \bibAnnoteFile{NoStop}{Gao:2010bb}%
\bibitem{Note2}%
  \BibitemOpen
  \bibinfo {note} {A recent full NLO calculation~\cite {Caola:2012rs} gives
  smaller $K$-factors, which can partly be attributed to the use of different
  parameters than those in Ref.~\cite {Gao:2010bb}. The parameters used for
  signal generation in this paper corresponds more closely to Ref.~\cite
  {Gao:2010bb}.}%
  \bibAnnoteFile{Stop}{Note2}%
\bibitem{Alwall:2007st}%
  \BibitemOpen
  \bibfield{author}{%
  \bibinfo {author} {\bibfnamefont{J.}~\bibnamefont{Alwall}} \emph{et~al.},\ }%
  \bibfield{journal}{%
  \Doi{10.1088/1126-6708/2007/09/028}{\bibinfo {journal} {J. High Energy
  Phys.}}\ }%
  \textbf{\bibinfo {volume} {0709}},\ \bibinfo {pages} {028} (\bibinfo {year}
  {2007}),\ \Eprint{http://arxiv.org/abs/0706.2334}{arXiv:0706.2334}%
  \bibAnnoteFile{NoStop}{Alwall:2007st}%
\bibitem{Aad:2012qf}%
  \BibitemOpen
  \bibfield{author}{%
  \bibinfo {author} {\bibnamefont{\mbox{ATLAS Collaboration}}},\ }%
  \bibfield{journal}{%
  \Doi{10.1016/j.physletb.2012.03.083}{\bibinfo {journal} {Phys. Lett. B}}\ }%
  \textbf{\bibinfo {volume} {711}},\ \bibinfo {pages} {244} (\bibinfo {year}
  {2012}),\ \Eprint{http://arxiv.org/abs/1201.1889}{arXiv:1201.1889}%
  \bibAnnoteFile{NoStop}{Aad:2012qf}%
\bibitem{Cacciari:2008gp}%
  \BibitemOpen
  \bibfield{author}{%
  \bibinfo {author} {\bibfnamefont{M.}~\bibnamefont{Cacciari}}, \bibinfo
  {author} {\bibfnamefont{G.~P.}\ \bibnamefont{Salam}},\ and\ \bibinfo {author}
  {\bibfnamefont{G.}~\bibnamefont{Soyez}},\ }%
  \bibfield{journal}{%
  \Doi{10.1088/1126-6708/2008/04/063}{\bibinfo {journal} {J. High Energy
  Phys.}}\ }%
  \textbf{\bibinfo {volume} {0804}},\ \bibinfo {pages} {063} (\bibinfo {year}
  {2008}),\ \Eprint{http://arxiv.org/abs/0802.1189}{arXiv:0802.1189}%
  \bibAnnoteFile{NoStop}{Cacciari:2008gp}%
\bibitem{Cacciari:2011ma}%
  \BibitemOpen
  \bibfield{author}{%
  \bibinfo {author} {\bibfnamefont{M.}~\bibnamefont{Cacciari}}, \bibinfo
  {author} {\bibfnamefont{G.~P.}\ \bibnamefont{Salam}},\ and\ \bibinfo {author}
  {\bibfnamefont{G.}~\bibnamefont{Soyez}},\ }%
  \bibfield{journal}{%
  \bibinfo {journal} {Eur. Phys. J. C}\ }%
  \textbf{\bibinfo {volume} {72}},\ \bibinfo {pages} {1896} (\bibinfo {year}
  {2012}),\ \Eprint{http://arxiv.org/abs/1111.6097}{arXiv:1111.6097}%
  \bibAnnoteFile{NoStop}{Cacciari:2011ma}%
\bibitem{Lampl:2008zz}%
  \BibitemOpen
  \bibfield{author}{%
  \bibinfo {author} {\bibfnamefont{W.}~\bibnamefont{Lampl}} \emph{et~al.},\ }%
  \bibfield{journal}{%
  \bibinfo {journal} {{ATL-LARG-PUB-2008-002}}}%
   (\bibinfo {year} {2008}),\ \bibinfo {note}
  {{https://cdsweb.cern.ch/record/1099735}}%
  \bibAnnoteFile{NoStop}{Lampl:2008zz}%
\bibitem{Aad:2011he}%
  \BibitemOpen
  \bibfield{author}{%
  \bibinfo {author} {\bibnamefont{\mbox{ATLAS Collaboration}}},\ }%
  \bibfield{journal}{%
  \Doi{10.1140/epjc/s10052-013-2304-2}{\bibinfo {journal} {Eur. Phys. J. C}}\
  }%
  \textbf{\bibinfo {volume} {73}},\ \bibinfo {pages} {2304} (\bibinfo {year}
  {2013}),\ \Eprint{http://arxiv.org/abs/1112.6426}{arXiv:1112.6426}%
  \bibAnnoteFile{NoStop}{Aad:2011he}%
\bibitem{ISS-0501}%
  \BibitemOpen
  \bibfield{author}{%
  \bibinfo {author} {\bibfnamefont{{\c C}.}~\bibnamefont{{\.I}{\c s}sever}},
  \bibinfo {author} {\bibfnamefont{K.}~\bibnamefont{Borras}},\ and\ \bibinfo
  {author} {\bibfnamefont{D.}~\bibnamefont{Wegener}},\ }%
  \bibfield{journal}{%
  \bibinfo {journal} {Nucl. Instrum. Methods Phys. Res., Sect. A}\ }%
  \textbf{\bibinfo {volume} {545}},\ \bibinfo {pages} {803} (\bibinfo {year}
  {2005}),\
  \Eprint{http://arxiv.org/abs/physics/0408129}{arXiv:physics/0408129}%
  \bibAnnoteFile{NoStop}{ISS-0501}%
\bibitem{Barillari:2009zza}%
  \BibitemOpen
  \bibfield{author}{%
  \bibinfo {author} {\bibfnamefont{T.}~\bibnamefont{Barillari}} \emph{et~al.},\
  }%
  \bibfield{journal}{%
  \bibinfo {journal} {{ATL-LARG-PUB-2009-001-2}}}%
   (\bibinfo {year} {2009}),\ \bibinfo {note}
  {{https://cdsweb.cern.ch/record/1112035}}%
  \bibAnnoteFile{NoStop}{Barillari:2009zza}%
\bibitem{ATLAS-CONF-2012-065}%
  \BibitemOpen
  \bibfield{author}{%
  \bibinfo {author} {\bibnamefont{\mbox{ATLAS Collaboration}}},\ }%
  \bibfield{journal}{%
  \bibinfo {journal} {{ATLAS-CONF-2012-065~}}}%
   (\bibinfo {year} {2012}),\ \bibinfo {note}
  {{https://cdsweb.cern.ch/record/1459530}}%
  \bibAnnoteFile{NoStop}{ATLAS-CONF-2012-065}%
\bibitem{ATLAS-CONF-2012-043}%
  \BibitemOpen
  \bibfield{author}{%
  \bibinfo {author} {\bibnamefont{\mbox{ATLAS Collaboration}}},\ }%
  \bibfield{journal}{%
  \bibinfo {journal} {{ATLAS-CONF-2012-043~}}}%
   (\bibinfo {year} {2012}),\ \bibinfo {note}
  {{https://cdsweb.cern.ch/record/1435197}}%
  \bibAnnoteFile{NoStop}{ATLAS-CONF-2012-043}%
\bibitem{Aad:2011mk}%
  \BibitemOpen
  \bibfield{author}{%
  \bibinfo {author} {\bibnamefont{\mbox{ATLAS Collaboration}}},\ }%
  \bibfield{journal}{%
  \bibinfo {journal} {Eur. Phys. J. C}\ }%
  \textbf{\bibinfo {volume} {72}},\ \bibinfo {pages} {1909} (\bibinfo {year}
  {2012}),\ \Eprint{http://arxiv.org/abs/1110.3174}{arXiv:1110.3174}%
  \bibAnnoteFile{NoStop}{Aad:2011mk}%
\bibitem{Rehermann:2010vq}%
  \BibitemOpen
  \bibfield{author}{%
  \bibinfo {author} {\bibfnamefont{K.}~\bibnamefont{Rehermann}}\ and\ \bibinfo
  {author} {\bibfnamefont{B.}~\bibnamefont{Tweedie}},\ }%
  \bibfield{journal}{%
  \Doi{10.1007/JHEP03(2011)059}{\bibinfo {journal} {J. High Energy Phys.}}\ }%
  \textbf{\bibinfo {volume} {1103}},\ \bibinfo {pages} {059} (\bibinfo {year}
  {2011}),\ \Eprint{http://arxiv.org/abs/1007.2221}{arXiv:1007.2221}%
  \bibAnnoteFile{NoStop}{Rehermann:2010vq}%
\bibitem{ATL-PHYS-PUB-2010-008}%
  \BibitemOpen
  \bibfield{author}{%
  \bibinfo {author} {\bibnamefont{\mbox{ATLAS Collaboration}}},\ }%
  \bibfield{journal}{%
  \bibinfo {journal} {{ATL-PHYS-PUB-2010-008 }}}%
   (\bibinfo {year} {2010}),\ \bibinfo {note}
  {{https://cdsweb.cern.ch/record/1278454}}%
  \bibAnnoteFile{NoStop}{ATL-PHYS-PUB-2010-008}%
\bibitem{Aad:2012re}%
  \BibitemOpen
  \bibfield{author}{%
  \bibinfo {author} {\bibnamefont{\mbox{ATLAS Collaboration}}},\ }%
  \bibfield{journal}{%
  \Doi{10.1140/epjc/s10052-011-1844-6}{\bibinfo {journal} {Eur. Phys. J. C}}\
  }%
  \textbf{\bibinfo {volume} {72}},\ \bibinfo {pages} {1844} (\bibinfo {year}
  {2012}),\ \Eprint{http://arxiv.org/abs/1108.5602}{arXiv:1108.5602}%
  \bibAnnoteFile{NoStop}{Aad:2012re}%
\bibitem{Aad:2012ef}%
  \BibitemOpen
  \bibfield{author}{%
  \bibinfo {author} {\bibnamefont{\mbox{ATLAS Collaboration}}},\ }%
  \bibfield{journal}{%
  \bibinfo {journal} {J. High Energy Phys.}\ }%
  \textbf{\bibinfo {volume} {1205}},\ \bibinfo {pages} {128} (\bibinfo {year}
  {2012}),\ \Eprint{http://arxiv.org/abs/1203.4606}{arXiv:1203.4606}%
  \bibAnnoteFile{NoStop}{Aad:2012ef}%
\bibitem{Catani:1993hr}%
  \BibitemOpen
  \bibfield{author}{%
  \bibinfo {author} {\bibfnamefont{S.}~\bibnamefont{Catani}}, \bibinfo {author}
  {\bibfnamefont{Y.~L.}\ \bibnamefont{Dokshitzer}}, \bibinfo {author}
  {\bibfnamefont{M.~H.}\ \bibnamefont{Seymour}},\ and\ \bibinfo {author}
  {\bibfnamefont{B.~R.}\ \bibnamefont{Webber}},\ }%
  \bibfield{journal}{%
  \Doi{10.1016/0550-3213(93)90166-M}{\bibinfo {journal} {Nucl. Phys. B}}\ }%
  \textbf{\bibinfo {volume} {406}},\ \bibinfo {pages} {187} (\bibinfo {year}
  {1993})%
  \bibAnnoteFile{NoStop}{Catani:1993hr}%
\bibitem{Ellis:1993tq}%
  \BibitemOpen
  \bibfield{author}{%
  \bibinfo {author} {\bibfnamefont{S.~D.}\ \bibnamefont{Ellis}}\ and\ \bibinfo
  {author} {\bibfnamefont{D.~E.}\ \bibnamefont{Soper}},\ }%
  \bibfield{journal}{%
  \Doi{10.1103/PhysRevD.48.3160}{\bibinfo {journal} {Phys. Rev. D}}\ }%
  \textbf{\bibinfo {volume} {48}},\ \bibinfo {pages} {3160} (\bibinfo {year}
  {1993}),\ \Eprint{http://arxiv.org/abs/hep-ph/9305266}{arXiv:hep-ph/9305266}%
  \bibAnnoteFile{NoStop}{Ellis:1993tq}%
\bibitem{Aaltonen:2010jr}%
  \BibitemOpen
  \bibfield{author}{%
  \bibinfo {author} {\bibnamefont{{CDF Collaboration, T.~Aaltonen
  \emph{et~al}.}}},\ }%
  \bibfield{journal}{%
  \Doi{10.1103/PhysRevD.82.112005}{\bibinfo {journal} {Phys. Rev. D}}\ }%
  \textbf{\bibinfo {volume} {82}},\ \bibinfo {pages} {112005} (\bibinfo {year}
  {2010}),\ \Eprint{http://arxiv.org/abs/1004.1181}{arXiv:1004.1181}%
  \bibAnnoteFile{NoStop}{Aaltonen:2010jr}%
\bibitem{Aad:2012ux}%
  \BibitemOpen
  \bibfield{author}{%
  \bibinfo {author} {\bibnamefont{\mbox{ATLAS Collaboration}}},\ }%
  \bibfield{journal}{%
  \bibinfo {journal} {Phys. Lett. B}\ }%
  \textbf{\bibinfo {volume} {717}},\ \bibinfo {pages} {330} (\bibinfo {year}
  {2012}),\ \Eprint{http://arxiv.org/abs/1205.3130}{arXiv:1205.3130}%
  \bibAnnoteFile{NoStop}{Aad:2012ux}%
\bibitem{Note3}%
  \BibitemOpen
  \bibinfo {note} {The values used are $m_W = 83.2$~\protect \textrm {Ge\kern
  -0.1em V}, $m_{t_{h}-W} = 90.9$~\protect \textrm {Ge\kern -0.1em V},
  $m_{t_{\ell }} = 167.6$~\protect \textrm {Ge\kern -0.1em V}, $\sigma _W =
  10.7$~\protect \textrm {Ge\kern -0.1em V}, $\sigma _{t_{h}-W} =
  12.8$~\protect \textrm {Ge\kern -0.1em V}, $\sigma _{t_{\ell }} =
  20.5$~\protect \textrm {Ge\kern -0.1em V}, $p_{\protect \mathrm
  T,t_{h}}-p_{\protect \mathrm T,t_{\ell }} =-7.4$~\protect \textrm {Ge\kern
  -0.1em V}\ and $\sigma _{\protect \mathrm {diff}\protect \ensuremath
  {p_{\protect \mathrm {T}}}} =64.0$~\protect \textrm {Ge\kern -0.1em V}.}%
  \bibAnnoteFile{Stop}{Note3}%
\bibitem{ATLAS:2012an}%
  \BibitemOpen
  \bibfield{author}{%
  \bibinfo {author} {\bibnamefont{\mbox{ATLAS Collaboration}}},\ }%
  \bibfield{journal}{%
  \Doi{10.1140/epjc/s10052-012-2039-5}{\bibinfo {journal} {Eur. Phys. J. C}}\
  }%
  \textbf{\bibinfo {volume} {72}},\ \bibinfo {pages} {2039} (\bibinfo {year}
  {2012}),\ \Eprint{http://arxiv.org/abs/1203.4211}{arXiv:1203.4211}%
  \bibAnnoteFile{NoStop}{ATLAS:2012an}%
\bibitem{Aad:2012hg}%
  \BibitemOpen
  \bibfield{author}{%
  \bibinfo {author} {\bibnamefont{\mbox{ATLAS Collaboration}}},\ }%
  \bibfield{journal}{%
  \Doi{10.1140/epjc/s10052-012-2261-1}{\bibinfo {journal} {Eur. Phys. J. C}}\
  }%
  \textbf{\bibinfo {volume} {73}},\ \bibinfo {pages} {2261} (\bibinfo {year}
  {2013}),\ \Eprint{http://arxiv.org/abs/1207.5644}{arXiv:1207.5644}%
  \bibAnnoteFile{NoStop}{Aad:2012hg}%
\bibitem{Aad:2011kp}%
  \BibitemOpen
  \bibfield{author}{%
  \bibinfo {author} {\bibnamefont{\mbox{ATLAS Collaboration}}},\ }%
  \bibfield{journal}{%
  \Doi{10.1016/j.physletb.2011.12.046}{\bibinfo {journal} {Phys. Lett. B}}\ }%
  \textbf{\bibinfo {volume} {707}},\ \bibinfo {pages} {418} (\bibinfo {year}
  {2012}),\ \Eprint{http://arxiv.org/abs/1109.1470}{arXiv:1109.1470}%
  \bibAnnoteFile{NoStop}{Aad:2011kp}%
\bibitem{Martin:2009iq}%
  \BibitemOpen
  \bibfield{author}{%
  \bibinfo {author} {\bibfnamefont{A.}~\bibnamefont{Martin}}, \bibinfo {author}
  {\bibfnamefont{W.}~\bibnamefont{Stirling}}, \bibinfo {author}
  {\bibfnamefont{R.}~\bibnamefont{Thorne}},\ and\ \bibinfo {author}
  {\bibfnamefont{G.}~\bibnamefont{Watt}},\ }%
  \bibfield{journal}{%
  \Doi{10.1140/epjc/s10052-009-1072-5}{\bibinfo {journal} {Eur. Phys. J. C}}\
  }%
  \textbf{\bibinfo {volume} {63}},\ \bibinfo {pages} {189} (\bibinfo {year}
  {2009}),\ \Eprint{http://arxiv.org/abs/0901.0002}{arXiv:0901.0002}%
  \bibAnnoteFile{NoStop}{Martin:2009iq}%
\bibitem{Martin:2009bu}%
  \BibitemOpen
  \bibfield{author}{%
  \bibinfo {author} {\bibfnamefont{A.}~\bibnamefont{Martin}}, \bibinfo {author}
  {\bibfnamefont{W.}~\bibnamefont{Stirling}}, \bibinfo {author}
  {\bibfnamefont{R.}~\bibnamefont{Thorne}},\ and\ \bibinfo {author}
  {\bibfnamefont{G.}~\bibnamefont{Watt}},\ }%
  \bibfield{journal}{%
  \Doi{10.1140/epjc/s10052-009-1164-2}{\bibinfo {journal} {Eur. Phys. J. C}}\
  }%
  \textbf{\bibinfo {volume} {64}},\ \bibinfo {pages} {653} (\bibinfo {year}
  {2009}),\ \Eprint{http://arxiv.org/abs/0905.3531}{arXiv:0905.3531}%
  \bibAnnoteFile{NoStop}{Martin:2009bu}%
\bibitem{Cacciari:2011hy}%
  \BibitemOpen
  \bibfield{author}{%
  \bibinfo {author} {\bibfnamefont{M.}~\bibnamefont{Cacciari}} \emph{et~al.},\
  }%
  \bibfield{journal}{%
  \Doi{10.1016/j.physletb.2012.03.013}{\bibinfo {journal} {Phys. Lett. B}}\ }%
  \textbf{\bibinfo {volume} {710}},\ \bibinfo {pages} {612} (\bibinfo {year}
  {2012}),\ \Eprint{http://arxiv.org/abs/1111.5869}{arXiv:1111.5869}%
  \bibAnnoteFile{NoStop}{Cacciari:2011hy}%
\bibitem{Czakon:2011xx}%
  \BibitemOpen
  \bibfield{author}{%
  \bibinfo {author} {\bibnamefont{{M.~Czakon and A.~Mitov,}}}\ }%
  \Eprint{http://arxiv.org/abs/1112.5675}{arXiv:1112.5675}%
  \bibAnnoteFile{NoStop}{Czakon:2011xx}%
\bibitem{Aad:2013ucp}%
  \BibitemOpen
  \bibfield{author}{%
  \bibinfo {author} {\bibnamefont{\mbox{ATLAS Collaboration.}}}\ }%
  \bibinfo {note} {{Submitted to Eur. Phys. J. C}},\
  \Eprint{http://arxiv.org/abs/1302.4393}{arXiv:1302.4393}%
  \bibAnnoteFile{NoStop}{Aad:2013ucp}%
\bibitem{Ball:2012cx}%
  \BibitemOpen
  \bibfield{author}{%
  \bibinfo {author} {\bibfnamefont{R.~D.}\ \bibnamefont{Ball}} \emph{et~al.},\
  }%
  \bibfield{journal}{%
  \Doi{10.1016/j.nuclphysb.2012.10.003}{\bibinfo {journal} {Nucl. Phys. B}}\ }%
  \textbf{\bibinfo {volume} {867}},\ \bibinfo {pages} {244} (\bibinfo {year}
  {2013}),\ \Eprint{http://arxiv.org/abs/1207.1303}{arXiv:1207.1303}%
  \bibAnnoteFile{NoStop}{Ball:2012cx}%
\bibitem{Note4}%
  \BibitemOpen
  \bibinfo {note} {The CT10 PDF uncertainties are scaled down by a factor
  1.6645 to reach the 68\% confidence-level.}%
  \bibAnnoteFile{Stop}{Note4}%
\bibitem{Botje:2011sn}%
  \BibitemOpen
  \bibfield{author}{%
  \bibinfo {author} {\bibnamefont{{M.~Botje \emph{et al.},}}}\ }%
  \Eprint{http://arxiv.org/abs/1101.0538}{arXiv:1101.0538}%
  \bibAnnoteFile{NoStop}{Botje:2011sn}%
\bibitem{ATLAS-CONF-2012-040}%
  \BibitemOpen
  \bibfield{author}{%
  \bibinfo {author} {\bibnamefont{\mbox{ATLAS Collaboration}}},\ }%
  \bibfield{journal}{%
  \bibinfo {journal} {{ATLAS-CONF-2012-040~}}}%
   (\bibinfo {year} {2012}),\ \bibinfo {note}
  {{https://cdsweb.cern.ch/record/1435194}}%
  \bibAnnoteFile{NoStop}{ATLAS-CONF-2012-040}%
\bibitem{Note5}%
  \BibitemOpen
  \bibinfo {note} {The additional $b$-tagging uncertainty is an extrapolation
  of uncertainty from regions of lower \protect \ensuremath {p_{\protect
  \mathrm {T}}}, and it is approximately 12\% for $b$-jets and 17\% for
  $c$-jets, added in quadrature with the jet efficiency correction factor for
  the 140--200 \protect \textrm {Ge\kern -0.1em V}\ region.}%
  \bibAnnoteFile{Stop}{Note5}%
\bibitem{ATLAS:2012al}%
  \BibitemOpen
  \bibfield{author}{%
  \bibinfo {author} {\bibnamefont{\mbox{ATLAS Collaboration}}},\ }%
  \bibfield{journal}{%
  \Doi{10.1140/epjc/s10052-012-2043-9}{\bibinfo {journal} {Eur. Phys. J. C}}\
  }%
  \textbf{\bibinfo {volume} {72}},\ \bibinfo {pages} {2043} (\bibinfo {year}
  {2012}),\ \Eprint{http://arxiv.org/abs/1203.5015}{arXiv:1203.5015}%
  \bibAnnoteFile{NoStop}{ATLAS:2012al}%
\bibitem{Manohar:2012rs}%
  \BibitemOpen
  \bibfield{author}{%
  \bibinfo {author} {\bibfnamefont{A.~V.}\ \bibnamefont{Manohar}}\ and\
  \bibinfo {author} {\bibfnamefont{M.}~\bibnamefont{Trott}},\ }%
  \bibfield{journal}{%
  \Doi{10.1016/j.physletb.2012.04.013}{\bibinfo {journal} {Phys. Lett. B}}\ }%
  \textbf{\bibinfo {volume} {711}},\ \bibinfo {pages} {313} (\bibinfo {year}
  {2012}),\ \Eprint{http://arxiv.org/abs/1201.3926}{arXiv:1201.3926}%
  \bibAnnoteFile{NoStop}{Manohar:2012rs}%
\bibitem{Choudalakis:2011qn}%
  \BibitemOpen
  \bibfield{author}{%
  \bibinfo {author} {\bibnamefont{{G.~Choudalakis,}}}\ }%
  \Eprint{http://arxiv.org/abs/1101.0390}{arXiv:1101.0390}%
  \bibAnnoteFile{NoStop}{Choudalakis:2011qn}%
\bibitem{Bertram:2000br}%
  \BibitemOpen
  \bibfield{author}{%
  \bibinfo {author} {\bibfnamefont{I.}~\bibnamefont{Bertram}} \emph{et~al.},\
  }%
  \bibfield{journal}{%
  \bibinfo {journal} {FERMILAB-TM-2104, D0-NOTE-3476, D0-NOTE-2775-A}}%
   (\bibinfo {year} {2000}),\ \bibinfo {note} {{FERMILAB-TM-2104}}%
  \bibAnnoteFile{NoStop}{Bertram:2000br}%
\bibitem{Berger_prior:2009}%
  \BibitemOpen
  \bibfield{author}{%
  \bibinfo {author} {\bibfnamefont{J.~O.}\ \bibnamefont{Berger}}, \bibinfo
  {author} {\bibfnamefont{J.~M.}\ \bibnamefont{Bernardo}},\ and\ \bibinfo
  {author} {\bibfnamefont{D.}~\bibnamefont{Sun}},\ }%
  \bibfield{journal}{%
  \Doi{10.1214/07-AOS587}{\bibinfo {journal} {Ann. Stat.}}\ }%
  \textbf{\bibinfo {volume} {37}},\ \bibinfo {pages} {905} (\bibinfo {year}
  {2009}),\ \Eprint{http://arxiv.org/abs/0904.0156}{arXiv:0904.0156}%
  \bibAnnoteFile{NoStop}{Berger_prior:2009}%
\bibitem{Casadei:2011hx}%
  \BibitemOpen
  \bibfield{author}{%
  \bibinfo {author} {\bibfnamefont{D.}~\bibnamefont{Casadei}},\ }%
  \bibfield{journal}{%
  \Doi{10.1088/1748-0221/7/01/P01012}{\bibinfo {journal} {JINST}}\ }%
  \textbf{\bibinfo {volume} {7}},\ \bibinfo {pages} {P01012} (\bibinfo {year}
  {2012}),\ \Eprint{http://arxiv.org/abs/1108.4270}{arXiv:1108.4270}%
  \bibAnnoteFile{NoStop}{Casadei:2011hx}%
\bibitem{Junk:1999kv}%
  \BibitemOpen
  \bibfield{author}{%
  \bibinfo {author} {\bibfnamefont{T.}~\bibnamefont{Junk}},\ }%
  \bibfield{journal}{%
  \Doi{10.1016/S0168-9002(99)00498-2}{\bibinfo {journal} {Nucl. Instrum.
  Methods Phys. Res., Sect. A}}\ }%
  \textbf{\bibinfo {volume} {434}},\ \bibinfo {pages} {435} (\bibinfo {year}
  {1999}),\ \Eprint{http://arxiv.org/abs/hep-ex/9902006}{arXiv:hep-ex/9902006}%
  \bibAnnoteFile{NoStop}{Junk:1999kv}%
\bibitem{Read:2002hq}%
  \BibitemOpen
  \bibfield{author}{%
  \bibinfo {author} {\bibfnamefont{A.~L.}\ \bibnamefont{Read}},\ }%
  \bibfield{journal}{%
  \Doi{10.1088/0954-3899/28/10/313}{\bibinfo {journal} {J. Phys. G}}\ }%
  \textbf{\bibinfo {volume} {28}},\ \bibinfo {pages} {2693} (\bibinfo {year}
  {2002})%
  \bibAnnoteFile{NoStop}{Read:2002hq}%
\bibitem{Caola:2012rs}%
  \BibitemOpen
  \bibfield{author}{%
  \bibinfo {author} {\bibnamefont{{F.~Caola, K.~Melnikov, and M.~Schulze,}}}\
  }%
  \Eprint{http://arxiv.org/abs/1211.6387}{arXiv:1211.6387}%
  \bibAnnoteFile{NoStop}{Caola:2012rs}%
\end{thebibliography}%

\clearpage

\onecolumngrid

\begin{flushleft}
{\Large The ATLAS Collaboration}

\bigskip

G.~Aad$^{\rm 48}$,
T.~Abajyan$^{\rm 21}$,
B.~Abbott$^{\rm 111}$,
J.~Abdallah$^{\rm 12}$,
S.~Abdel~Khalek$^{\rm 115}$,
A.A.~Abdelalim$^{\rm 49}$,
O.~Abdinov$^{\rm 11}$,
R.~Aben$^{\rm 105}$,
B.~Abi$^{\rm 112}$,
M.~Abolins$^{\rm 88}$
% ...
\bigskip

% ...
$^{11}$ Institute of Physics, Azerbaijan Academy of Sciences, Baku, Azerbaijan\\
$^{12}$ Institut de F\'{i}sica d'Altes Energies and Departament de F\'{i}sica de la Universitat Aut\`{o}noma de Barcelona and ICREA, Barcelona, Spain\\
% ...
$^{21}$ Physikalisches Institut, University of Bonn, Bonn, Germany\\
% ...
$^{48}$ Fakult\"{a}t f\"{u}r Mathematik und Physik, Albert-Ludwigs-Universit\"{a}t, Freiburg, Germany\\
$^{49}$ Section de Physique, Universit\'{e} de Gen\`{e}ve, Geneva, Switzerland\\
% ...
$^{88}$ Department of Physics and Astronomy, Michigan State University, East Lansing MI, United States of America\\
% ...
$^{105}$ Nikhef National Institute for Subatomic Physics and University of Amsterdam, Amsterdam, Netherlands\\
% ...
$^{111}$ Homer L. Dodge Department of Physics and Astronomy, University of Oklahoma, Norman OK, United States of America\\
$^{112}$ Department of Physics, Oklahoma State University, Stillwater OK, United States of America\\
% ...
$^{115}$ LAL, Universit\'{e} Paris-Sud and CNRS/IN2P3, Orsay, France\\
% ...
\end{flushleft}


\end{document}
%